\documentclass[aps,pra,reprint,showkeys,amssymb,floatfix,longbibliography,superscriptaddress]{revtex4-2}
\usepackage{graphicx}
\usepackage{amsmath,amsfonts,amssymb}
\usepackage{braket}
\usepackage{xcolor}
\usepackage{tikz}
\usepackage{color}
\usepackage{enumerate}
\usepackage[caption=false]{subfig}
\usepackage{multirow}
\usepackage{qcircuit}
\usepackage[normalem]{ulem}
\usepackage{placeins}
\usepackage{url}
\usepackage{dsfont}
\usepackage{braket}
\usepackage[multidot]{grffile}
\usepackage{float}
\usepackage{lipsum}
\usepackage{afterpage}
\usepackage{epstopdf}
\usepackage{silence}
\usepackage{longtable}
\usepackage{capt-of}
\usepackage{tikz-cd}
\usepackage{epigraph}
\usepackage{csquotes}
\usepackage[RPvoltages]{circuitikz}
\usepackage{adjustbox}
\usetikzlibrary{positioning, fit, calc}
\usetikzlibrary{shapes}
\usetikzlibrary{plotmarks}
\usepackage[colorlinks=true,urlcolor=blue,linkcolor=blue,citecolor=blue,breaklinks=true,pdfauthor={Hannes Lagemann}]{hyperref}
\newcommand{\appref}[1]{Appendix~\ref{#1}}
\newcommand{\secref}[1]{Sec.~\ref{#1}}

\newcommand{\secsref}[2]{Secs.~\ref{#1}--\ref{#2}}

\newcommand{\figref}[1]{Fig.~\ref{#1}}
\newcommand{\figsref}[1]{Figs.~\ref{#1}}
\newcommand{\tabref}[1]{Table~\ref{#1}}
\newcommand{\tabaref}[2]{Tables~\ref{#1} and \ref{#2}}
\newcommand{\tabaaref}[3]{Tables~\ref{#1}, \ref{#2} and \ref{#3}}
\newcommand{\tabsref}[2]{Tables~\ref{#1}--\ref{#2}}
\newcommand{\equref}[1]{Eq.~\eqref{#1}}
\newcommand{\equaref}[2]{Eqs.~\eqref{#1} and \eqref{#2}}

\newcommand{\ketbra}[2]{\ket{#1}\!\!\bra{#2}}

\newcommand{\change}[1]{\textcolor{black}{#1}}
\newcommand{\changetwo}[1]{\textcolor{black}{#1}}
\newcommand{\changethree}[1]{\textcolor{black}{#1}}

\newcommand{\ie}{i.e.~}
\newcommand{\eg}{e.g.~}

\newcommand{\REF}{Ref.~}
\newcommand{\REFS}{Refs.~}
\newcommand{\brr}[1]{\left(#1\right)}

\newcommand{\var}{(t)}
\newcommand{\op}[1]{\hat{#1}}
\newcommand{\state}{\ket{\Psi\var}}
\newcommand{\norm}[1]{\lVert#1\lVert}
\newcommand{\tens}[1]{\mathbin{\mathop{\otimes}\limits_{#1}}}

\newcommand{\ith}{$i^{\text{th}}$}
\newcommand{\trace}{\text{Tr}}

\newcommand{\hold}{0}

\newcommand{\FQ}{\Phi_{0}}
\newcommand{\ROT}{R^{(x)}}
\newcommand{\CNOT}{\text{CNOT}}
\newcommand{\CZ}{\text{CZ}}
\newcommand{\TP}{2 \pi}
\newcommand{\HA}{\text{H}}

\begin{document}
\title{On the fragility of gate-error metrics in simulation models of flux-tunable transmon quantum computers}
\author{H. Lagemann}
\email[Corresponding author: ]{hannes.a.l@me.com}
\thanks{\\Published as \href{https://journals.aps.org/pra/abstract/10.1103/PhysRevA.108.022604}{Phys. Rev. A 108, 022604 (2023)};}
\affiliation{J\"ulich Supercomputing Centre, Institute for Advanced Simulation, Forschungszentrum J\"ulich, 52425 J\"ulich, Germany}
\affiliation{RWTH Aachen University, D-52056 Aachen, Germany}
\author{D. Willsch}
\affiliation{J\"ulich Supercomputing Centre, Institute for Advanced Simulation, Forschungszentrum J\"ulich, 52425 J\"ulich, Germany}
\author{M. Willsch}
\affiliation{J\"ulich Supercomputing Centre, Institute for Advanced Simulation, Forschungszentrum J\"ulich, 52425 J\"ulich, Germany}
\affiliation{AIDAS, 52425 J\"ulich, Germany}
\author{F. Jin}
\affiliation{J\"ulich Supercomputing Centre, Institute for Advanced Simulation, Forschungszentrum J\"ulich, 52425 J\"ulich, Germany}
\author{H. De Raedt}
\affiliation{J\"ulich Supercomputing Centre, Institute for Advanced Simulation, Forschungszentrum J\"ulich, 52425 J\"ulich, Germany}
\affiliation{Zernike Institute for Advanced Materials,\\
University of Groningen, Nijenborgh 4, NL-9747 AG Groningen, The Netherlands}
\author{K. Michielsen}
\affiliation{J\"ulich Supercomputing Centre, Institute for Advanced Simulation, Forschungszentrum J\"ulich, 52425 J\"ulich, Germany}
\affiliation{RWTH Aachen University, D-52056 Aachen, Germany}
\affiliation{AIDAS, 52425 J\"ulich, Germany}

\date{\today}

\begin{abstract}
 Constructing a quantum computer requires immensely precise control over a quantum system. A lack of precision is often quantified by gate-error metrics, such as the average infidelity or the diamond distance. However, usually such gate-error metrics are only considered for individual gates, and not the errors that accumulate over consecutive gates. Furthermore, it is not well known how susceptible the metrics are to the assumptions which make up the model. Here, we investigate these issues using realistic simulation models of quantum computers with flux-tunable transmons and coupling resonators. \changethree{Our main findings reveal that (1) gate-error metrics are indeed affected by the many assumptions of the model, (2) consecutive gate errors do not accumulate linearly, and (3) gate-error metrics are poor predictors for the performance of consecutive gates. Additionally, we discuss a potential limitation in the scalability of the studied device architecture.}
\end{abstract}

\maketitle

\section{Introduction}\label{sec:Introduction}
The realisation of a gate-based quantum computer in the real world is an engineering task which requires a tremendous amount of precision in terms of control over a quantum system. In this work we will clearly differentiate between the following three concepts: the ideal gate-based quantum computer (IGQC), the prototype gate-based quantum computers (PGQCs) and the non-ideal gate-based quantum computers (NIGQCs).

The state of an IGQC with $N \in \mathbb{N}$ qubits is described by a time-independent state vector $\ket{\psi}$ in a $2^{N}$ dimensional Hilbert space as described in \REFS\cite{Nielsen:2011:QCQ:1972505,Watrous2018}. In this model changes of the state vector occur instantaneously by applying unitary operators $\op{U}$ to the state vector $\ket{\psi}$. Therefore, all time-dependent aspects of gate errors which are omnipresent on PGQCs, see \REF\cite{Wi17,Michielsen17}, are neglected in this model. One often uses quantum operations, see \REFS\cite{Nielsen:2011:QCQ:1972505,Watrous2018}, to introduce gate errors to the IGQC model. Many gate-error metrics like the average gate infidelity, see \REFS\cite{Nielsen2002,Jin21} and the diamond distance, see \REFS\cite{Kitaev1997,Sanders2015}, can be expressed in terms of quantum operations. However, also quantum operations do not describe the time-dependent aspects of gate errors.

In this work, we are interested in modelling the appearance of gate errors as a real-time process. This type of modelling is motivated by the fact that superconducting PGQCs are physical systems and consequently one finds that these devices are inherently dynamic and affected by various internal and external factors. A review of the literature suggests that we have to take into account the variable control signals, see \REF\cite{Rol19,Werninghaus2021}, the temperature of the device, see \REF\cite{Krinner2019}, temporal stability of device parameters like the qubit frequency, see \REF\cite{Burnett2019}, cosmic radiation, see \REF\cite{McEwen22} and so on. Therefore, a complete mathematical description of these devices which takes into account all the relevant factors in one mathematical model seems prohibitive. However, we may be able to describe certain aspects of a superconducting PGQC by making use of simplified models. The simulations we perform to obtain the results presented in this work are intended to describe certain aspects of specific two-qubit, three-qubit and four-qubit superconducting PGQCs. The device architecture, device parameters and control pulses we use are similar to the ones used in experiments described in \REFS\cite{Lacroix2020,Krinner2020,Andersen20}.

We define a NIGQC model as a model where the state of a gate-based quantum computer is described as a real-time entity. In this work, the state of a NIGQC is by assumption completely determined by the state vector $\state$ and the dynamics of the system are generated by the \changetwo{time-dependent Schr\"odinger equation} with $\hbar=1$
\begin{equation}\label{eq:TDSE}
  i\partial_{t}\state=\op{H}\var\state,
\end{equation}
where $\op{H}\var=\op{H}^{\dagger}\var$ denotes a circuit Hamiltonian which is obtained by means of the lumped-element approximation, see \REF\cite[Section 1.4]{Balanis12} and an associated effective Hamiltonian, see \REF\cite{Lagemann21}. Note that the NIGQC models that we consider are idealised versions of the PGQCs used in experiments.

Our simulation software solves the \changetwo{time-dependent Schr\"odinger equation} numerically with the product-formula algorithm, see \REFS\cite{DeRaedt87,Huyghebaert90}, for a given Hamiltonian $\op{H}\var$ and a sequence of control pulses which allows us to implement a sequence of gates with our NIGQC. This enables us to compute the gate errors for the individual gates in the program sequence. In the following, we call the sequence of gate errors which arises from this procedure a \emph{gate-error trajectory}.

This manuscript is structured as follows. First, in \secref{sec:ModelsAndDeviceParameters} we specify the \changetwo{Hamiltonians} $\op{H}\var$, the device parameters and the control pulses we use to implement our NIGQC models. Second, in \secref{sec:ErrorMetrics} we discuss the gate-error metrics and how we compute them. Third, in \secref{sec:Results} we present our findings. In \secref{sec:SpectrumAnalysis}, we discuss the spectrum of a four-qubit NIGQC and its relevance for the gate-error metrics we compute. Next, in \secref{sec:GateMetricsAndControlPulseParameters}, we discuss the gate-error metrics we obtained by optimising the parameters of the control pulses. Then, in \secref{sec:Influence of higher states on gate-error trajectories} we show how gate-errors and the corresponding gate-error trajectories which arise over time are affected by modelling the dynamics of the system with different numbers of basis states. The gate errors in this section are obtained with the circuit Hamiltonian. Next, in \secref{sec:Influence of small pulse parameter deviations on gate-error trajectories} we show how deviations in the control pulse parameters affect the gate-error trajectories. The gate errors in this section are also obtained with the circuit Hamiltonian. Then, in \secref{sec:InfluenceOfTheAdiabaticApproximationOnGate-errorTrajectories} we show how the modelling of flux-tunable transmons as adiabatic and \changetwo{nonadiabatic} anharmonic oscillators affects the gate errors over time. The gate errors in this section are obtained with an effective Hamiltonian. A summary, discussion and the conclusions drawn are presented in \secref{sec:SummaryAndConclusions}. Note that we use $\hbar=1$ throughout this work.

To assist the reader in navigating through the material, we list the main findings:
\begin{enumerate}
  \item The authors of \REFS\cite{Wi17} concluded that the initial values for gate-error metrics like the diamond distance and the average infidelity are poor predictors of the future gate-error trajectories which emerge over time. In this work we provide new evidence which supports and strengthens this conclusion. Furthermore, we provide a concise theoretical explanation for why this is the case and how we should interpret gate-error metrics like the diamond distance and the average infidelity, see \secsref{sec:Influence of higher states on gate-error trajectories}{sec:InfluenceOfTheAdiabaticApproximationOnGate-errorTrajectories}.

  \item By analysing the spectrum of a four-qubit NIGQC, we discuss a problem which potentially limits the up-scaling capabilities of device architectures which implement two-qubit gates by tuning the energies of basis states into resonance with one another. The issues we discuss implicate an exponentially large optimisation problem for future quantum computer engineers, see \secref{sec:SpectrumAnalysis}.

  \item By simulating the time evolution of different NIGQC models, we explicitly show that even seemingly small changes in the assumptions which make up the underlying model can substantially affect the gate-error metrics. The fact itself is not surprising. However, the extent to which the changes affect the gate-error metrics during the course of the time evolution of the system is something worth knowing, see \secsref{sec:Influence of higher states on gate-error trajectories}{sec:InfluenceOfTheAdiabaticApproximationOnGate-errorTrajectories}.

  \item By executing simple gate repetition programs on different NIGQCs, we show that gate-error metrics in NIGQC models do not behave linearly. Furthermore, we also show that gate-error metrics respond to changes in the model in a non-linear manner, see \secsref{sec:Influence of higher states on gate-error trajectories}{sec:InfluenceOfTheAdiabaticApproximationOnGate-errorTrajectories}. Given these findings we conjecture that gate errors for consecutive gates are not simply given by the sum of the gate errors for the individual gates in the program sequence but emerge due to a complex interplay of small deviations with respect to the target gates which occur over time.
\end{enumerate}


\section{Models and device parameters}\label{sec:ModelsAndDeviceParameters}
\begin{figure*}[tbp!]
\renewcommand{\hold}{1.5}
\centering
\begin{minipage}{\textwidth}
    \centering
    \includegraphics[scale=\hold]{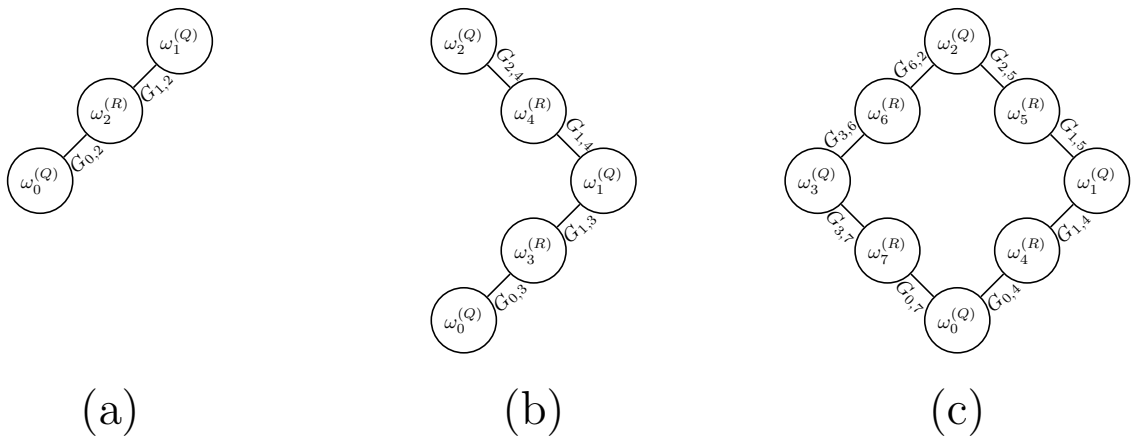}
  \end{minipage}
\caption{Illustrations of flux-tunable transmon qubits with park qubit frequencies $\omega^{(Q)}$ and resonators with resonance frequencies $\omega^{(R)}$ which constitute non-ideal gate-based transmon quantum computers (NIGQCs) with \changetwo{$N=2$(a), $N=3$(b) and $N=4$(c)} qubits. The different subsystems are indexed with the discrete variable $i \in \{0, ..,7\}$ such that the bare basis states of the system can be expressed as $\ket{\mathbf{z}}$, where $\mathbf{z}=(z_{|I|-1}, ..., z_{0})$ is an n-tuple  . The Hamiltonians we use to model the dynamics of our NIGQCs are given by \equaref{eq:Circuit_Ham_def}{eq:Eff_Ham_def}. The device parameters we use to specify the Hamiltonians are listed in \tabref{tab:device_parameters}. The control pulses we use to implement the gates are displayed in \figsref{fig:PulseTimeEvolution}(a-c).\label{fig:device_sketch}}
\end{figure*}
\begin{table*}[!tbp]
\caption{Device parameters for a four-qubit virtual chip. The device architecture and device parameters are motivated by experiments described in \REFS\cite{Lacroix2020,Krinner2020,Andersen20}. The index $i$ denotes the different circuit elements which are part of the system (in total there are eight circuit elements). The parameter $\omega_{i}^{(R)}$ denotes the coupler frequency, \ie the transmission line resonator frequency. Since all couplers have the same frequency, we only show one row for all coupling resonators, see the last row where $i\in\{4,5,6,7\}$. The constants $\omega_{i}^{(Q)}$ and $\alpha_{i}^{(Q)}$ denote the qubit frequency and anharmonicity, respectively. $E_{C_{i}}$ denotes the capacitive energy of the \ith qubit. Similarly, $E_{J_{i},r}$ and $E_{J_{i},l}$ denote the corresponding Josephson energies. \changetwo{Note that in this work we express the Hamiltonian in \equref{eq:Circuit_Ham} with a factor of $E_{C}$ instead of $4 E_{C}$. This means we adopt the original convention used in \REF\cite{Vion2002CPBqubitsQuantronium} and not the one of \REF\cite{Koch}}. The parameter $\varphi_{0,i}$ denotes the so-called flux offset, \ie the external flux $\varphi_{0,i}(t)$ at time $t=0$. The interaction strength $G_{i,j}$ between the different circuit elements is set to $300$ MHz for all $i,j \in \mathbb{N}^{0}$. All units (except the one of the flux offset) are in GHz. The flux offset is given in units of the flux quantum $\FQ$. We use these device parameters to model the two-qubit, three-qubit and four-qubit NIGQCs illustrated in \figsref{fig:device_sketch}(a-d). Note that we use $\hbar=1$ throughout this work.\label{tab:device_parameters}}
\begin{ruledtabular}
\begin{tabular}{c c c c c c c c c}
$i$&              $\omega_{i}^{(R)}/2\pi$&              $\omega_{i}^{(Q)}/2\pi$&              $\alpha_{i}^{(Q)}/2\pi$&              $E_{C_{i}}/2\pi$&              $E_{J_{i},l}/2\pi$&              $E_{J_{i},r}/2\pi$&              $\varphi_{0,i}/2\pi$&                             \\
\hline
0              &              $\text{n/a}$   &              $4.200$        &              $-0.320$       &              $1.068$        &              $2.355$        &              $7.064$        &              $0$            &                             \\

1              &              $\text{n/a}$   &              $5.200$        &              $-0.295$       &              $1.037$        &              $3.612$        &              $10.837$       &              $0$            &                             \\

2              &              $\text{n/a}$   &              $5.700$        &              $-0.285$       &              $1.017$        &              $4.374$        &              $13.122$       &              $0$            &                             \\

3              &              $\text{n/a}$   &              $4.960$        &              $-0.300$       &              $1.045$        &              $3.281$        &              $9.843$        &              $0$            &                             \\

$4-7$              &              $45.000$       &              $\text{n/a}$   &              $\text{n/a}$   &              $\text{n/a}$   &              $\text{n/a}$   &              $\text{n/a}$   &              $\text{n/a}$   &                             \\
\end{tabular}
\end{ruledtabular}
\end{table*}
Figures \ref{fig:device_sketch}(a-c) show illustrations of the two-qubit(a), three-qubit(b) and four-qubit(c) NIGQCs we consider in this work. The parameters $\omega^{(Q)}$ denote the park qubit frequencies for the different flux-tunable transmon qubits. The park qubit frequency is the frequency at which a transmon qubit resides if no external flux is applied. Similarly, the parameters $\omega^{(R)}$ refer to LC resonator frequencies. In the following, we will simply call these systems transmon qubits and resonators. The interactions between the different transmon qubits are conveyed by the resonators. We model the interactions between the different subsystems as dipole-dipole interactions and use the constants $G_{i,j}$, where $i,j \in \mathbb{N}^{0}$, to control the interaction strength. The device parameters listed in \tabref{tab:device_parameters} are used for all simulations in this work.

The remainder of this section is devoted to the models we use to describe the dynamics of the NIGQCs in \figsref{fig:device_sketch}(a-c). In \secref{sec:TheCircuitHamiltonianModel} we introduce a circuit Hamiltonian. In \secref{sec:TheEffectiveHamiltonianModel} we introduce an effective Hamiltonian which is related to the circuit Hamiltonian. Finally, in \secref{sec:TheControlPulses}, we define the control pulses we use to implement the different gates.

The device architecture, device parameters and control pulses we use are motivated by experiments described in \REFS\cite{Lacroix2020,Krinner2020,Andersen20}.

\subsection{The circuit Hamiltonian model}\label{sec:TheCircuitHamiltonianModel}
\newcommand{\idxCIR}{\text{Cir.}}
\newcommand{\idxFFT}{\text{Fix.}}
\newcommand{\idxFTT}{\text{Tun.}}
\newcommand{\idxRES}{\text{Res.}}
\newcommand{\idxINT}{\text{Int.}}
The circuit Hamiltonian we use to model our NIGQCs, is defined as
\begin{equation}\label{eq:Circuit_Ham_def}
  \op{H}_{\idxCIR} =\op{H}_{\idxRES,\Sigma}+\op{H}_{\idxFTT,\Sigma}+\op{V}_{\idxINT}.
\end{equation}
The first term,
\begin{equation}\label{eq:Res_Ham}
    \op{H}_{\idxRES,\Sigma} =\sum_{k\in K} \omega_{k}^{(R)} \op{a}_{k}^{\dagger}\op{a}_{k},
\end{equation}
describes a collection of non-interacting resonators. Here $K \subseteq \mathbb{N}^{0}$ denotes an index set for the resonators and $\omega_{k}^{(R)}$ refers to the different resonator frequencies. The operators $\op{a}$ and $\op{a}^{\dagger}$ are the bosonic annihilation and creation  operators, respectively.

Similarly, the second term,
\begin{align}\label{eq:Circuit_Ham}
  \begin{split}
  \op{H}_{\idxFTT,\Sigma} &=\sum_{j\in J} \left( \right. E_{C_{j}} \brr{\op{n}_{j}-n_{j}(t) }^{2} + (\frac{1}{2}-\beta_{j}) \dot{\varphi}_{j}(t) \op{n}_{j}\\
    &- E_{J_{l,j}} \cos\brr{\op{\varphi}_{j}+ \beta_{j} \varphi_{j}(t)} \\
    &- E_{J_{r,j}} \cos\brr{\op{\varphi}_{j}+(\beta_{j}-1)\varphi_{j}(t)} \left. \right),
  \end{split}
\end{align}
describes a collection of non-interacting flux-tunable transmons. Here $J \subseteq \mathbb{N}^{0}$ denotes an index set for the transmon qubits. Furthermore, the parameters $E_{C_{j}}$, $E_{J_{l,j}}$ and $E_{J_{r,j}}$ denote the capacitive energies, the left Josephson energies and the right Josephson energies. \changetwo{Note that the Hamiltonian in \equref{eq:Circuit_Ham} is often expressed with a factor $4 E_{C}$ instead of $E_{C}$. In this work we adopt the original convention used in \REF\cite{Vion2002CPBqubitsQuantronium} and not the one of \REF\cite{Koch}}. The parameters $\beta_{j}$ are not device parameters but determine the variables we use for the quantisation of the circuit. We use $\beta_{j}=1/2$ for all simulations in this work, cf.~\REFS\cite{You,Riwar21}. In \appref{app:DerivationOfACircuitHamiltonianForFluxTunableTransmons} we provide a detailed derivation of the Hamiltonian for a single flux-tunable transmon with a charge drive term. This derivation is motivated by the work in \REF\cite{You}.

The device parameters in \tabref{tab:device_parameters} were obtained as follows. The Quantum Device Lab which is affiliated with the ETH Zurich provided us with the qubit frequencies $\omega_{j}^{(Q)}$, anharmonicities $\alpha_{j}^{(Q)}$ and asymmetry factors $d_{j}$ for the various transmon qubits in the system as well as the coupling resonator frequencies $\omega_{k}^{(R)}=45$ GHz. We then used the relations
\begin{subequations}
   \begin{align}
       \omega_{j}^{(Q)}&=(E_{j}^{(1)}-E_{j}^{(0)}),\\
       \alpha_{j}^{(Q)}&=(E_{j}^{(2)}-E_{j}^{(0)})-2 \omega_{j}^{(Q)},\\
       d_{j}&=\frac{ E_{J_{r,j}} -E_{J_{l,j}} }{ E_{J_{r,j}} + E_{J_{l,j}} },
   \end{align}
\end{subequations}
and Hamiltonian \equref{eq:Circuit_Ham} to fit the energy levels $E_{j}^{(0)}$, $E_{j}^{(1)}$ and $E_{j}^{(2)}$ to the provided data.

The third term
\begin{equation}\label{eq:Circuit_Int}
  \op{V}_{\idxINT}=\sum_{(k,j)\in K \times J} G_{k,j} \brr{\op{a}_{k}+\op{a}_{k}^{\dagger}} \tens{} \op{n}_{j},
\end{equation}
describes dipole-dipole interactions between resonators and transmon qubits. Here $G_{k,j}$ is a real-valued constant which is set to $300$ MHz for all simulations in this work. This value for interaction strength constants $G_{k,j}$ roughly reproduces the gate durations of around $100$ ns which were found to be appropriate in the experiment.

For our simulations of Hamiltonian \equref{eq:Circuit_Ham_def} we use the product-formula algorithm, see \REF\cite{DeRaedt87,Huyghebaert90} and what we call a bare basis. This bare basis is formed by the tensor product states of the harmonic oscillator and the transmon basis states, which are obtained for the external flux $\varphi_{j}(t)$ at time $t=0$. \changetwo{In this manuscript, we model the dynamics of Hamiltonian \equref{eq:Circuit_Ham_def} with four basis states for all resonators and most transmon qubits are modelled with sixteen basis states. Only the transmon qubit with the index $i=0$, see \figsref{fig:device_sketch}(a-c), is modelled with four basis states. Note that this transmon does not experience a flux drive. Therefore, we can model this transmon with four basis states only. There is one exception to this rule, namely the results in \secref{sec:Influence of higher states on gate-error trajectories} are obtained with four and sixteen basis states for all transmon qubits.}

\subsection{The effective Hamiltonian model}\label{sec:TheEffectiveHamiltonianModel}
\newcommand{\idxEFF}{\text{Eff.}}
\newcommand{\idxFTE}{\text{Tun.eff.}}
\newcommand{\idxFD}{\text{Flux}}
\newcommand{\idxCD}{\text{Charge}}
The approximations which are needed to obtain the effective Hamiltonian defined in this section \changetwo{are} discussed in detail in \REF\cite{Lagemann21}. Furthermore, the full effective model is defined in terms of twelve equations. In order to provide a concise discussion of the effective model, we only discuss the most relevant relations. A more detailed discussion of the effective model with all twelve equations can be found in \appref{app:TheEffectiveHamiltonianModel}.

The effective Hamiltonian we use to model our NIGQCs is defined by
\begin{equation}\label{eq:Eff_Ham_def}
  \op{H}_{\text{Eff.}}=\op{H}_{\idxRES,\Sigma}+\op{H}_{\idxFTE,\Sigma}+\op{D}_{\idxCD,\Sigma}+\op{\mathcal{D}}_{\idxFD,\Sigma}+\op{W}_{\idxINT}.
\end{equation}
The second term,
\begin{equation}\label{eq:Eff_Ham}
  \op{H}_{\idxFTE,\Sigma}=\sum_{j \in J} \omega_{j}^{(q)}(t) \op{b}_{j}^{\dagger}\op{b}_{j} + \frac{\alpha_{j}^{(q)}(t)}{2} \brr{\op{b}_{j}^{\dagger}\op{b}_{j}\brr{\op{b}_{j}^{\dagger}\op{b}_{j}-\op{I}}},
\end{equation}
describes a collection of non-interacting flux-tunable transmons which are modelled as adiabatic, anharmonic oscillators in a time-dependent basis. The operators $\op{b}$ and $\op{b}^{\dagger}$ are the bosonic annihilation and creation operators, respectively. The functions $\omega_{j}^{(q)}(t)$ and $\alpha_{j}^{(q)}(t)$, see \equaref{eq:tunable_freq}{eq:tunable_anharm}, denote the flux-tunable qubit frequency and anharmonicity, respectively. Note that the time dependence in both these functions is given by the external flux $\varphi_{j}(t)$. In this work, we use the series expansions obtained by the authors of \REF\cite{Didier} to approximate the lowest three energy levels of Hamiltonian \equref{eq:Circuit_Ham} for the external flux values $\varphi_{j}(t)$ at any given point in time.

The third term,
\begin{equation}\label{eq:drive_charge}
  \op{D}_{\idxCD,\Sigma}=\sum_{j \in J} \Omega_{j}(t) \brr{\op{b}_{j}^{\dagger} + \op{b}_{j}},
\end{equation}
describes a charge drive. Here $\Omega_{j}(t) \propto -2 E_{C_{j}} n_{j}(t)$ and we approximate the charge operators $\op{n}_{j}$ by effective charge operators $\op{n}_{j,\text{eff.}}$ which can be expressed in terms of the bosonic operators, see \REF\cite{Koch}.

The fourth term,
\begin{align}\label{eq:drive_flux}
  \begin{split}
    \op{\mathcal{D}}_{\idxFD,\Sigma}&=\sum_{j \in J}  \left( \right. - i \sqrt{\frac{\xi_{j}(t)}{2}} \dot{\varphi}_{\text{eff.},j}(t) \brr{\op{b}_{j}^{\dagger}- \op{b}_{j}}\\
    &+ \frac{i}{4} \frac{\dot{\xi}_{j}(t)}{\xi_{j}(t)} \brr{\op{b}_{j}^{\dagger} \op{b}_{j}^{\dagger} - \op{b}_{j} \op{b}_{j}} \left.\right) ,
  \end{split}
\end{align}
describes a \changetwo{nonadiabatic} flux drive. Here $\dot{\varphi}_{\text{eff.},j}(t) \propto \dot{\varphi}_{j}(t)$ and  $\dot{\xi}_{j}(t)/\xi_{j}(t) \propto \dot{\varphi}_{j}(t)$. Note that $\dot{\varphi}_{\text{eff.},j}(t)$ and $\xi_{j}(t)$ are given by \equref{eq:aux_sub}(a,c). This term results from the fact that we model the effective flux-tunable transmon in a time-dependent basis. Consequently, for the \changetwo{time-dependent Schr\"odinger equation} to stay form invariant, see \REF\cite{Lagemann21}, a time-dependent basis transformation term is needed.

The fifth term,
\begin{equation}\label{eq:Eff_Int}
    \op{W}_{\idxINT}=\sum_{(k,j)\in K \times J} g_{k,j}^{(a, b)}(t) \brr{\op{a}_{k}^{\dagger} + \op{a}_{k}} \tens{} \brr{\op{b}_{j}^{\dagger} + \op{b}_{j}},
\end{equation}
describes time-dependent dipole-dipole interactions. The time dependence of the interaction strength is a result of the fact that we model the effective flux-tunable transmon in a time-dependent basis, see \REF\cite{Lagemann21}. The function $g_{k,j}^{(a, b)}(t)$ is given by \equref{eq:time_dep_int_strength}. This time-dependent interaction strength model is motivated by the work in \REF\cite{Koch}.

As before, we use the product-formula algorithm, see \REF\cite{DeRaedt87,Huyghebaert90}, to solve the \changetwo{time-dependent Schr\"odinger equation} for the Hamiltonian \equref{eq:Eff_Ham_def}. Here we use a bare basis that is formed by the tensor product states of the time-independent (for the resonators) and time-dependent (for the transmon qubits) harmonic oscillator for the simulations. \changetwo{The dynamics of Hamiltonian \equref{eq:Eff_Ham_def} are modelled with four basis states for all resonators and transmons}.

\subsection{The control pulses and gate implementations}\label{sec:TheControlPulses}
\newcommand{\Cctl}{n(t)}
\newcommand{\CctlAp}{a}
\newcommand{\CctlDr}{b}
\newcommand{\CctlDf}{\omega^{(D)}}
\newcommand{\CctlSg}{\sigma}
\newcommand{\CctlTd}{T_{d}}
\newcommand{\Fctl}{\varphi(t)}
\newcommand{\FctlAp}{\delta}
\newcommand{\FctlSg}{\sigma}
\newcommand{\FctlTp}{T_{p}}
\newcommand{\FctlTd}{T_{d}}

\renewcommand{\hold}{0.5}
\begin{figure}[!tbp]
  \centering
  \begin{minipage}{0.495\textwidth}
      \centering
      \includegraphics[scale=\hold]{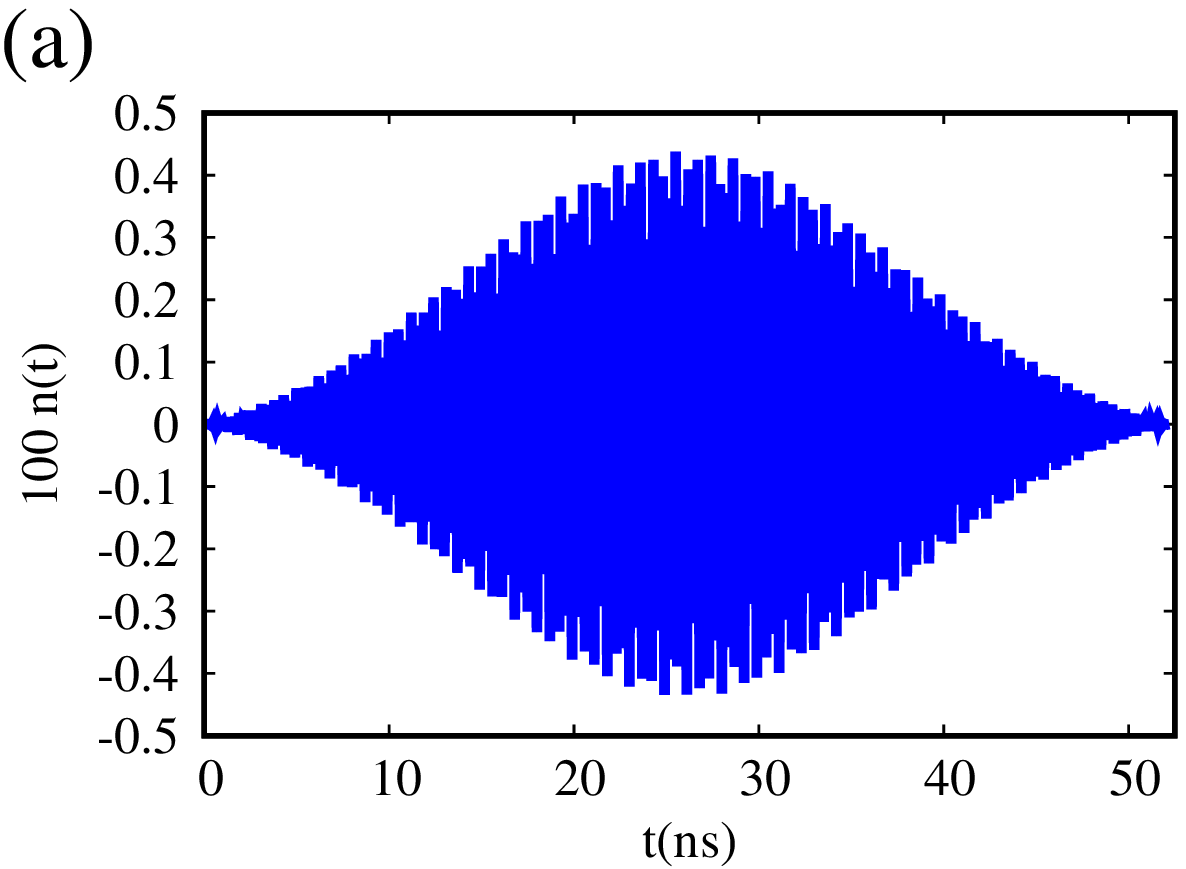}
      \includegraphics[scale=\hold]{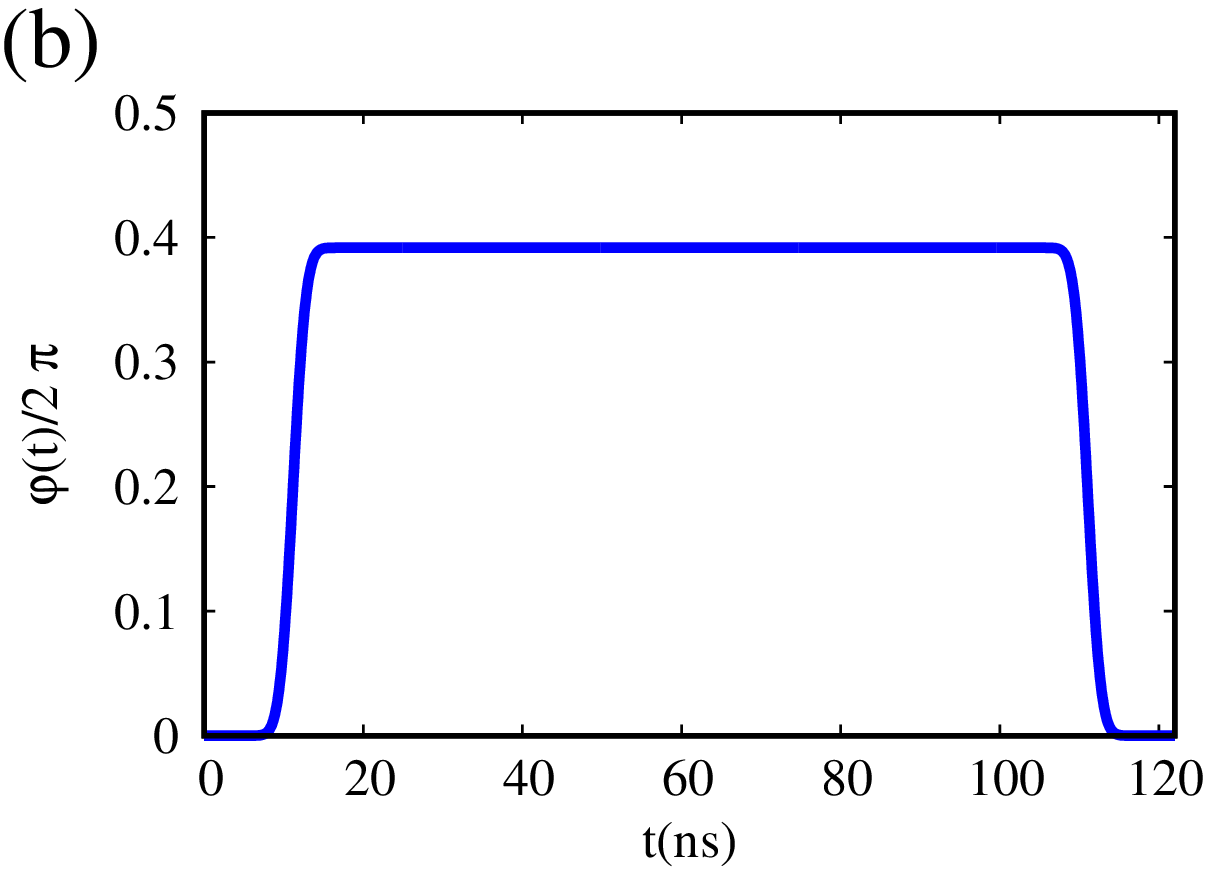}
      \includegraphics[scale=\hold]{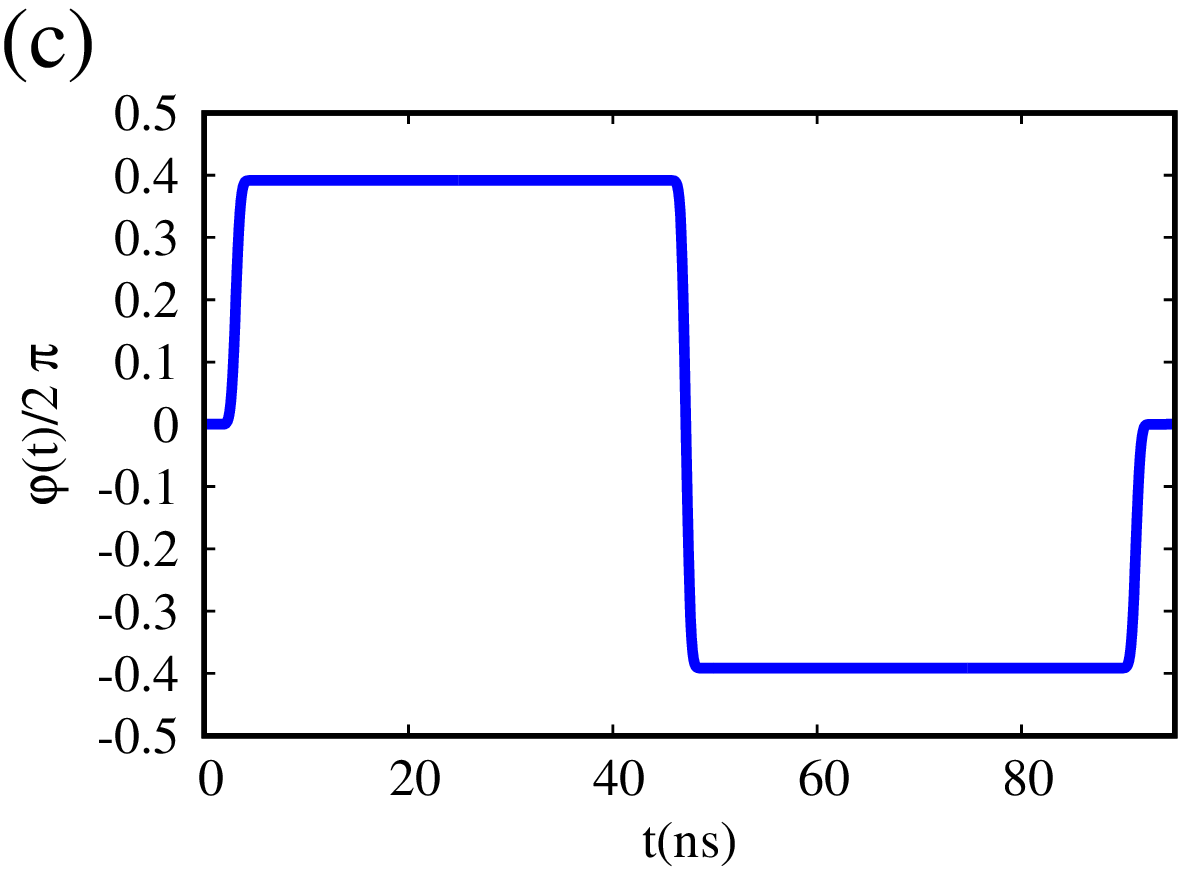}
  \end{minipage}
  \caption{(Color online) External charge $n(t)$(a) given by \equref{eq:charge_ctl} and external fluxes $\varphi(t)$(b-c) given by \equaref{eq:flux_ctl_ump}{eq:flux_ctl_bmp}, respectively, as a functions of time $t$. (a) microwave pulse (MP); (b) unimodal pulse (UMP); (c) bimodal pulse (BMP). In the circuit model both functions $n(t)$ and $\varphi(t)$ enter via the Hamiltonian \equref{eq:Circuit_Ham}. In the effective model both functions $n(t)$ and $\varphi(t)$ enter via the driving terms \equaref{eq:drive_charge}{eq:drive_flux}, respectively. \changethree{Moreover}, in the effective model the external flux $\varphi(t)$ also enters the time-dependent interaction strength in \equref{eq:Eff_Int} and the tunable qubit frequency and anharmonicity in \equref{eq:Eff_Ham}, see \appref{app:TheEffectiveHamiltonianModel} for more details.}\label{fig:PulseTimeEvolution}
\end{figure}

Figures \ref{fig:PulseTimeEvolution}(a-c) show the external charge $\Cctl$(a) and the external flux $\Fctl$(b-c) as functions of time $t$. In (a) we show a microwave pulse (MP), in (b) we show a unimodal pulse (UMP) and in (c) we show a bimodal pulse (BMP). In this section we introduce the functions which are used to obtain the data we show in \figsref{fig:PulseTimeEvolution}(a-c). We use these functions as control pulses for our NIGQC models.

For the single-qubit gates we use the external charge
\begin{equation}\label{eq:charge_ctl}
  \begin{split}
    \Cctl&= \CctlAp G(t,\CctlSg,\CctlTd) \cos\brr{\CctlDf t-\phi}\\
        &+ \CctlDr \dot{G}(t,\CctlSg,\CctlTd) \sin\brr{\CctlDf t-\phi},
  \end{split}
\end{equation}
to drive transitions that can be used to implement $\ROT(\pi/2)$ rotations with a microwave control pulse. Here $G(t,\CctlSg,\CctlTd)$ is a Gaussian envelope function centered around half of the pulse duration $\CctlTd$. The width and the shape of the Gaussian are determined by the parameters $\CctlSg$ and $\CctlTd$. Furthermore, $\CctlAp$ denotes the pulse amplitude, $\CctlDf$ refers to the drive frequency, $\CctlDr$ is the amplitude of the DRAG pulse component, see \REF\cite{Motzoi09}, and $\phi$ is the phase of the pulse. We use the phase $\phi$ in \equref{eq:charge_ctl} to implement virtual $Z$ gates, see \REF\cite{McKay17}, on our NIGQCs.

Two-qubit $\CZ$ gates can be implemented with an external flux of the form
\begin{equation}\label{eq:flux_ctl_ump}
  \Fctl= \frac{\FctlAp}{2}\brr{\text{erf}\brr{\frac{t}{\sqrt{2}\FctlSg}} - \text{erf}\brr{\frac{(t-\FctlTp)}{\sqrt{2}\FctlSg}}},
\end{equation}
where $\FctlAp$ denotes the pulse amplitude, $\FctlSg$ is a parameter which allows us to control how fast the pulse flanks rise and fall, $\FctlTp$ is the pulse time, and $\mathrm{erf}$ denotes the Gauss error function. Note that in the computer program we add an additional free time evolution (the buffer time) to the pulse such that the complete pulse duration $\FctlTd$ is longer than $\FctlTp$. We denote this pulse as the UMP.

We can also use the external flux
\begin{equation}\label{eq:flux_ctl_bmp}
  \begin{split}
    \Fctl= \frac{\FctlAp}{2}  &\left( \right. \text{erf}\brr{\frac{t}{\sqrt{2}\FctlSg}} - 2\text{erf}\brr{\frac{(t-\FctlTp/2)}{\sqrt{2}\FctlSg}}\\
    &+\text{erf}\brr{\frac{(t-\FctlTp)}{\sqrt{2}\FctlSg}} \left. \right),
  \end{split}
\end{equation}
to implement two-qubit $\CZ$ gates. \changetwo{This pulse is referred to as the BMP}. Note that the BMP pulse is sometimes referred to as net-zero flux pulse. The UMP and the BMP are also used to implement gates in experiments, see \REFS\cite{Lacroix2020,Krinner2020,Andersen20,Rol19}.

Every $\CZ$ gate in our NIGQC model is implemented by means of a flux pulse $\varphi(t)$ followed by single-qubit z-axis rotations $R_{i}^{(z)}(\phi_{i})$ for every qubit in the NIGQC model, see \REF\cite[Section VII B 2]{Blais2020circuit}. Since the z-axis rotation parameters $\phi_{i}$ are quite numerous, \eg a four-qubit system with four $\CZ$ gates has sixteen of these parameters, we omit the $\phi_{i}$ phases from the control pulse parameter \tabsref{tab:CtlTqgIIMB}{tab:CtlTqgIVHB} in \appref{app:ControlPulseParameters}. Finally, we execute these z-axis rotations by means of virtual $Z$ gates in combination with a transformation of the frame of reference which only affects the phases of the state vector but not the state vector amplitudes, see \REF\cite[Section 3.3.2]{Willsch2020}.



\section{Computation of gate-error quantifiers}\label{sec:ErrorMetrics}

\newcommand{\DNV}{\mu_{\diamond}}
\newcommand{\FV}{\mu_{\text{F}_{\text{avg}}}}
\newcommand{\IFV}{\mu_{\text{IF}_{\text{avg}}}}
\newcommand{\LNV}{\mu_{\text{Leak}}}
\newcommand{\SDV}{\mu_{\text{SD}}}
\newcommand{\SDAV}{\mu_{\text{SD}_{\text{avg}}}}

\newcommand{\labeldianorm}{\diamond}
\newcommand{\err}[1]{\mu_{#1}}
\newcommand{\idlabel}{\text{U}}
\newcommand{\aclabel}{\text{M}}
\newcommand{\SUPOP}[1]{\check{#1}}
\newcommand{\dx}[1]{d\!#1}
\begin{figure*}[tbp!]
\renewcommand{\hold}{1.20}
\centering
\includegraphics[scale=\hold]{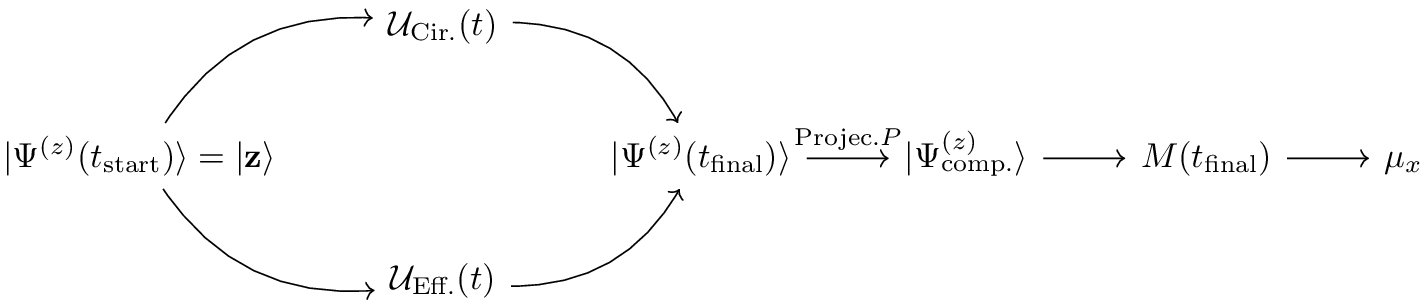}
\caption{Illustration of the computational process we use to obtain the gate-error quantifiers $\mu_{x}$. Here $x$ is a label for an arbitrary metric \change{and $\op{\mathcal{U}}_{\idxCIR}$ ($\op{\mathcal{U}}_{\idxEFF}$) is the time-evolution operator in \equref{eq:TDSEOperator} for the circuit (effective) Hamiltonian given by \equref{eq:Circuit_Ham_def} (\equref{eq:Eff_Ham_def})}. We compute the state vectors \change{$\ket{\Psi^{(z)}(t_{\text{final}})}$} with a product-formula algorithm, see \REFS\cite{DeRaedt87,Huyghebaert90}, for all computational basis states $2^{N}$ of a NIGQC and then store the data in the matrix \change{$M(t_{\text{final}})$}. In this matrix, every column corresponds to the evolution of one computational state \change{$\ket{\Psi^{(z)}_{\text{comp.}}}=P \ket{\Psi^{(z)}(t_{\text{final}})}$}, where $P$ \change{denotes a projection matrix onto the computational subspace}. We use the Message Passing Interface (MPI) to parallelise the $2^{N}$ independent tasks. The computations are performed on the supercomputer JUWELS, see \REF\cite{JUWELS}. Finally, we compute the metric $\mu_{x}$ with regard to a target operation $U$. The metrics we compute provide a measure for how close $M$ and $U$ are.}\label{fig:comp_sketch}
\end{figure*}

In this section we discuss the gate-error quantifiers we compute as well as how we determine the numerical values.

A computation in the IGQC model can be understood as a mapping
\begin{equation}
  \ket{\psi} \mapsto \ket{\psi^{\prime}},
\end{equation}
between an initial state $\ket{\psi}$ and a final state $\ket{\psi^{\prime}}$ and the state vector of an $N$-qubit IGQC can be expressed as
\begin{equation}
  \ket{\psi}=\sum_{\mathbf{z} \in \{0,1\}^{N}} c_{\mathbf{z}} \ket{\mathbf{z}},
\end{equation}
where $\{ \ket{\mathbf{z}} \}$ are the $2^{N}$ computational basis states of the IGQC. All state vectors $\ket{\psi}$ are normalised complex vectors in a finite dimensional Hilbert space $\mathcal{H}^{2^{N}}$. The computation is described by a unitary operator $\op{U}$.

One simple error measure is the statistical distance
\begin{equation}\label{eq:stat_dis}
  \mu_{\text{SD}}\brr{\mathbf{p},\mathbf{\tilde{p}}}=\frac{1}{2} \sum_{\mathbf{z} \in \{0,1\}^{N}} \norm{p_{\mathbf{z}}-\tilde{p}_{\mathbf{z}}}_{1},
\end{equation}
where\changetwo{,} by definition $p_{\mathbf{z}}=\norm{\braket{\mathbf{z}|\psi}}_{1}^{2}$ are the probability amplitudes of the IGQC (which we use as a reference distribution to compare against) and $\tilde{p}_{\mathbf{z}}$ denote the actual distribution that is being evaluated. This distribution can either be generated by a PGQC or a NIGQC. Furthermore, $\norm{\cdot}_{1}$ denotes the absolute value. The advantage of the statistical distance is that we can more or less easily measure the relative frequencies $p_{\mathbf{z}}$ in an experiment. While such an error measure is useful in practice, see \REF\cite{Michielsen17,Wi17,Willsch2020}, it neglects the phase information of the state vector $\ket{\psi}$. \changethree{Moreover}, the statistical distance only provides a measure of closeness for one particular input state. For this reason, usually more sophisticated gate-error metrics are considered.

Most gate metrics are defined in terms of quantum operations. From a mathematical point of view, quantum operations are superoperators $\SUPOP{\mathcal{E}}(\op{\rho})$ which act on the space of density operators $\op{\rho} \in \mathcal{P}$. Additionally, we require $\SUPOP{\mathcal{E}}$ to be linear, Hermiticity preserving and completely positive, see \REFS\cite{Nielsen:2011:QCQ:1972505,Watrous2018}. Note that if one can also show that $\SUPOP{\mathcal{E}}$ is trace preserving, $\SUPOP{\mathcal{E}}$ is usually referred to as quantum or error channel.

The error metrics we consider in this work can be expressed in terms of the two quantum operations
\begin{equation}\label{eq:QO_ideal}
  \SUPOP{\mathcal{E}}_{\idlabel}(\op{\rho})=\op{U} \op{\rho} \op{U}^{\dagger}
\end{equation}
and
\begin{equation}\label{eq:QO_act}
  \SUPOP{\mathcal{E}}_{\aclabel}(\op{\rho})=\op{M} \op{\rho} \op{M}^{\dagger},
\end{equation}
where $\op{\rho}=\ketbra{\psi}{\psi}$ and $\ket{\psi} \in \mathcal{H}^{2^{N}}$. The operator $\op{M}$ in \equref{eq:QO_act} is defined as
\begin{equation}
  \op{M}= \op{P}\op{\mathcal{U}}\brr{t,t_{0}} \op{P},
\end{equation}
where $\op{P}$ is a projection operator \change{onto the computational space} and $\op{\mathcal{U}}$ denotes the formal solution of the \changetwo{time-dependent Schr\"odinger equation}
\begin{equation}\label{eq:TDSEOperator}
    \hat{\mathcal{U}}(t,t_{0}) = \mathcal{T} \exp\left( -i \int_{t_{0}}^{t} \hat{H}(t^{\prime}) dt^{\prime} \right).
\end{equation}
Here $\mathcal{T}$ is the time-ordering symbol. \changethree{In addition}, the Hamiltonian $\op{H}$ is either given by \equref{eq:Circuit_Ham_def} or \equref{eq:Eff_Ham_def}.

The first error metric that we consider is the average fidelity
\begin{equation}
  \FV=\int \braket{\psi| \SUPOP{\mathcal{E}}_{\aclabel} \SUPOP{\mathcal{E}}_{\idlabel}^{-1} \brr{\ketbra{\psi}{\psi}}|\psi} \dx{\ket{\psi}},
\end{equation}
where the integral is taken over all states $\ket{\psi} \in \mathcal{H}^{2^{N}}$ in the Hilbert space. If we define the auxiliary operator
\begin{equation}
  \op{V}=\op{U}\op{M}^{\dagger},
\end{equation}
we can express the average fidelity as
\begin{equation}\label{eq:fid_avg}
  \FV=\frac{\norm{\trace(\op{V})}_{1}^{2}+\trace(\op{M} \op{M}^{\dagger})}{D\brr{D+1}},
\end{equation}
where $D=2^{N}$. This is expression is derived in \REF\cite[Section 7]{Jin21}. We can use the average fidelity to define the average infidelity
\begin{equation}\label{eq:in_fid_avg}
  \IFV=1-\FV.
\end{equation}

In order to define a leakage measure for our NIGQCs, we make use of the second term in the numerator of \equref{eq:fid_avg}. We define the leakage measure as
\begin{equation}\label{eq:leak}
  \LNV=1-\brr{\frac{\trace(\op{M} \op{M}^{\dagger})}{D}},
\end{equation}
where $\trace(\op{M} \op{M}^{\dagger})$ can be expressed as the sum of probability amplitudes. Hence, we find $0 \leq \trace(\op{M} \op{M}^{\dagger}) \leq D$\changetwo{,} and therefore it follows $\LNV \in [0,1]$.

The second error metric we consider is the diamond distance
\begin{equation}
  \DNV= \frac{1}{2} \norm{ \SUPOP{\mathcal{E}}_{\aclabel} \SUPOP{\mathcal{E}}_{\idlabel}^{-1} - \op{I}}_{\labeldianorm},
\end{equation}
where $\norm{\cdot}_{\labeldianorm}$ denotes the diamond norm, see \REFS\cite{Kitaev1997,Sanders2015}. We can express the diamond distance in terms of an infimum
\renewcommand{\hold}{Q}
\begin{equation}\label{eq:diamond_norm_inf}
  \begin{split}
    \DNV=\frac{1}{2}\inf_{\hold \in \text{GL}_{4}\brr{\mathbb{C}}} \left\{\right. &\norm{\brr{\op{V},\op{I}} \op{\hold}^{-\dagger}\op{\hold}^{-1} \brr{\op{V},\op{I}}^{T}}_{2}^{\frac{1}{2}} \\
    &\norm{\brr{\op{V},-\op{I}} \op{\hold}^{\dagger}\op{\hold}^{1} \brr{\op{V},-\op{I}}^{T}}_{2}^{\frac{1}{2}} \left. \right\},
  \end{split}
\end{equation}
over all complex, invertible, two by two matrices, see \REF\cite{Johnston09}. Similarly, we can express the diamond distance in terms of a supremum
\renewcommand{\hold}{\psi}
\begin{equation}\label{eq:diamond_norm_sup}
  \begin{split}
    \DNV=\frac{1}{2}\sup_{\ket{\hold} \in \mathcal{H}^{2^{N}}}  \left\{\right. &\norm{\brr{\op{V}^{\dagger}\otimes\op{I}} \ketbra{\hold}{\hold} \brr{\op{V}^{\dagger}\otimes\op{I}}^{\dagger}\\
     &- \ketbra{\hold}{\hold}}_{\trace} \left. \right\},
  \end{split}
\end{equation}
over all state vectors of the Hilbert space of the IGQC, see \REF\cite{Watrous2018}. We obtain the diamond distance up to the fourth decimal by cornering
\begin{equation}
  \DNV^{(\text{inf})} \leq \DNV \leq \DNV^{(\text{sup})},
\end{equation}
the value $\DNV$ with the infimum $\DNV^{(\text{inf})}$ and supremum $\DNV^{(\text{sup})}$ expressions in \equaref{eq:diamond_norm_inf}{eq:diamond_norm_sup}. The algorithms we use are discussed in \REF\cite[Section 6.1.2]{Willsch2020}.

Figure \ref{fig:comp_sketch} illustrates the computational process we use to obtain the matrix $M$ and in turn the various gate-error quantifiers $\mu_{x}$ in this work. The label $x$ denotes an arbitrary metric or measure. First, we simulate the time evolution $\ket{\Psi^{(z)}(t_{final})}$ of the system for all $2^{N}$ basis states $\ket{\Psi^{(z)}(t_{start})}=\ket{\mathbf{z}}$ of the NIGQC. Note that we simulate the $2^{N}$ time evolutions $\ket{\Psi^{(z)}(t)}$ in parallel on the supercomputer JUWELS, see \REF\cite{JUWELS}. Then we make use of a projection matrix $P$ and map the state vectors to the computational states $\ket{\Psi_{\text{comp.}}^{(z)}}=P \ket{\Psi^{(z)}(t_{final})}$. Finally, we store the data by building the matrix
\begin{equation}\label{eq:prop_matrix}
   M=\sum_{\mathbf{z} \in \{0,1\}^{N}} \ketbra{\Psi_{\text{comp.}}^{(z)}}{\mathbf{z}},
\end{equation}
where $\ket{\mathbf{z}}$ are the cartesian unit vectors, in the computer program. Finally, we compute the gate-error quantifiers $\mu_{x}$ by means of the matrix $V=U M^{\dagger}$.

\section{Results}\label{sec:Results}
In this section we present our findings. First, in \secref{sec:SpectrumAnalysis} we discuss the spectrum of the four-qubit NIGQC illustrated in \figsref{fig:device_sketch}(c) and its relevance for the gate-error metrics we compute. Here we model the system with the circuit Hamiltonian \equref{eq:Circuit_Ham_def}. Next, in \secref{sec:GateMetricsAndControlPulseParameters} we discuss the results of the calibration process for the different NIGQCs illustrated in \figsref{fig:device_sketch}(a-c). Here we consider both the circuit and the effective model. Then, in \secref{sec:Influence of higher states on gate-error trajectories} we show how modelling the time evolution of NIGQCs with more and less basis states affects the gate-error quantifiers. Here we use the circuit Hamiltonian \equref{eq:Circuit_Ham_def} to model the dynamics of the four-qubit NIGQC illustrated in \figsref{fig:device_sketch}(c). Next, in \secref{sec:Influence of small pulse parameter deviations on gate-error trajectories} we show how small deviations in a single control pulse parameter can affect gate-error metrics. Here we use the circuit Hamiltonian \equref{eq:Circuit_Ham_def} to model the dynamics of the two-qubit, three-qubit and four-qubit NIGQCs illustrated in \figsref{fig:device_sketch}(a-c). Finally, in \secref{sec:InfluenceOfTheAdiabaticApproximationOnGate-errorTrajectories} we show how the commonly used adiabatic approximation for flux-tunable transmons affects gate-error metrics. Here we use the adiabatic and \changetwo{nonadiabatic} effective Hamiltonian \equref{eq:Circuit_Ham_def} to model the dynamics of the two-qubit, three-qubit and four-qubit NIGQCs illustrated in \figsref{fig:device_sketch}(a-c).

\subsection{Spectrum for a four-qubit NIGQC}\label{sec:SpectrumAnalysis}
\renewcommand{\hold}{0.725}
\begin{figure*}[!tbp]
  \centering
  \begin{minipage}{1.0\textwidth}
      \centering
      \includegraphics[scale=\hold]{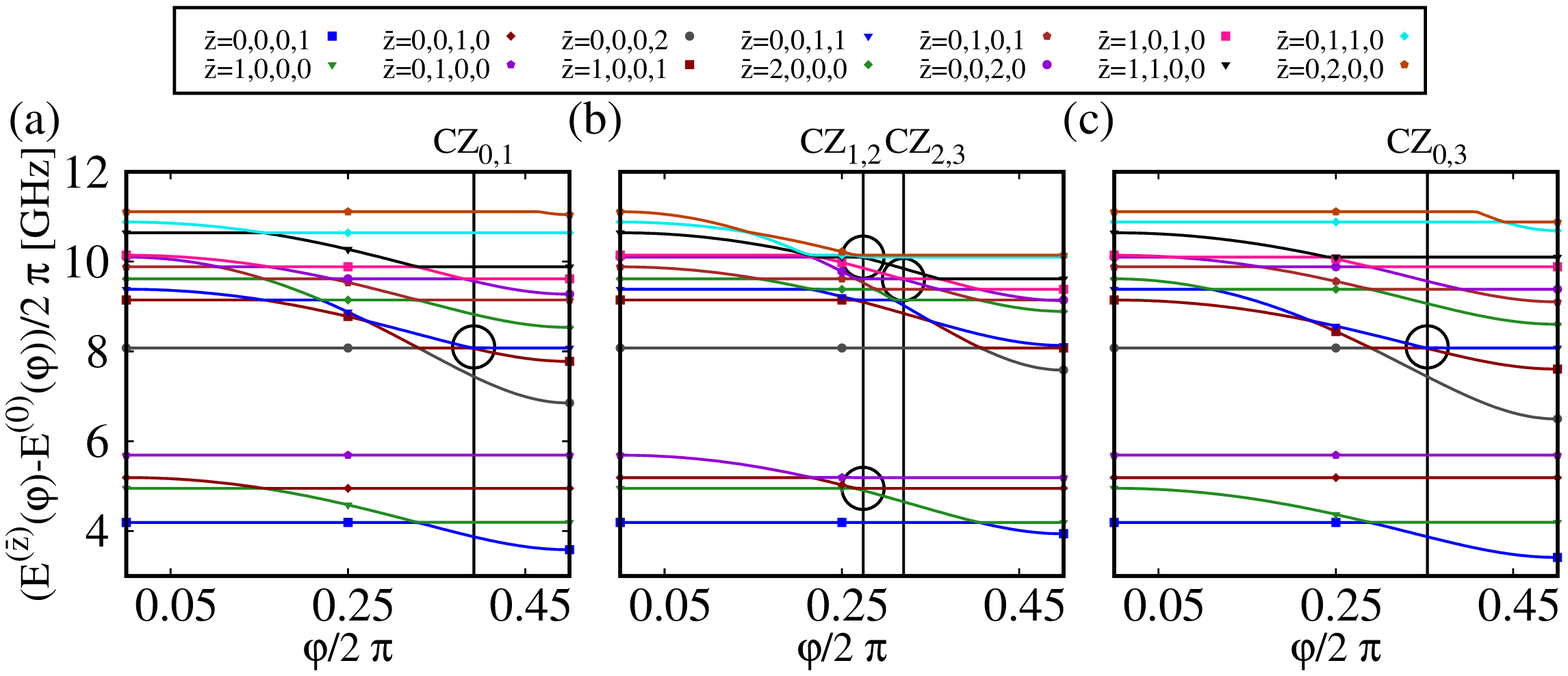}
  \end{minipage}
  \caption{(Color Online) The lowest fourteen energy levels of the four-qubit NIGQC illustrated in \figref{fig:device_sketch}(c) as functions of the external flux offset $\varphi$ for the second ($i=1$) in (a), the third ($i=2$) and the fourth ($i=3$) transmon qubit. We use the \textbf{circuit Hamiltonian} \equref{eq:Circuit_Ham_def} and the device parameters listed in \tabref{tab:device_parameters} to obtain the results with a standard diagonalisation algorithm, see \REF\cite{MKL09}. All transmons are modelled with three basis states and all resonators are modelled with two basis states. We also mark \change{(vertical lines)} the pulse amplitudes \change{$\FctlAp$} used to implement the $\CZ$ gates on the four-qubit NIGQC. \change{Furthermore, we mark important energy level repulsions (ELRs) by means of black circles}. In (b) we can observe that the pulse amplitude for the $\CZ_{1,2}$ gate is near two \change{ELRs}. The first one is in the energy band between $4$ and $6$ GHz. This \change{ELR} leads to unwanted transitions between the first excited states of two qubits. The second one is in the energy band between $9$ and $11$ GHz. This \change{ELR} is used to implement the $\CZ_{1,2}$ gate. In (a-c) we can observe that the other $\CZ$ gates do not suffer from the same problem. \changethree{Moreover}, we can observe that driving different transmons ($i=1$) in (a), ($i=2$) in (b) and ($i=3$) in (c) leads to different spectral patterns.\label{fig:device_spectrum}}
\end{figure*}
In this section we discuss the spectrum of the four-qubit NIGQC illustrated in \figref{fig:device_sketch}(c) and its relevance for the implementation of two-qubit gates.

Broadly speaking, the general idea of implementing two-qubit $\CZ$ gates with a UMP or BMP is to tune a target computational state like $\ket{0,0,1,1}$ into resonance with a non-computational state like $\ket{0,0,0,2}$, wait some time until the population \changetwo{returns} to the computational target state, and hope that this state has gained an additional phase of $e^{i \pi}$ with respect to all the other computational states of the NIGQC. The word ``tune'' in this context refers to tuning the energies of the instantaneous eigenstates of the system, \ie the discussion is implicitly carried out in the instantaneous basis of the system. Additionally, single-qubit z-axis rotations $R_{i}^{(z)}(\phi_{i})$ are used to improve the performance, see \REF\cite[Section VII B 2]{Blais2020circuit}.

The time it takes for our devices to swap the population back and forth between a computational and a non-computational state of the NIGQC lies between $75$ and $100$ ns. This time corresponds to the plateau time in \figsref{fig:PulseTimeEvolution}(b-c). We fix these times by choosing the interaction strength constant $G$ such that we approximately obtain the same gate times as in the experiment. Furthermore, the pulse amplitude $\FctlAp$ or the plateau height is fixed by the condition that the states involved in this process should have close-by energies. The time it takes to reach the plateau is crucial; we discuss this part of the gate implementation later in this section.

In \REF\cite[Section VII B 2]{Blais2020circuit}, the authors differentiate between gates that are implemented adiabatically, see \REF\cite{DiCarlo2009} and gates that are implemented \changetwo{nonadiabatic}ally, see \REF\cite{DiCarlo2010}. However, in the context of quantum theory the words adiabatic and \changetwo{nonadiabatic} are usually associated with the adiabatic approximation, see \REF\cite[Section 6.6]{Weinberg2015}. Note that the process described above is clearly not an adiabatic one, \ie the probability amplitudes of the state vector change over time. This and the fact that in almost all studies, the transmon qubit is modelled as an adiabatic anharmonic oscillator or as an adiabatic two-level system, leads us to the conclusion that in the context of the gate implementation protocol described above the words adiabatic and \changetwo{nonadiabatic} are simply used to differentiate between gates with short and long pulse durations. Note that in \secref{sec:InfluenceOfTheAdiabaticApproximationOnGate-errorTrajectories} we investigate a related issue. There are two additional problems with the picture described above. First, assigning labels to the energies $E_{z}(\varphi)$ and eigenstates $\ket{\phi_{z}(\varphi)}$ of a continuous set of Hermitian matrices $\{H(\varphi)\}$, where $z\in \mathbb{N}^{0}$ and $\varphi/\TP \in [0,1]$, is a non-trivial problem in itself, see \REFS\cite{Hund1927,vonNeumann1993,Uhlig2020,Srinivasan2020}. Second, once we apply a UMP or BMP flux drive, we cannot guarantee that only the desired transitions occur.

In order to provide clarity on these issues, in the discussion of the four-qubit NIGQC spectrum, we do not adopt the same nomenclature used in \REF\cite[Section VII B 2]{Blais2020circuit}.

Figures~\ref{fig:device_spectrum}(a-c) show the lowest fourteen energy levels of the four-qubit NIGQC illustrated in \figref{fig:device_sketch}(c) as functions of the external flux offset $\varphi$ for the second ($i=1$) in (a), the third ($i=2$) and the fourth ($i=3$) transmon qubit. This means that each panel is obtained by repeatedly diagonalising the matrix for different flux offset values $\varphi/\TP \in [0,0.5]$. The results are obtained with the circuit Hamiltonian \equref{eq:Circuit_Ham_def}, the device parameters listed in \tabref{tab:device_parameters} and a standard diagonalisation algorithm, see \REF\cite{MKL09}. Here we use two basis states for the coupling resonators and three basis states for the different transmon qubits in the system. Note that the circuit Hamiltonian \equref{eq:Circuit_Ham} has two symmetry points, one at $\varphi/\TP=0.5$ and one at $\varphi/\TP=1$.

We label the energies $E_{\bar{z}}$ of the interacting system according to the sorted energies $E_{z}$ of the non-interacting system for the flux offset $\varphi=0$. The markers in \figsref{fig:device_spectrum}(a-c) are there to guide the eye. Additionally, we employ black vertical lines to mark the flux offset values $\varphi$ that correspond to pulse amplitudes $\FctlAp$ for the $\CZ$ gates in \tabref{tab:CtlTqgIVMB}. Furthermore, we employ black circles to mark the energy level repulsions (\change{ELR}s) that we use to implement the $\CZ$ gates and the \change{ELR}s that are problematic for the implementation of $\CZ$ gates with low gate-error metrics, see the line with two circles in \figsref{fig:device_spectrum}(b).

If we start at the flux offset value $\varphi=0$ and then move to a value $\varphi=\FctlAp$ by driving a transmon qubit, we usually pass through several unused \change{ELR}s, \ie not used to implement the $\CZ$ gates, with some of the computational basis states of the NIGQC before we reach the \change{ELR} or the flux offset value that we use to implement the $\CZ$ gate. Here we have to pass through the unused \change{ELR}s sufficiently fast because otherwise we can observe population exchange between the two states involved. This can be clearly observed (data not shown) if one studies the matrix in \equref{eq:prop_matrix} while optimising the control pulse parameters or the probability amplitudes themselves during the time evolution of the system. \changethree{Also}, one cannot move the system too fast, otherwise one observes (data not shown) all sorts of other transitions, \eg the excited coupler states suddenly become populated. The pulse optimisation algorithm that we employ finds a balance between these two mechanisms and fine tunes the pulse amplitude $\delta$ and the remaining pulse parameters such that nearly perfect population exchange occurs and the phases are aligned properly. Note that we have to take all $2^{N}$ computational basis states into account, see \figref{fig:comp_sketch}.

In \figref{fig:device_spectrum}(b), we can also identify a problem that the optimisation cannot solve on its own, see the line with two circles. The \change{ELR} in the energy band between $4$ GHz and $6$ GHz induces a population exchange between two computational states of the NIGQC and is therefore unwanted. \changethree{In addition}, in the energy band between $9$ GHz and $12$ GHz for the same flux offset value, we can observe a dense structure of energy levels with many nearby \change{ELR}s.

Obviously, this leads to the question whether or not such a gate implementation scheme can be scaled up. To the best our knowledge, the largest PGQC to date with the device architecture discussed in this work has seventeen qubits, see \REF\cite{Krinner21}. In order to avoid the problem just discussed, \ie frequency collisions during the gate, some authors suggest to apply additional flux pules to non-interacting transmon qubits to mitigate the problem, see \REF\cite[Supplementary Material Section 1]{Andersen20}. Other authors mitigate the issue by redefining the target two-qubit gate, see \REF\cite[Supplementary Material Section VI C 1]{Arute19}, so that the time evolution of the system fits more naturally to the target two-qubit gate.

We \changetwo{emphasise} that driving different transmon qubits results in different energy levels being populated and different \change{ELR}s being active, see \figsref{fig:device_spectrum}(a-c). \change{Consequently, before we can even start building a device, we have to solve an optimisation problem of exponential size, i.e. we have to avoid \change{ELR}s between all the $2^{N}$ computational basis states of the NIGQC or PGQC. Note that here we have to take into account different pulse amplitudes $\FctlAp$ for the different $\CZ$ gates we implement and the different spectra and ELRs which result from driving different transmon qubits. Strictly speaking, our conclusions only apply to the frequency-tunable transmon device architecture studied in this work. However, the consequences of the problem are much more general and are imprinted on many other systems (cf.~\REF\cite{Baker22,Cohen2022}).}

We also \changetwo{emphasise} that using microwave pulses (MPs) to implement single-qubit gates leads to a similar problem, \ie the MPs are not monochromatic and contain frequency components that can cause unwanted transitions. The more transmon qubits we employ, the more frequency components we have to take into account. \change{If we consider an N-qubit transmon computer, we have to take into account that there are $2^{N}$ relevant energy levels and the associated energy differences between these states. Each additional energy difference, i.e. transition frequency, reduces the available frequencies and thus adds to the frequency crowding problem (cf.~\REF\cite{Hertzberg2021FrequencyCrowdingProblem})}.

\subsection{Gate metrics for the elementary gate set}\label{sec:GateMetricsAndControlPulseParameters}
In this section we discuss the results of the control pulse optimisation procedure for the three different NIGQCs illustrated in \figsref{fig:device_sketch}(a-c).

\renewcommand{\hold}{0.75}
\begin{figure*}[!tbp]
  \centering
  \includegraphics[scale=\hold]{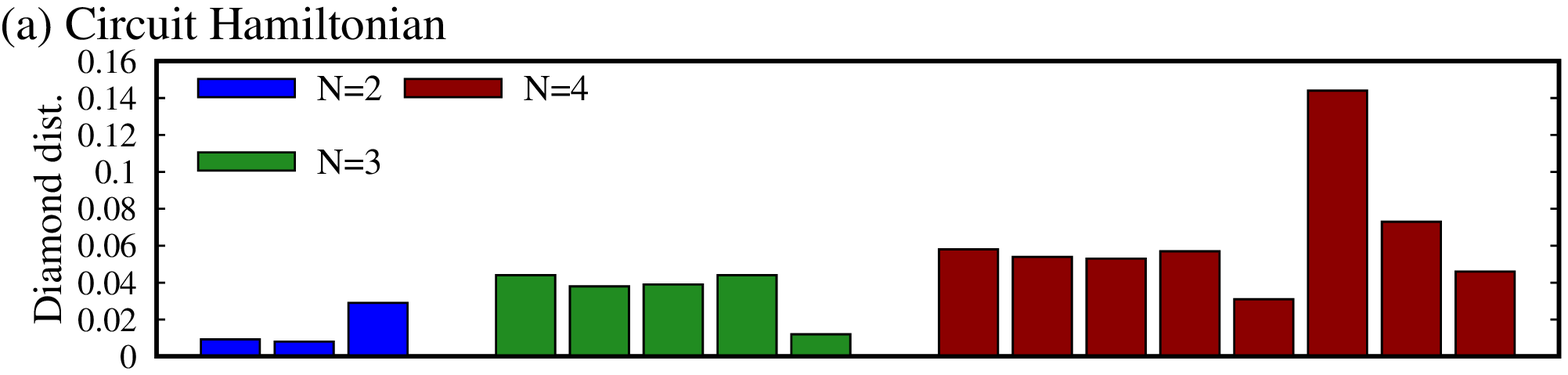}
  \includegraphics[scale=\hold]{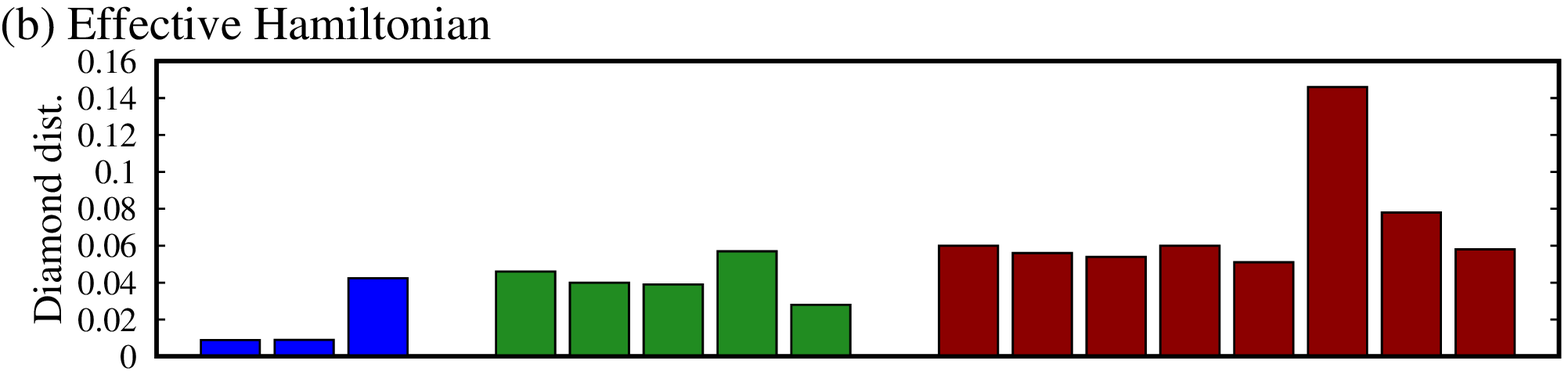}
  \includegraphics[scale=\hold]{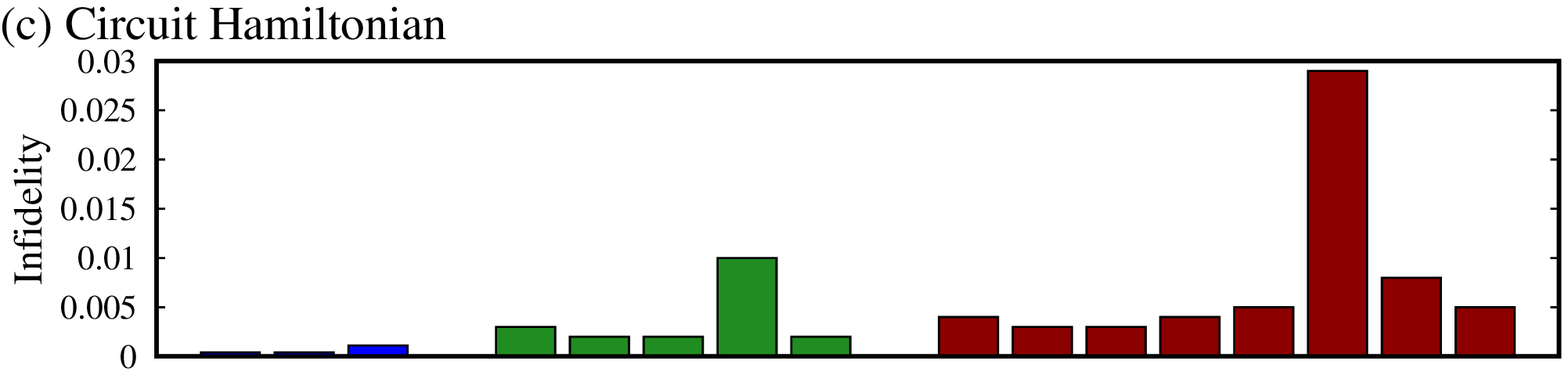}
  \includegraphics[scale=\hold]{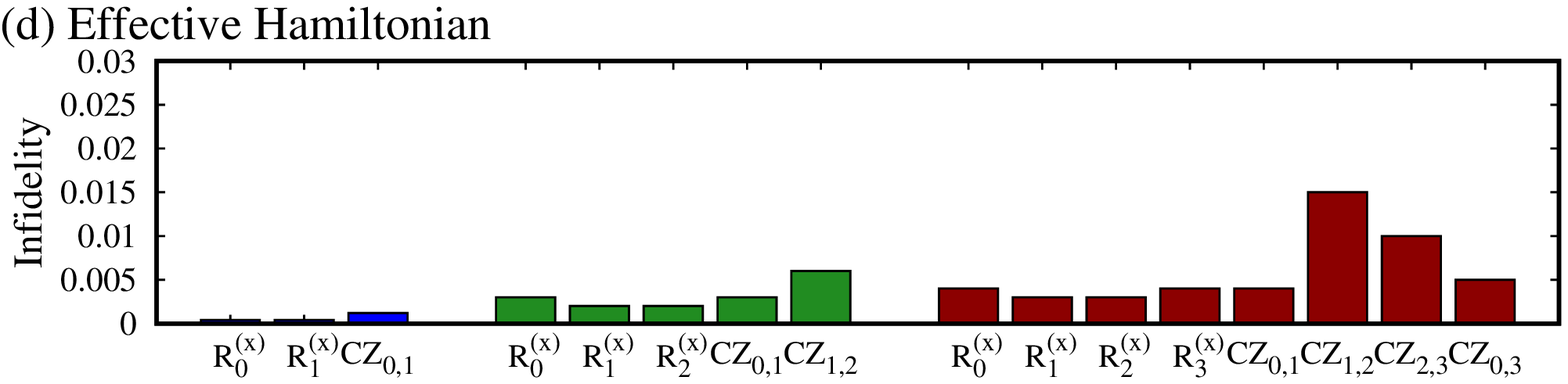}
  \caption{(Color Online) Diamond distances $\DNV$(a,b) given by \equaref{eq:diamond_norm_inf}{eq:diamond_norm_sup} and average infidelity $\IFV$\change{(c,d)} given by \equref{eq:fid_avg} for $\ROT(\pi/2)$ rotations and $\CZ$ gates for the NIGQCs illustrated in \figsref{fig:device_sketch}(a-c), where $N=2$ \change{in \figref{fig:device_sketch}(a)}, $N=3$ \change{in \figref{fig:device_sketch}(b)} and $N=4$ \change{in \figref{fig:device_sketch}(c)}. The error metrics are obtained with the \textbf{circuit Hamiltonian} \equref{eq:Circuit_Ham_def} (a,c) and the \textbf{effective Hamiltonian} \equref{eq:Eff_Ham_def}(b,d). We use the device parameters listed in \tabref{tab:device_parameters} and the pulse parameters listed in \tabsref{tab:CtlSqgIIMB}{tab:CtlTqgIVHB} to obtain the results. The single-qubit $\ROT(\pi/2)$ rotations are obtained with a microwave pulse, see \figref{fig:PulseTimeEvolution}(a). The two-qubit $\CZ$ gates are obtained with the unimodal pulse, see \figref{fig:PulseTimeEvolution}(b).\label{fig:metrics}}
\end{figure*}

Figures \ref{fig:metrics}(a-d) show the diamond distance $\DNV$(a,b) and the average infidelity $\IFV$(c,d) for the two-qubit (in blue), the three-qubit (in green) and the four-qubit (in red) NIGQCS modelled with the circuit Hamiltonian \equref{eq:Circuit_Ham_def}(a,c) and the effective Hamiltonian \equref{eq:Eff_Ham_def}(b,d). The results are also listed in \tabsref{tab:EMIIMB}{tab:EMIVHB}, see \appref{app:ControlPulseParameters}. The single-qubit $\ROT(\pi/2)$ rotations are implemented with the MP in \equref{eq:charge_ctl}. The two-qubit $\CZ$ gates are implemented with the UMP in \equref{eq:flux_ctl_ump}. We use the device parameters listed in \tabref{tab:device_parameters} and the pulse parameters listed in \tabsref{tab:CtlSqgIIMB}{tab:CtlTqgIVHB} to obtain the results, see \appref{app:ControlPulseParameters}. We employ the algorithms in the open-source Nlopt library, see \REF\cite{NLopt}, to perform the optimisation of the control pulse parameters.

\change{The smallest values for the distance $\DNV$ and the average infidelity $\IFV$ obtained with the circuit Hamiltonian \equref{eq:Circuit_Ham_def}(effective Hamiltonian \equref{eq:Eff_Ham_def}) are $\DNV=0.0080$ and $\IFV=0.0004$ ($\DNV=0.0089$ and $\IFV=0.0004$) for the $\ROT_{1}(\pi/2)$ ($\ROT_{0}(\pi/2)$) gate and the two-qubit NIGQC. Note that the same gate modelled with the circuit Hamiltonian \equref{eq:Circuit_Ham_def}(effective Hamiltonian \equref{eq:Eff_Ham_def}) on the four-qubit NIGQC yields $\DNV=0.058$ and $\IFV=0.004$ ($\DNV=0.060$ and $\IFV=0.004$). Consequently, we have lost about one order of magnitude in accuracy by adding additional circuit elements to the system.}

\change{The largest values for the distance $\DNV$ and the average infidelity $\IFV$ obtained with the circuit Hamiltonian \equref{eq:Circuit_Ham_def}(effective Hamiltonian \equref{eq:Eff_Ham_def}) are $\DNV=0.144$ and $\IFV=0.029$ ($\DNV=0.146$ and $\IFV=0.015$) for the $\CZ_{1,2}$ gate and the four-qubit NIGQC.} 

We can potentially explain the large values for the $\CZ_{1,2}$ gate-error metrics in both models by means of the energy level repulsions (\change{ELR}s) in \figsref{fig:device_spectrum}(b). The \change{ELR} used to implement the $\CZ_{1,2}$ gate is too close to other \change{ELR}s which lead to additional transitions between the states of the system. The optimisation algorithm cannot solve this problem. Note that we employ the quite accurate (see \REF\cite{Didier} and \REF\cite[Appendix B]{Lagemann21}) series expansions in \equaref{eq:tunable_freq}{eq:tunable_anharm} to model the tunable qubit frequency and anharmonicity, respectively. If we employ the less precise first-order expansions, we cannot reproduce the results for the $\CZ_{1,2}$ gate with the effective model.

Furthermore, in \figsref{fig:metrics}(a-d) we can clearly observe a trend to larger gate-error metrics for larger systems, this is the case for both models. Note that the smallest system contains three circuit elements and the largest system consists of eight circuit elements.

While optimising the control pulse parameters for the two-qubit gates, we noticed that with increasing system size, \change{\ie the number of transmon qubits and couplers}, it becomes more difficult to pass though the various \change{ELR}s in the spectrum sufficiently fast (slowly), see \figsref{fig:device_spectrum}(a-c). If we let the pulse flanks, see \figref{fig:PulseTimeEvolution}(b), rise (fall) too slowly, we pass though various \change{ELR}s so slowly that unwanted population exchanges occur. If we let the pulse flanks rise (fall) too fast, we observe what are probably \changetwo{nonadiabatic} transitions which can even excite the resonators. Obviously, there exists an analogous problem for drive frequencies $\CctlDf$ of the single-qubit $\ROT(\pi/2)$ rotations since the MP pulses are not strictly monochromatic. Both these problems can potentially explain the tendency to larger gate-error metrics in larger NIGQCs.

Furthermore, we also need to consider the difficult task given to the optimisation algorithm. Optimising the control pulse parameters for an $N$-qubit NIGQC amounts to aligning the time evolution of the system such that the matrix $M$ given by \equref{eq:prop_matrix} with $2^{2N+1}$ double precision numbers is quasi perfectly aligned with the target matrix $U$. We do not know of an optimisation algorithm that can solve such a task with guarantee of success.

We also find (data not shown) that we cannot simply use the same optimised control pulse parameters for both the circuit model and the effective model. If we use the parameters of the circuit model for the effective model, we can observe diamond distances and average infidelities close to one. The reason for this is that parameters like the drive frequency $\CctlDf$ ($\hbar \CctlDf$) of the MP pulse and the pulse amplitude $\FctlAp$ ($\hbar \omega^{(q)}(\delta)$) of the UMP must be at least fine tuned up to the sixth decimal (a couple of \change{kHz}). Also, the values that the optimisation algorithm obtains are very sensitive to changes in the model and the model parameters. For example, if the interaction strength $G$ is changed from $300$ MHz to $301$ MHz, we would already need to restart the whole optimisation of the control pulse parameters to obtain gate-error metrics that are not significantly worse than the ones shown in \figsref{fig:metrics}(a-d). Obviously, this lack of robustness can be expected to become even more severe when an actual experiment, instead of a simulation, is conducted.

\subsection{Influence of higher states on gate-error trajectories obtained with the circuit Hamiltonian}\label{sec:Influence of higher states on gate-error trajectories}
In this section, we discuss simulation results for the implementation of $\ROT_{0}(\pi/2)$ gates on the four-qubit NIGQC illustrated in \figsref{fig:device_sketch}(c), modelled with the circuit Hamiltonian \equref{eq:Circuit_Ham_def} using four and sixteen basis states for the transmon qubits. All resonators are modelled with four basis states. \changethree{Also}, we use a fixed set of control pulse parameters for all simulations, see \tabref{tab:CtlSqgIVMB} row one in \appref{app:ControlPulseParameters}.

\change{In \tabref{tab:states} we show the results for two different simulations. In the first (second) case, we simulate the $\ROT_{0}(\pi/2)$ gate with four (sixteen) basis states for all transmon qubits in the system.}

On the one hand, we can observe that the diamond distance $\DNV$ exhibits an increase in the third decimal. Similarly, the average infidelity $\IFV$ increases in the fourth decimal. On the other hand, we can see that the leakage measure $\LNV$ and the statistical distance $\SDV$ (obtained for the NIGQC computational basis state $\ket{0,0,0,0}$) are the same, up the fourth decimal. Note that the leakage measure $\LNV$ and statistical distance $\SDV$ are computed from the squares of the state vector amplitudes only. Consequently, the phase of the system is neglected completely. This makes both these quantifiers less susceptible to changes in the number of basis states.
\begin{table}[tbp!]
\caption{Error metrics for a four-qubit NIGQC as illustrated in \change{\figref{fig:device_sketch}(c)}. The error metrics are obtained with the \textbf{circuit Hamiltonian} \equref{eq:Eff_Ham_def}, the device parameters listed in \tabref{tab:device_parameters} and the pulse parameters listed in \tabref{tab:CtlSqgIVMB}. First column: target gate; second column: number of basis states used to model the dynamics of the transmons; third column: diamond distance $\DNV$ given by \equaref{eq:diamond_norm_inf}{eq:diamond_norm_sup}; fourth column: average infidelity $\IFV$ given by \equref{eq:fid_avg}; fifth column: leakage measure $\LNV$ given by \equref{eq:leak}; sixth column: statistical distance given by \equref{eq:stat_dis}. The statistical distance is obtained for the ground state of the NIGQC.\label{tab:states}}
\begin{ruledtabular}
\begin{tabular}{c c c c c c }
$\text{Gate}$& $\text{States}$&     $\DNV$&              $\IFV$&              $\LNV$&              $\SDV$\\

\hline

$\ROT_{0}(\pi/2)$ & $4$   &              $0.0505$       &              $0.0037$       &              $0.0024$       &              $0.0014$       \\

$\ROT_{0}(\pi/2)$ & $16$  &          $0.0584$       &              $0.0040$       &              $0.0024$       &              $0.0014$       \\

\end{tabular}
\end{ruledtabular}
\end{table}

\renewcommand{\hold}{0.74}
\begin{figure}[!tbp]
  \centering
  \begin{minipage}{0.495\textwidth}
      \centering
      \includegraphics[scale=\hold]{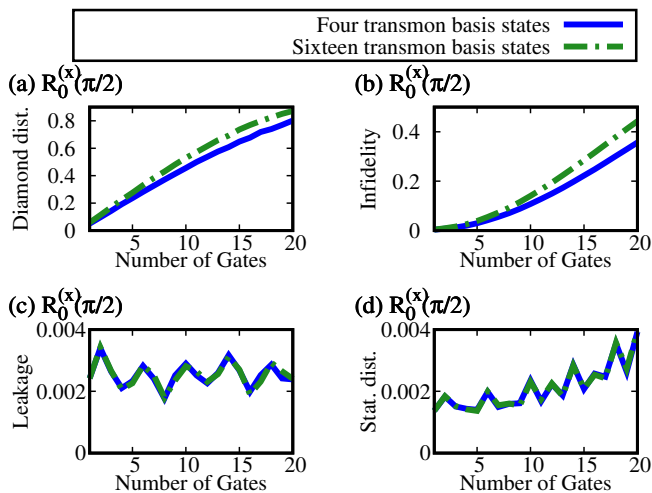} %
  \end{minipage}
  \caption{(Color Online) Gate errors as functions of the number of gates for a program which executes twenty $\ROT(\pi/2)$ in a row. (a) diamond distance; (b) average infidelity; (c) leakage measure; (d) statistical distance for the initial state $\ket{0,0,0,0}$. We ran the gate sequence on the four-qubit NIGQC illustrated in \figref{fig:device_sketch}(c). The results are obtained with the \textbf{circuit Hamiltonian} \equref{eq:Circuit_Ham_def} and the device parameters listed in \tabref{tab:device_parameters}. We ran the program twice. The first time with four basis states for every flux-tunable transmon in the system. These results are displayed with blue solid lines. The second time with sixteen basis states for every flux-tunable transmon in the system. These results are displayed with green dash-dotted lines. We observe that the gate-error metrics in (a-b) deviate by about 10\% after twenty repetitions. The deviations for the gate-error metrics in (c-d) are smaller by more than a factor of one hundred.\label{fig:STATES}}
\end{figure}

Figures~\ref{fig:STATES}(a-d) show the diamond distance $\DNV$(a), the average infidelity $\IFV$(b), the leakage measure $\LNV$(c) and the statistical distance $\SDV$(d) as functions of the number of consecutively executed $\ROT_{0}(\pi/2)$ gates. Furthermore, we obtain the blue line when all transmon qubits are modelled with four basis states only. The green line is obtained when all transmon qubits are modelled with sixteen basis states. The statistical distance $\SDV$ is obtained for the initial state $\ket{0,0,0,0}$ of the NIGQC.

The diamond distance $\DNV$(a) and the average infidelity $\IFV$(b) modelled with four (in blue) and sixteen (in green) basis states start to deviate after the execution of a couple of $\ROT_{0}(\pi/2)$ gates. Finally, after twenty $\ROT_{0}(\pi/2)$ gates we find that the diamond distance $\DNV$(a) and the average infidelity $\IFV$(b) deviate by about 10\% for both cases. We also observe, that the leakage measure $\LNV$ and the statistical distance $\SDV$ are less affected by changing the number of basis states, \ie both gate-error quantifiers are roughly the same up to the fourth decimal. Note that usually gates are modelled with two or three basis states only, see for example \REFS\cite{Gu21,Wittler21,McKay16,Roth19,Rol19,Yan18}.

\changetwo{Finally, we conclude that gate-error metrics obtained with a fixed number of basis states are only valid if the corresponding numerical values have actually converged. Changing the number of basis states can result in a new NIGQC model and there is no guarantee that the gate-error metrics obtained with the new model are the same as for the old model. Thereby we exclude the unlikely case that one can explicitly show that the truncated time-evolution operators, see \equaref{eq:TDSEOperator}{eq:prop_matrix}, for both models are the same. In practice, one has to increase the number of basis states until the results converge to a stable value that is independent of further increase.
We followed this procedure for all the results presented in the rest of the manuscript. However, since in this section the intention was to highlight the relevance of this procedure, we intentionally showed results for both four and sixteen basis states in \tabref{tab:states} and \figref{fig:STATES}. These results additionally show that some gate-error metrics are less susceptible to changes in the number of basis states and that a growing number of gates might require us to use more basis states, to model the dynamics of the system. Interestingly, this type of resilience might also be relevant for experiments, in which the higher states may also be substantially different than expected from theory~\cite{willsch2023josephsonharmonics}}. 

\subsection{Influence of parameter changes on gate-error trajectories obtained with the circuit Hamiltonian}\label{sec:Influence of small pulse parameter deviations on gate-error trajectories}
In this section we discuss simulation results for the implementation of $\CNOT_{0,1}=\HA_{0}\CZ_{0,1}\HA_{0}$ gate repetition programs executed with the two-qubit, three-qubit and four-qubit NIGQCs illustrated in \figsref{fig:device_sketch}(a-c), modelled with the circuit Hamiltonian \equref{eq:Circuit_Ham_def} and slightly different $\delta + \Delta\delta$ control pulse parameters for the $\CZ_{0,1}$ gates implemented with the UMP given by \equref{eq:flux_ctl_ump}. In the following, $\Delta\delta$ denotes the offset value. The original control pulse parameters for the UMPs we use to implement the $\CZ_{0,1}$ gates are listed in \tabref{tab:CtlTqgIIMB} (\tabref{tab:CtlTqgIIIMB}, \tabref{tab:CtlTqgIVMB}) row one for the two-qubit (three-qubit, four-qubit) NIGQC. All $\HA_{0}$ gates are implemented with $\ROT(\pi/2)$ rotations and virtual $Z$ gates. We use the MP given by \equref{eq:charge_ctl} and the control pulse parameters listed in \tabref{tab:CtlSqgIIMB} (\tabref{tab:CtlSqgIIIMB}, \tabref{tab:CtlSqgIVMB}) row one for the two-qubit (three-qubit, four-qubit) NIGQC. Note that for these simulations we do not optimise the circuit by eliminating the $\HA_{0}$ gates. We are interested in how the errors caused by different $\Delta\delta$ interact with the single-qubit rotations. \changetwo{In this section, we model the dynamics of Hamiltonian \equref{eq:Circuit_Ham_def} with sixteen basis states for all transmon qubits, except the one with the index $i=0$, cf.~\figsref{fig:device_sketch}(a-c), where we use four basis states. Additionally all resonators are modelled with four basis states only}.

\begin{table}[tbp!]
\caption{Error metrics for a two-qubit NIGQC as illustrated in \figsref{fig:device_sketch}(a). The error metrics are obtained with the \textbf{circuit Hamiltonian} \equref{eq:Eff_Ham_def}, the device parameters listed in \tabref{tab:device_parameters} and the pulse parameters listed in \tabaref{tab:CtlSqgIIMB}{tab:CtlTqgIIMB}. First column: target gate; second column: offset value $\Delta\delta$ we use to implement the $\CZ$ gates; third column: diamond distance $\DNV$ given by \equaref{eq:diamond_norm_inf}{eq:diamond_norm_sup}; fourth column: average infidelity $\IFV$ given by \equref{eq:fid_avg}; fifth column: leakage measure $\LNV$ given by \equref{eq:leak}; sixth column: statistical distance given by \equref{eq:stat_dis}. The statistical distance is obtained for the ground states of the NIGQCs. We employ the UMP given by \equref{eq:flux_ctl_ump} to obtain the results. \label{tab:para_devi}}
\begin{ruledtabular}
\begin{tabular}{c c c c c c }
$\text{Gate}$& $\Delta\delta/\TP$&              $\DNV$&              $\IFV$&              $\LNV$&              $\SDV$\\

\hline

$\CNOT_{0,1}$& $0$      &       $0.0386$       &              $0.0018$       &              $0.0012$       &              $0.0013$       \\

$\CNOT_{0,1}$& $10^{-6}$          &   $0.0390$       &              $0.0018$       &              $0.0012$       &              $0.0013$       \\

$\CNOT_{0,1}$& $10^{-5}$          &   $0.0456$       &              $0.0022$       &              $0.0012$       &              $0.0013$       \\

$\CNOT_{0,1}$& $10^{-4}$           &   $0.1594$       &              $0.0156$       &              $0.0018$       &              $0.0038$       \\

\end{tabular}
\end{ruledtabular}
\end{table}
\tabref{tab:para_devi} shows the gate-error quantifiers for the execution of a single $\CNOT_{0,1}$ on the two-qubit NIGQC illustrated in \figsref{fig:device_sketch}(a). The results are obtained with four slightly different pulse amplitudes $\delta+\Delta \delta$ for the UMP which implements the $\CZ_{0,1}$ gate. We use $\Delta \delta/\TP=0$ in row one, $\Delta \delta/\TP=10^{-6}$ in row two, $\Delta \delta/\TP=10^{-5}$ in row three and $\Delta \delta/\TP=10^{-4}$ in row four.

\change{We observe that the diamond distance $\DNV$ is affected by the third decimal if we change the pulse amplitude by $\Delta \delta/\TP=10^{-6}$. However, for this case the average infidelity $\IFV$ is the same up to the fourth decimal. If we consider the offset value $\Delta \delta/\TP=10^{-5}$ ($\Delta \delta/\TP=10^{-4}$), we find that the diamond distance $\DNV$ is affected by the second (first) decimal and the average infidelity $\IFV$ is affected by the third (second) decimal.} 

\change{If we consider the flux-tunable qubit frequency of the transmon qubit we drive, see $\omega_{1}^{(Q)}$ in \figref{fig:device_sketch}(a), we find that keeping the pulse amplitude stable up to the sixth (fourth) decimal, for the offset $\Delta \delta/\TP=10^{-6}$, means controlling the flux-tunable qubit frequency up to a couple of kHz (MHz)}. For this estimate of the energy scale we considered the spectrum of the corresponding circuit Hamiltonian with the device parameters listed in \tabref{tab:device_parameters} row $i=1$.

\renewcommand{\hold}{0.74}
\begin{figure}[!tbp]
  \centering
  \begin{minipage}{0.495\textwidth}
      \centering
      \includegraphics[scale=\hold]{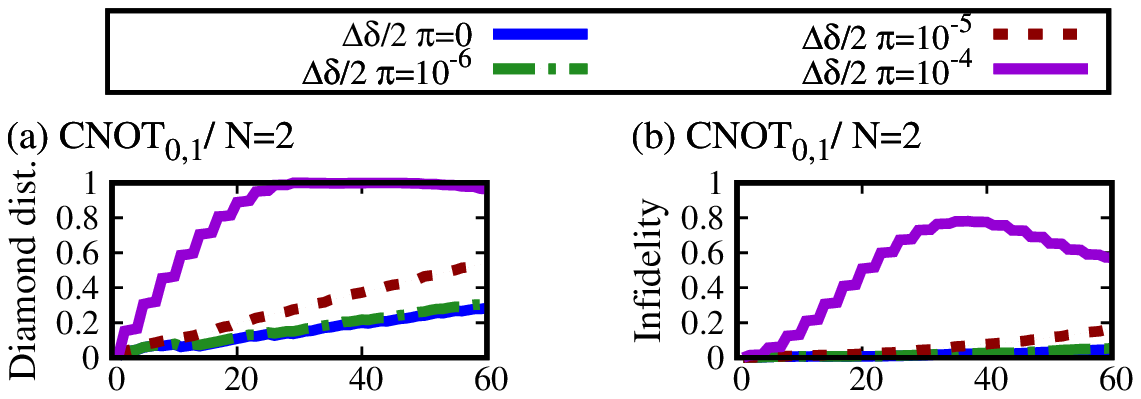}
      \includegraphics[scale=\hold]{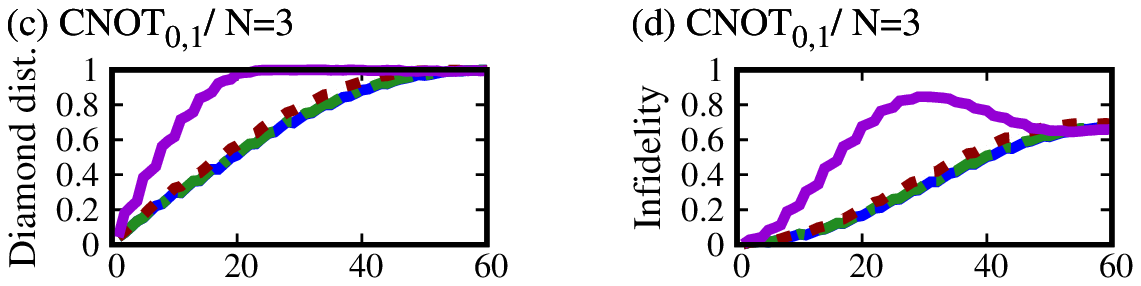}
      \includegraphics[scale=\hold]{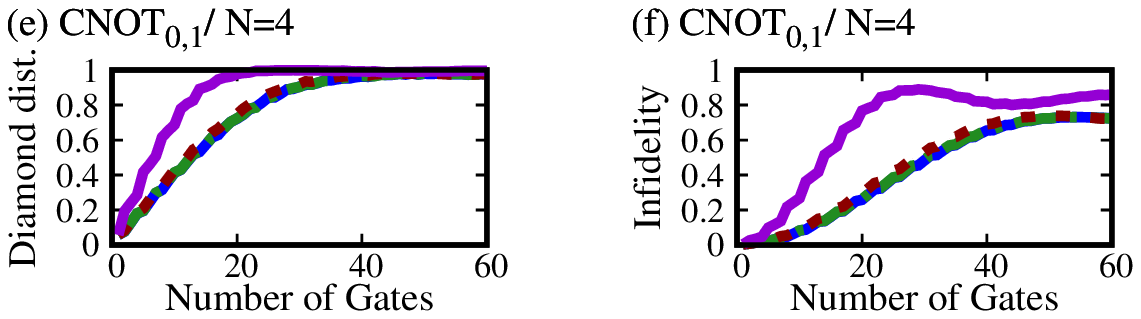}
  \end{minipage}
  \caption{(Color Online) Gate errors as functions of the number of gates for a program which executes twenty $\CNOT$ gates in a row. Note that every $\CNOT$ is implemented with one UMP pulse and two MP pulses. (a,c,e) diamond distance; (b,d,f) average infidelity. In (a,b) we run the program on the two-qubit NIGQC; in (c,d) we run the program on the three-qubit NIGQC and in (e,f) we run the program on the four-qubit NIGQC. The two-, three- and four-qubit systems are illustrated in \figref{fig:device_sketch}(a-c), respectively. The results are obtained with the \textbf{circuit Hamiltonian} \equref{eq:Circuit_Ham_def} and the device parameters listed in \tabref{tab:device_parameters}. We run the program four times on each NIGQC. Each time we added an offset $\Delta\FctlAp$ to the pulse amplitude $\FctlAp$ of the UMP which implements the $\CZ$ gates. The results for the offset $\Delta\FctlAp/\TP=0$ ($\Delta\FctlAp/\TP=10^{-6}$, $\Delta\FctlAp/\TP=10^{-5}$, $\Delta\FctlAp/\TP=10^{-4}$) are shown in blue (green, red, violet) with solid (dashed, dashed, solid) lines. For the offset $\Delta\FctlAp/\TP=10^{-4}$ we can observe some type of tipping behaviour, \ie the gate-error trajectory changes its form completely, see (a,b,d,f). \changethree{Also}, we observe interesting non-linear behaviour for the gate-error trajectories which are generated by the $\CNOT$ sequence.\label{fig:PARA}}
\end{figure}

Figures~\ref{fig:PARA}(a-f) show the diamond distance $\DNV$(a,c,e) and the infidelity $\IFV$(b,d,f) as functions of the number of gates for the two-qubit (a-b), three-qubit (c-d) and four-qubit (e-f) NIGQCs illustrated in \figsref{fig:device_sketch}(a-c). Here we executed a gate sequence which contains twenty $\CNOT$ gates in a row on the different NIGQCs. In each panel we show the results for four different offsets; $\Delta \delta/\TP=0$ (blue solid), $\Delta \delta/\TP=10^{-6}$ (green dashed), $\Delta \delta/\TP=10^{-5}$ (red dashed) and $\Delta \delta/\TP=10^{-4}$ (violet solid). Note that each $\CNOT$ is implemented with two MPs and one UMP and we do not remove the $\HA_{0}$ gates from the circuit.

We observe that the qualitative and quantitative behaviour of the gate-error trajectories barely changes for the offset $\Delta \delta/\TP=10^{-6}$, small deviations only become noticeable at the end of the repetition program. Once we increase the offset to $\Delta \delta/\TP=10^{-5}$, we find that the qualitative and quantitative behaviour of the gate-error trajectories can be affected after a couple of gates. If we consider the offset to $\Delta \delta/\TP=10^{-4}$, we can see some type of tipping behaviour in all panels, in the sense that the qualitative and quantitative behaviour of the gate-error trajectories can change in a non-linear manner. Note that one can obtain similar results by adding a frequency offset $\Delta \omega^{(D)}$ to the drive frequency $\omega^{(D)}$ of the single-qubit gate control pulse in \equref{eq:charge_ctl}.

Finally, we conclude that the stability of the gate-error metrics in our circuit Hamiltonian NIGQC model depends on our ability to control the flux-tunable qubit frequencies (the pulse amplitudes) up to a couple of \change{kHz} $(\Delta\delta/\TP=10^{-6})$. \change{Note that if we execute circuits which contain $\CZ$ gates, \eg $\CZ_{0,1}$, $\CZ_{1,2}$, $\CZ_{2,3}$ and $\CZ_{0,3}$, on different qubits and the corresponding UMPs are affected by the same offset $\Delta\delta$, then the system can become much more sensitive with regard to the offset value $\Delta \delta$, in comparison with the data, see \figsref{fig:PARA}(a-f) we discussed before}.

\change{We can convert the offset factor $\Delta\delta/\TP=10^{-6}$ for the flux $\Delta\Phi=(\Phi_{0}\Delta\varphi)/\TP$ from Weber to Tesla for an area of ten by ten micrometer which is characteristic for a flux-tunable transmon, see \REF\cite{Roth22}. Using the circuit Hamiltonian \equref{eq:Circuit_Ham_def} to simulate NIGQCs, we find that the gate-error metrics are sensitive to field strengths of about $10^{-11}$ Tesla. Here we assume that the external flux is given by $\Phi=|B| A$, where $|B|$ is the magnetic field strength and $A$ is the area of the surface, where the flux is threading through. For comparison, the earth's magnetic field strength is about $10^{-5}$ Tesla.}

Furthermore, the data presented in this section suggests that, in our NIGQC model the gate-error metrics obtained for a single gate cannot be used to predict how the gate-error metrics behave in the future. Consequently, the future behaviour of a gate-error trajectory is not only determined by its initial value. We can explain this finding in very simple terms. The future state of the system $\ket{\Psi(t)}$ is governed by the \changetwo{time-dependent Schr\"odinger equation} given by \equref{eq:TDSE} and not by the value of the gate-error metric $\mu_{x}$ itself ($x$ is a label for an arbitrary gate-error metric). Therefore, there is no reason why gate-error metrics suffice to allow a prediction of the time evolution. They are at most a measure of closeness at one particular point in time, i.e. a snapshot of the state of the system.

\subsection{Influence of the adiabatic approximation on gate-error trajectories obtained with the effective Hamiltonian}\label{sec:InfluenceOfTheAdiabaticApproximationOnGate-errorTrajectories}
\newcommand{\ListFastFluxPulses}{\cite{Foxen20,Gu21,Rol19,Roth19,McKay16,Baker22,Yan18,Blais2020circuit} }

In this section we discuss simulation results for the implementation of $\CZ_{0,1}$ and $\CNOT_{0,1}=\HA_{0}\CZ_{0,1}\HA_{0}$ gate repetition programs executed with the two-qubit, three-qubit and four-qubit NIGQCs illustrated in \figsref{fig:device_sketch}(a-c), modelled with the effective Hamiltonian \equref{eq:Eff_Ham_def} in the adiabatic and the \changetwo{nonadiabatic} regime. In the adiabatic regime we set the time derivative $\dot{\varphi}(t)$ of the external flux to zero such that the drive term in \equref{eq:drive_flux} is set to zero too. In the \changetwo{nonadiabatic} regime, we do not make this assumption, \ie we model the flux-tunable transmons as \changetwo{nonadiabatic} anharmonic oscillators in the harmonic basis. \changetwo{For our simulations, we use four basis states for all transmons and resonators in the model}.

For our simulations we employ the UMP, see \figref{fig:PulseTimeEvolution}(b), given by \equref{eq:flux_ctl_ump} and the BMP, see \figref{fig:PulseTimeEvolution}(c), given by \equref{eq:flux_ctl_bmp} to test whether or not it is appropriate to model flux-tunable transmons in the adiabatic regime as it is commonly done in the literature, see e.g., \REFS\ListFastFluxPulses.

The results for the UMP (BMP) are presented in \tabref{tab:EMAPPROXUMPHB} (\tabref{tab:EMAPPROXBMPHB}) and \figsref{fig:APPROX_UMP}(a-l) (\figsref{fig:APPROX_BMP}(a-l)). Note that the BMP in \figref{fig:PulseTimeEvolution}(c) shows a fast falling flank at around half of the pulse duration. Consequently, we expect that the deviations between the adiabatic and \changetwo{nonadiabatic} case are larger for the BMP. Furthermore, the UMP and BMP are characterised by long time intervals of about $80$ ns where the derivative $\dot{\varphi}(t)$ of the external flux is zero in both models. Consequently, the deviations between the adiabatic and \changetwo{nonadiabatic} model should originate from the pulse flanks we can see in \figsref{fig:PulseTimeEvolution}(b-c).

The control pulse parameters for the UMPs and BMPs we use to implement the $\CZ_{0,1}$ gates are listed in \tabref{tab:CtlTqgIIHB} (\tabref{tab:CtlTqgIIIHB}, \tabref{tab:CtlTqgIVHB}) row one and two for the two-qubit (three-qubit, four-qubit) NIGQC. All $\HA_{0}$ gates are implemented with $\ROT(\pi/2)$ rotations and virtual $Z$ gates. We employ the MP given by \equref{eq:charge_ctl} and the control pulse parameters listed in \tabref{tab:CtlSqgIIHB} (\tabref{tab:CtlSqgIIIHB}, \tabref{tab:CtlSqgIVHB}) row one for the two-qubit (three-qubit, four-qubit) NIGQC. 

\begin{table}[tbp!]
\caption{Error metrics for a two-qubit, three-qubit and four-qubit NIGQC as illustrated in \figsref{fig:device_sketch}(a-c). The error metrics are obtained with the \textbf{effective Hamiltonian} \equref{eq:Eff_Ham_def}, the device parameters listed in \tabref{tab:device_parameters} and the pulse parameters listed in \tabaaref{tab:CtlTqgIIHB}{tab:CtlTqgIIIHB}{tab:CtlTqgIVHB}. First column: target gate; second column: model we use to describe the flux-tunable transmons; \ie the adiabatic or the \changetwo{nonadiabatic} model; third column: diamond distance $\DNV$ given by \equaref{eq:diamond_norm_inf}{eq:diamond_norm_sup}; fourth column: average infidelity $\IFV$ given by \equref{eq:fid_avg}; fifth column: leakage measure $\LNV$ given by \equref{eq:leak}; sixth column: statistical distance given by \equref{eq:stat_dis}. The statistical distance is obtained for the ground states of the NIGQCs. Here we use the unimodal pulse (UMP) given by \equref{eq:flux_ctl_ump} to obtain the results. \label{tab:EMAPPROXUMPHB}}
\begin{ruledtabular}
\begin{tabular}{c c c c c c c }
$\text{Gate}$& $\text{Adiabatic}$& $\text{System}$ &      $\DNV$&              $\IFV$&              $\LNV$&              $\SDV$\\

\hline

$\CZ_{0,1}$& $\text{Yes}$   & \figsref{fig:device_sketch}(a) &              $0.0424$       &              $0.0012$       &              $0.0005$       &              $0.0012$       \\

$\CZ_{0,1}$& $\text{No}$    & \figsref{fig:device_sketch}(a) &             $0.0425$       &              $0.0012$       &              $0.0005$       &              $0.0012$       \\

$\CZ_{0,1}$& $\text{Yes}$   & \figsref{fig:device_sketch}(b) &             $0.0569$       &              $0.0033$       &              $0.0017$       &              $0.0026$       \\

$\CZ_{0,1}$& $\text{No}$    & \figsref{fig:device_sketch}(b) &             $0.0574$       &              $0.0033$       &              $0.0017$       &              $0.0026$       \\

$\CZ_{0,1}$& $\text{Yes}$   & \figsref{fig:device_sketch}(c) &              $0.0514$       &              $0.0040$       &              $0.0028$       &              $0.0025$       \\

$\CZ_{0,1}$& $\text{No}$    & \figsref{fig:device_sketch}(c) &              $0.0509$       &              $0.0040$       &              $0.0028$       &              $0.0025$       \\

\end{tabular}
\end{ruledtabular}
\end{table}
\tabref{tab:EMAPPROXUMPHB} shows the gate-error quantifiers for the execution of a single $\CZ_{0,1}$ on the two-qubit (row one and two), three-qubit (row three and four) and four-qubit NIGQCs (row five and six) illustrated in \figsref{fig:device_sketch}(a-c). We implement the $\CZ$ gates with UMPs. \change{The odd (even) row numbers show the results for the model in the adiabatic (\changetwo{nonadiabatic}) regime}. If we compare both cases for the different systems, we observe that the numerical values for the gate-error metrics and measure are nearly all the same except the ones for the diamond distance $\DNV$, we observe changes in the fourth and third decimal.

\renewcommand{\hold}{0.85}
\begin{figure*}[!tbp]
  \centering
  \includegraphics[scale=\hold]{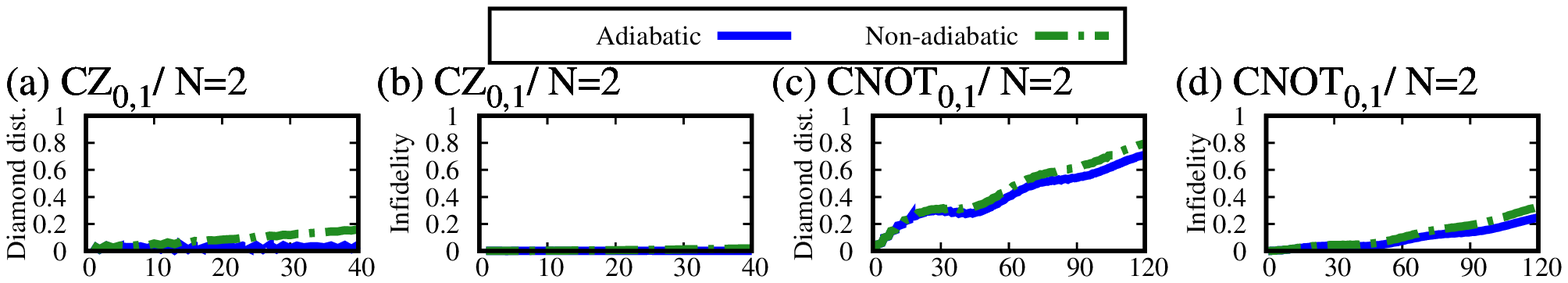}
  \includegraphics[scale=\hold]{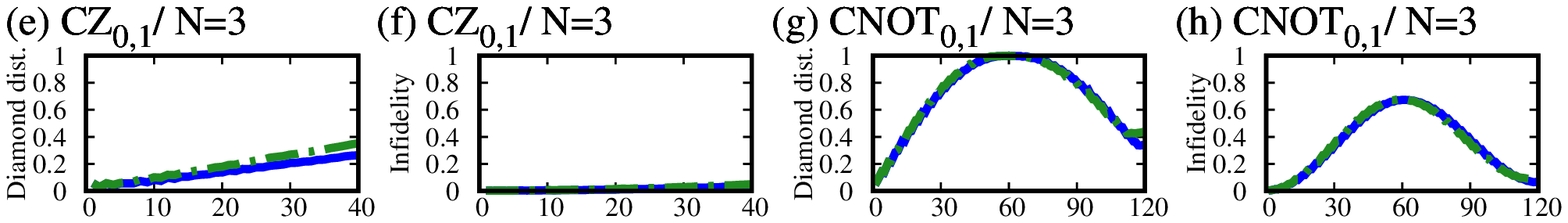}
  \includegraphics[scale=\hold]{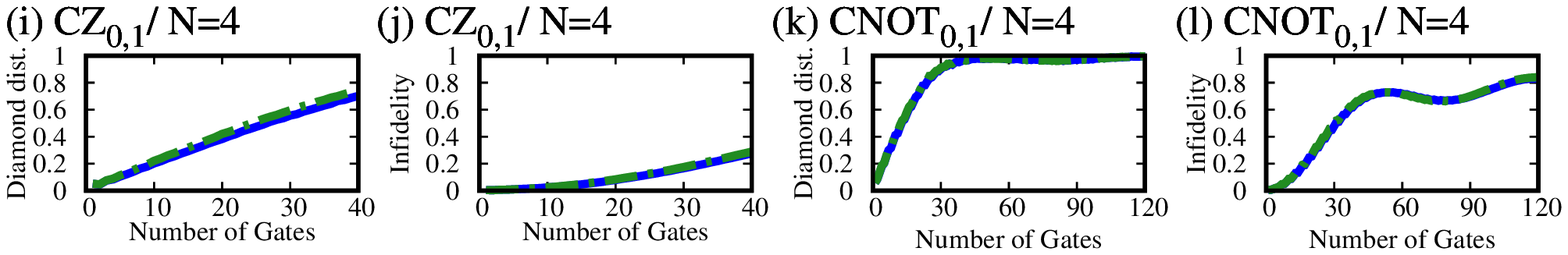}
  \caption{(Color Online) Gate errors as functions of the number of gates for a program which executes forty $\CZ$ in a row, see (a,b,e,f,i,j) and a program which executes forty $\CNOT$ in a row, see (c,d,g,h,k,l). Note that every $\CNOT$ is implemented with one UMP pulse and two MP pulses. (a,c,e,g,i,k) diamond distance; (b,d,f,h,j,l) average infidelity. In (a-d) we run the program on the two-qubit NIGQC, in (e-h) we run the program on the three-qubit NIGQC and in (i-l) we run the program on the four-qubit NIGQC. The two-, three- and four-qubit systems are illustrated in \figref{fig:device_sketch}(a-c), respectively. The results are obtained with the \textbf{effective Hamiltonian} \equref{eq:Circuit_Ham_def} and the device parameters listed in \tabref{tab:device_parameters}. We run the program two times on each NIGQC. On the first run, we model the flux-tunable transmons as adiabatic qubits. These results are displayed with blue solid lines. On the second run, we model the flux-tunable transmons as \changetwo{nonadiabatic} qubits. These results are displayed with green solid lines. We can observe that the gate-error trajectories for the two cases can deviate up to 20\% for the diamond distance and 10\% for the average infidelity. \changethree{Moreover}, we can observe a variety of interesting non-linear behaviour for the gate-error trajectories which are generated by the $\CNOT$ program. \label{fig:APPROX_UMP}}
\end{figure*}

Figures~\ref{fig:APPROX_UMP}(a-l) show the diamond distance $\DNV$(a,e,i,c,g,k) and the average infidelity $\IFV$(b,f,j,d,h,l) as functions of the number of gates for the two-qubit (a-d), three-qubit (e-h) and four-qubit (i-l) NIGQCs illustrated in \figsref{fig:device_sketch}(a-c). Here we executed programs which contain forty $\CZ$(a,b,e,f,i,j) and forty $\CNOT$(c,d,g,h,k,l) gates in a row on the different NIGQCs. As in \secref{sec:Influence of small pulse parameter deviations on gate-error trajectories}, we do not remove the $\HA_{0}$ gates from the circuit sequence to study how the errors of the $\CZ$ gates interact with the errors of the $\HA$ gates. In each panel we show the results for the two different case; adiabatic regime (blue solid) and \changetwo{nonadiabatic} regime (green dashed). Note that each $\CNOT$ is implemented with two MPs and one UMP this results in $120$ gates in total.

\renewcommand{\hold}{0.85}
\begin{figure*}[!tbp]
  \centering
  \includegraphics[scale=\hold]{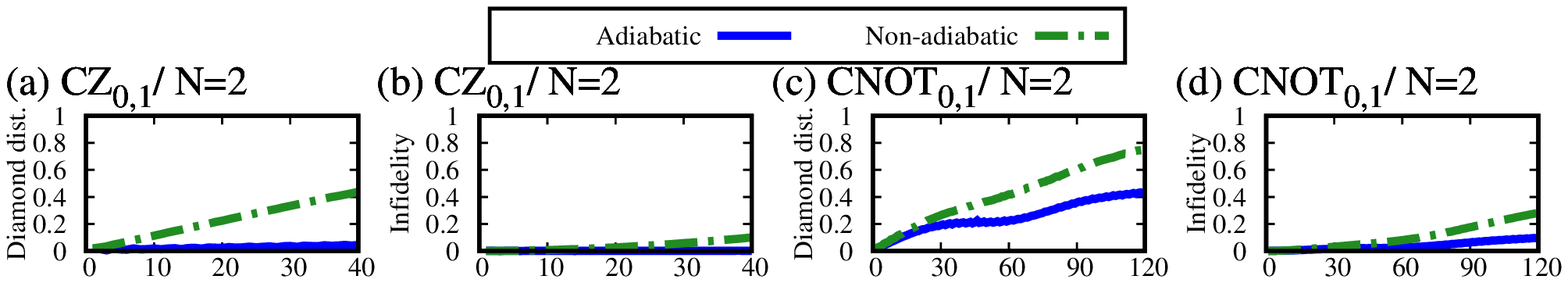}
  \includegraphics[scale=\hold]{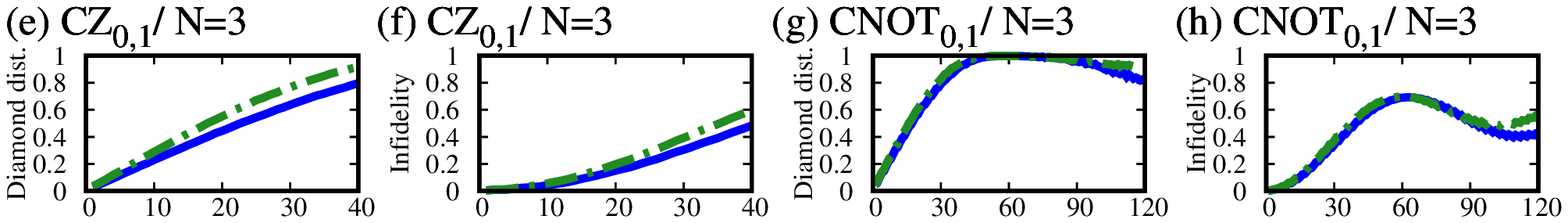}
  \includegraphics[scale=\hold]{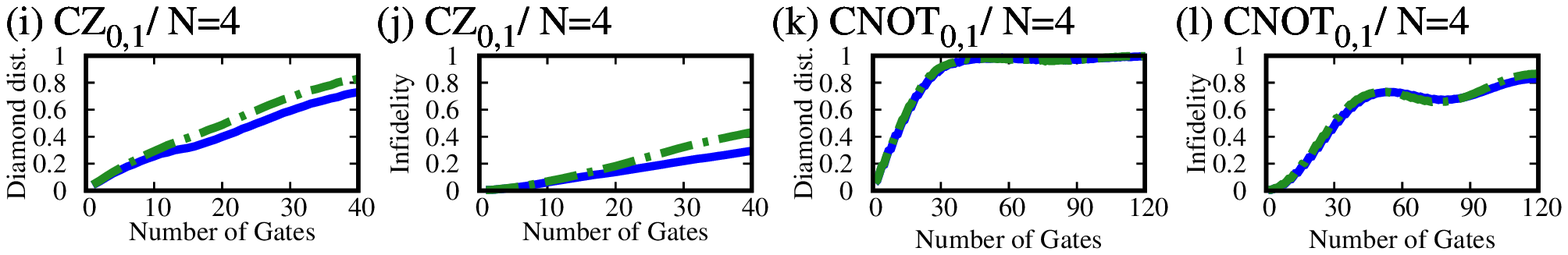}
  \caption{(Color Online) Gate errors as functions of the number of gates for a program which executes forty $\CZ$ in a row, see (a,b,e,f,i,j) and a program which executes forty $\CNOT$ in a row, see (c,d,g,h,k,l). Note that every $\CNOT$ is implemented with one BMP pulse and two MP pulses. (a,c,e,g,i,k) diamond distance; (b,d,f,h,j,l) average infidelity. In (a-d) we run the program on the two-qubit NIGQC, in (e-h) we run the program on the three-qubit NIGQC and in (i-l) we run the program on the four-qubit NIGQC. The two-, three- and four-qubit systems are illustrated in \figref{fig:device_sketch}(a-c), respectively. The results are obtained with the \textbf{effective Hamiltonian} \equref{eq:Circuit_Ham_def} and the device parameters listed in \tabref{tab:device_parameters}. We run the program two times on each NIGQC. On the first run, we model the flux-tunable transmons as adiabatic qubits. These results are displayed with blue solid lines. On the second run, we model the flux-tunable transmons as \changetwo{nonadiabatic} qubits. These results are displayed with green dashed lines. We can observe large qualitative and quantitative deviations for the gate-error trajectories which are generated by the two different models. \changethree{In addition}, we observe a variety of interesting non-linear behaviour for the gate-error trajectories which are generated by the $\CZ$ and $\CNOT$ programs.\label{fig:APPROX_BMP}}
\end{figure*}

If we consider \figsref{fig:APPROX_UMP}(a-d) for the two-qubit NIGQC described by the effective Hamiltonian \equref{eq:Eff_Ham_def}, we find that the small deviations between row one and two in \tabref{tab:EMAPPROXUMPHB} can still affect the time evolution of the system such that the gate-error trajectories for the adiabatic and \changetwo{nonadiabatic} model begin to diverge over time. Additionally, if we consider the results in \figsref{fig:APPROX_UMP}(e-l) we also observe that the qualitative and quantitative behaviour of the gate-error trajectories can simply be affected by the presence of additional circuit elements. Note that we execute the same programs on NIGQCs of increasing size and that the device parameters in \tabref{tab:device_parameters} for the first two transmon qubits and the coupling resonator do not change. However, each time we increase the size of the system we have to optimise the pulse parameters again. Therefore, if we consider the results presented in \secref{sec:Influence of small pulse parameter deviations on gate-error trajectories}, we expect a qualitative and quantitative change in the behaviour of the gate errors once we add additional circuit elements to the system and repeat the pulse optimisation. \changethree{Moreover}, we observe that the gate-error trajectories for the $\CZ$ and $\CNOT$ programs show a substantially different qualitative and quantitative behaviour. Although, the deviations between the adiabatic and \changetwo{nonadiabatic} case seem to be of the same order.

Overall, we find fair qualitative agreement for the adiabatic and \changetwo{nonadiabatic} case. The UMPs we model in this work are quite long compared to instances which can be found in the literature, see for example \REF\cite{Foxen20}. Therefore, we emphasise that one cannot generalise the results presented in this section and argue that neglecting the \changetwo{nonadiabatic} drive term in \equref{eq:drive_flux} is valid for all UMPs.

\begin{table}[tbp!]
\caption{Error metrics for a two-qubit, three-qubit and four-qubit NIGQC as illustrated in \figsref{fig:device_sketch}(a-c). The error metrics are obtain with the \textbf{effective Hamiltonian} \equref{eq:Eff_Ham_def}, the device parameters listed in \tabref{tab:device_parameters} and the pulse parameters listed in \tabaaref{tab:CtlTqgIIHB}{tab:CtlTqgIIIHB}{tab:CtlTqgIVHB}. The rows and columns show the same unit-less quantities as \tabref{tab:EMAPPROXUMPHB}. Here we use the bimodal pulse (BMP) given by \equref{eq:flux_ctl_bmp} to obtain the results.\label{tab:EMAPPROXBMPHB}}
\begin{ruledtabular}
\begin{tabular}{c c c c c c c }
$\text{Gate}$& $\text{Adiabatic}$& $\text{System}$ & $\DNV$&              $\IFV$&              $\LNV$&              $\SDV$\\

\hline

$\CZ_{0,1}$& $\text{Yes}$  & \figsref{fig:device_sketch}(a) &              $0.0167$       &              $0.0006$       &              $0.0005$       &              $0.0004$       \\

$\CZ_{0,1}$& $\text{No}$   & \figsref{fig:device_sketch}(a) &              $0.0195$       &              $0.0007$       &              $0.0005$       &              $0.0004$       \\

$\CZ_{0,1}$& $\text{Yes}$  & \figsref{fig:device_sketch}(b) &              $0.0306$       &              $0.0042$       &              $0.0036$       &              $0.0020$       \\

$\CZ_{0,1}$& $\text{No}$   & \figsref{fig:device_sketch}(b) &              $0.0336$       &              $0.0042$       &              $0.0035$       &              $0.0019$       \\

$\CZ_{0,1}$& $\text{Yes}$  & \figsref{fig:device_sketch}(c) &              $0.0415$       &              $0.0043$       &              $0.0035$       &              $0.0024$       \\

$\CZ_{0,1}$& $\text{No}$   & \figsref{fig:device_sketch}(c) &              $0.0435$       &              $0.0044$       &              $0.0035$       &              $0.0024$       \\

\end{tabular}
\end{ruledtabular}
\end{table}
\tabref{tab:EMAPPROXBMPHB} shows the gate-error quantifiers for the execution of a single $\CZ_{0,1}$ on the two-qubit (row one and two), three-qubit (row three and four) and four-qubit NIGQCs (row five and six) illustrated in \figsref{fig:device_sketch}(a-c). We implement the $\CZ$ gates with BMPs. \change{As before, the odd (even) row numbers show the results for the model in the adiabatic (\changetwo{nonadiabatic}) regime}. The even row number show the results for \changetwo{nonadiabatic} regime. If we compare both cases for the different systems, we notice that the numerical values for the diamond distance $\DNV$ show deviations in the third decimal. Furthermore, the numerical values for some of the average infidelities $\IFV$ show deviations in the fourth decimal. The leakage measure $\LNV$ and the statistical distance $\SDV$ are only affected in one case, see row three and four.

Figures~\ref{fig:APPROX_BMP}(a-l) show the diamond distance $\DNV$(a,e,i,c,g,k) and the infidelity $\IFV$(b,f,j,d,h,l) as functions of the number of gates for the two-qubit (a-d), three-qubit (e-h) and four-qubit (i-l) NIGQCs illustrated in \figsref{fig:device_sketch}(a-c). We executed programs which contain forty $\CZ$(a,b,e,f,i,j) and forty $\CNOT$(c,d,g,h,k,l) gates in a row on the different NIGQCs. As before, we do not remove the $\HA_{0}$ gates from the circuit sequence to study how the errors of the $\CZ$ gates interact with the errors of the $\HA$ gates. In each panel we show the results for the two different cases; adiabatic regime (blue solid) and \changetwo{nonadiabatic} regime (green dashed).

In \figsref{fig:APPROX_BMP}(a-l) we see that the small numerical deviations in the third and fourth decimal for the gate-error metrics listed in \tabref{tab:EMAPPROXBMPHB} can over time result in substantial changes in the qualitative and quantitative behaviour of the gate-error trajectories which result from modelling transmon qubits in the adiabatic regime. Furthermore, we find that overall the deviations between the adiabatic and \changetwo{nonadiabatic} model are much larger than for the case where we implement the $\CZ$ gates with UMPs instead of BMPs. Note that the BMP and UMP in \figsref{fig:PulseTimeEvolution}(b-c) are very similar except for the large pulse flank in the middle of the BMP pulse. If we compare \figsref{fig:APPROX_UMP}(a-d) and \figsref{fig:APPROX_BMP}(a-d), we also find that the qualitative and quantitative behaviour of the gate-error trajectories is not the same if we model them with different pulses. However, this is not surprising since in \secref{sec:Influence of small pulse parameter deviations on gate-error trajectories} we have already seen that gate-error trajectories are also very sensitive to changes in the model parameters and both sets of pulse parameters were optimised independently, \ie this can potentially explain why we see these substantial differences.

We emphasise that we have selected data which highlights the small border between the adiabatic and \changetwo{nonadiabatic} regime. In \appref{app:SimulationsOfTheParametricCouplerDeviceArchitecture}, we show the results for a different device architecture which is discussed in \REFS\cite{Gu21,Roth19,McKay16,Ganzhorn20} for which the results obtained with the two different models deviate much stronger. Furthermore, the authors of \REF\cite{Rol19} use an adiabatic effective model to describe a flux driven two-qubit system which consists of two transmon qubits and a coupling resonator. Here bimodal flux pulses with much shorter gate durations are considered. The data presented in this section suggests that the corresponding numerical results change once the flux drive term in \equref{eq:drive_flux} is part of the effective model. The reader may recall that in this section we only considered one approximation, \ie  assumption, which simplifies the model Hamiltonian. Most of the effective Hamiltonians used are the result of numerous approximations.

Finally, we conclude that it is not possible to decide beforehand whether or not an approximation, \ie an assumption which affects the time evolution of the system, is justified when it comes to modelling gate-error metrics like the diamond distance or the average infidelity. As before, we exclude the unlikely case that we can explicitly show that the truncated time-evolution operators, see \equaref{eq:TDSEOperator}{eq:prop_matrix}, for both models are the same. Consequently, every approximation (assumption) constitutes a new NIGQC model and we do not know whether or not the old reference model yields the same gate-error metrics. The data discussed in this section leads to the conjecture that the quantitative and qualitative behaviour of gate-error trajectories is not simply given by the sum of the individual gate-errors but emerges due to a complex interplay of small deviations with respect to the target gates which occur over time \change{(cf.~\REFS\cite{Alicki2002,Alicki2006,Terhal2005})}. Note that the Hamiltonians, see \equref{eq:Eff_Ham_def}, for the adiabatic and \changetwo{nonadiabatic} case only deviate for small periods of time and that the gate-errors for the different models gradually diverge over time and in the end might show a substantially different qualitative and quantitative behaviour.

Furthermore, as in \secref{sec:Influence of small pulse parameter deviations on gate-error trajectories}, the data presented in this section suggests that, in our NIGQC models the gate-error metrics obtained for a single gate cannot be used to predict how the gate-error metrics for the next gates in the gate sequence (the program) behave. Consequently, the future behaviour of a gate-error trajectory is not only determined by its initial value. As before, we can explain this finding as follows. The future state of the system $\ket{\Psi(t)}$ is still governed by the \changetwo{time-dependent Schr\"odinger equation} given by \equref{eq:TDSE} and not the value of the gate-error metric itself. Consequently, there is no reason why gate-error metrics should allow a prediction of the time evolution of the system beforehand. Note that we can clearly observe a divergence between the two types of NIGQC models we studied in this section.

\section{Summary, Discussion and Conclusion}\label{sec:SummaryAndConclusions}

We have studied the gate-error trajectories which arise if one repeats a gate several times in a row. For the simulations we modelled two-qubit, three-qubit and four-qubit superconducting non-ideal gate-based transmon quantum computers or non-ideal gate-based quantum computers (NIGQCs) for short, see \secref{sec:Introduction} and \figref{fig:device_sketch}(a-c). The time evolution of the state vector $\state$ which by assumption completely determines the state of a NIGQC is generated by the \changetwo{time-dependent Schr\"odinger equation} for a time-dependent Hamiltonian $\op{H}\var$. We used the circuit Hamiltonian \equref{eq:Circuit_Ham_def} and the associated effective Hamiltonian \equref{eq:Eff_Ham_def} to generate the dynamics of the systems. The control pulses discussed in \secref{sec:TheControlPulses} were used to implement two types of gates. We used the microwave pulse (MP) in \equref{eq:charge_ctl}, see \figref{fig:PulseTimeEvolution}(a), to implement the single-qubit $\ROT(\pi/2)$ rotations. We also implemented $\CZ$ gates. We either use the unimodal pulse (UMP) in \equref{eq:flux_ctl_ump}, see \figref{fig:PulseTimeEvolution}(b) or the bimodal pulse (BMP) in \equref{eq:flux_ctl_bmp}, see \figref{fig:PulseTimeEvolution}(c). This allowed us to compute various gate-error metrics like the diamond distance and the average infidelity as functions of the number of gates executed on the NIGQCs. For the computations we implemented the product-formula algorithms to solve the \changetwo{time-dependent Schr\"odinger equation}, see \REF\cite[Section 4.1-4.8]{Lagemann2022}, the algorithms discussed in \REF\cite{LagemannMSCThesis} to determine the state $\ket{\psi}$ of the ideal gate-based quantum computer (IGBC) and the open-source library Nlopt, see \REF\cite{NLopt}, to optimise the control pulse parameters. The complete simulation software with exception of the optimisation algorithms was developed and implemented in house. The main results in this manuscript are presented in \secref{sec:Results}. All results in this work are obtained with the device parameters listed in \tabref{tab:device_parameters}.

In \secref{sec:SpectrumAnalysis} we discussed the spectrum of the four-qubit NIGQC illustrated in \figref{fig:device_sketch}(c) and its relevance for the implementation of two-qubit $\CZ$ gates. We modelled the system with the circuit Hamiltonian \equref{eq:Circuit_Ham_def} and discussed how the complexity of the energy levels which increases with the system size, \change{\ie the number of transmon qubits and couplers}, can affect the gate-error metrics of the two-qubit $\CZ$ gates that we considered. Furthermore, this problem also affects the scaling-up capabilities of the device architecture discussed in this work.

In \secref{sec:GateMetricsAndControlPulseParameters} we discussed the results of the control pulse optimisation, see \figref{fig:metrics} and \tabsref{tab:EMIIMB}{tab:EMIVHB}. The corresponding control pulse parameters are listed in \tabsref{tab:CtlSqgIIMB}{tab:CtlTqgIVHB} in \appref{app:ControlPulseParameters}. We found that the gate-error metrics have a tendency to grow with the system size and that we can explain this tendency by studying the energy spectrum of the system, see for example \figref{fig:device_spectrum}(a-c) and by assessing the difficult task we hand over to the optimisation algorithms that we need to use.

In \secref{sec:Influence of higher states on gate-error trajectories} we studied a simple sequence which consists of twenty single-qubit $\ROT(\pi/2)$ gates on a four-qubit NIGQC, see \figref{fig:device_sketch}(c). We modelled the system by the circuit Hamiltonian \equref{eq:Circuit_Ham_def} and computed gate-error metrics with four and sixteen basis states for every flux-tunable transmon in the system. The results are displayed in \figsref{fig:STATES}(a-d). We found that after twenty repetitions the diamond distances and the average infidelities computed with different numbers of basis states deviate by about 10\%. \changethree{Moreover}, we also saw that the deviations for the leakage error measure and the statistical distances are smaller by about a factor of one hundred. This can potentially be explained by the fact that these gate-error quantifiers are computed from the squares of the state vector amplitudes only. Note that gates are often modelled with two or three basis states only, see for example \REFS\cite{Gu21,Wittler21,McKay16,Roth19,Rol19,Yan18}.

In \secref{sec:Influence of small pulse parameter deviations on gate-error trajectories} we studied a sequence which consists of twenty $\CNOT$ gates on a two-qubit, a three-qubit and a four-qubit NIGQC, see \figsref{fig:device_sketch}(a-c) respectively. The results are shown in \figsref{fig:PARA}(a-f). We modelled the system by the circuit Hamiltonian \equref{eq:Circuit_Ham_def} and used the UMP given by \equref{eq:flux_ctl_ump}, see \figsref{fig:PulseTimeEvolution}(b), to implement the $\CZ$ gates. We repeated the simulations four times and added the offsets $\Delta\delta/\TP \in \{0, 10^{-6}, 10^{-5}, 10^{-4} \}$ to the calibrated control pulse amplitudes $\delta$ which implement the $\CZ$ gates. The parameters can be found in the first rows of \tabaaref{tab:CtlTqgIIMB}{tab:CtlTqgIIIMB}{tab:CtlTqgIIIMB}. We found that the gate-error metrics are affected, to some extent, by all three non-zero offset values $\Delta\delta$. We also noticed some type of tipping behaviour for the offset factor $\Delta\delta/\TP=10^{-4}$, \ie the qualitative and qualitative behaviour of the gate-error trajectories changed drastically for this offset. This was the case for all three systems, see \figsref{fig:device_sketch}(a-c). If we convert the offset factor $\Delta\delta/\TP=10^{-6}$ for the flux $\Delta\Phi=(\Phi_{0}\Delta\varphi)/\TP$ from Weber to Tesla for an area of ten by ten micrometer which is characteristic for a flux-tunable transmon, see \REF\cite{Roth22}, we find that the gate-error metrics in NIGQCs modelled with the circuit Hamiltonian \equref{eq:Circuit_Ham_def} are sensitive to field strengths of about $10^{-11}$ Tesla. \changetwo{For reasons of comparison, we state that the earth's magnetic field strength is about $10^{-5}$ Tesla strong}. Additionally, we found that the gate-error trajectories which are generated by the circuit Hamiltonian \equref{eq:Circuit_Ham_def} and the \changetwo{time-dependent Schr\"odinger equation} exhibit interesting non-linear behaviour.

In \secref{sec:InfluenceOfTheAdiabaticApproximationOnGate-errorTrajectories} we present results for sequences of forty $\CZ$ and forty $\CNOT$ gates on  two-qubit, three-qubit and four-qubit NIGQCs. The results are shown in \figsref{fig:APPROX_UMP}(a-l) and \figsref{fig:APPROX_BMP}(a-l). We modelled the system with the effective Hamiltonian \equref{eq:Eff_Ham_def} and used the UMP to obtain the results in \figsref{fig:APPROX_UMP}(a-l) and the bimodal pulse to to obtain the results in \figsref{fig:APPROX_BMP}(a-l). We ran the $\CZ$ and $\CNOT$ sequences  on two different types of NIGQCs. On the first type of NIGQC the flux-tunable transmons are modelled adiabatically, \ie we set the time derivative $\dot{\varphi}(t)=0$ of the external flux, which is used to implement the $\CZ$ gates, to zero. On the second type of NIGQC we model all flux-tunable transmons non adiabatically. Although the UMP we used seemingly justifies the adiabatic approximation, see \figref{fig:PulseTimeEvolution}(b), we found that the results for the diamond distance can vary up to $0.2$ and the results for the average fidelity up to $0.1$. \changethree{In addition}, in most cases the qualitative behaviour of the gate-error trajectory is not affected by the adiabatic approximation for the UMP. The bimodal pulses we used do not seem to justify making the adiabatic approximation, see \figref{fig:PulseTimeEvolution}(c) and in fact we found that the corresponding gate-error trajectories for the adiabatic and \changetwo{nonadiabatic} case show strong qualitative and quantitative deviations. Additionally, we also observed that the gate-error trajectories which are generated by the effective Hamiltonian \equref{eq:Eff_Ham_def} and the \changetwo{time-dependent Schr\"odinger equation} show interesting non-linear behaviour. Furthermore, we found that the qualitative behaviour of the gate-error trajectories for the $\CZ$ and $\CNOT$ gates are usually not the same. In fact, often they show a completely different behaviour. Therefore, we suspect that the qualitative behaviour of a gate-error trajectory is not simply given by the sum of the individual errors but arises due to a complex interplay of small deviations with respect to the target gates.

The results in \secsref{sec:GateMetricsAndControlPulseParameters}{sec:InfluenceOfTheAdiabaticApproximationOnGate-errorTrajectories} show that even seemingly small changes in the model, \ie in the assumptions we make, can substantially affected the gate-error metrics we compute. The fact itself is not surprising and something one should expect. However, the extent to which the changes affect the gate-error metrics during the course of the time evolution is something worth knowing. Note that in each section we focused on one aspect of the model which affects the computation of gate-error metrics. One can easily imagine what happens if one begins to combine changes in the different aspects of the models, namely that we cannot determine a root cause anymore.

Based on the data presented in \secsref{sec:SpectrumAnalysis}{sec:InfluenceOfTheAdiabaticApproximationOnGate-errorTrajectories}, we conclude that almost all assumptions we make about the model can substantially affect the time-evolution of the systems and consequently the gate-error metrics we model. Therefore, we advocate the view that every assumption leads to a new independent NIGQC model and we simply cannot estimate how the different assumptions affect the gate-error metrics we model. Again we exclude the unlikely case that we can explicitly show that the truncated time-evolution operators, see \equaref{eq:TDSEOperator}{eq:prop_matrix}, for both models are the same. Therefore, the data presented in this work emphasises the narrow borders between certain NIGQC models. Note that we could have selected other data which shows much larger deviations between the various NIGQC models. However, emphasising the narrow path between two seemingly very similar models also has the benefit of adding evidence to our conjecture that in NIGQC models gate errors for consecutive gates are not simply given by the sum of the gate errors for the individual gates in the program sequence but emerge due to a complex interplay of small deviations with respect to the target gates which occur over time \change{(cf.~\REFS\cite{Alicki2002,Alicki2006,Terhal2005})}.

Since we found that nearby values for the diamond distance and the average infidelity for a given target gate can lead to very different qualitative behaviour for the gate-error trajectories which arise if we execute the target gate several times, we conclude that the gate-error metrics for a given target gate cannot be used to predict the behaviour of the gate-error sequence which emerges over time. Note that we showed this with two different generic model Hamiltonians \equaref{eq:Circuit_Ham_def}{eq:Eff_Ham_def}, \ie this is not a feature of one particular model. \changethree{Moreover}, if we take into account the results in \REFS\cite{Wi17,Willsch2020}, we also find that this not a feature of one particular device architecture. Consequently, we advocate the view that gate-error metrics are at most snapshots for the state of a system at one particular moment in time and not predictors of the gate errors which emerge over time and/or the performance for complete programs (algorithms) executed on the system. In fact, this shortcoming is something we should expect. The gate-error metrics we compute are derived in the context of the IGQC model. This model is inherently static, \ie changes in the state of the system are modelled as if they occur instantaneously. However, if the future state of the system is governed by the \changetwo{time-dependent Schr\"odinger equation}, there is no reason why gate-error metrics alone should be able to predict the time-evolution of the system. They are at most a measure of closeness at one particular point in time.

This then leads to the question: how can we actually study gate errors in PGQCs if, in most cases, we cannot conclusively compare two NIGQC models and determine where the differences in the gate-error metrics originate from? In \secref{sec:Introduction}, we briefly mentioned, see \REFS\cite{Murray2021,Martinis14,Rol19,Werninghaus2021,Krinner2019,Burnett2019,Burnett2019}, several problems which plague superconducting PGQCs. The seemingly more complex circuit Hamiltonian \equref{eq:Circuit_Ham_def} is derived on the basis of the lumped-element approximation, see \REF\cite[Section 1.4]{Balanis12} and neglects many of these problems by assumption. Additionally, it is not a trivial task to access the state vector and/or the density operator which is supposed to describes the complete state of a PGQC, also by assumption. Adding this uncertainty to all the other uncertain factors in an experiment, we find ourself in a position where studying individual gate errors in an actual experiment seems impossible. Therefore, we advocate the view that one should use benchmark protocols like the ones discussed in \REFS\cite{Wi17,Michielsen17} to assess and/or compare different PGQCs.

In this work we focused on the execution of simple gate sequences which generate interesting gate-error trajectories for various simulation settings. For future work, it might be interesting to see how variational hybrid algorithms like \change{for example the quantum approximate optimization algorithm, see \REFS\cite{farhi2014quantum,WILLSCHM2022}}, perform on the NIGQCs calibrated for this work.

\begin{acknowledgments}
The authors gratefully acknowledge the Gauss Centre for Supercomputing e.V.~(www.gauss-centre.eu) for funding this project by providing computing time on the GCS Supercomputer JUWELS \cite{JUWELS} at J\"ulich Supercomputing Centre (JSC).~H.L.~acknowledges support from the project OpenSuperQ (820363) of the EU Quantum Flagship.~D.W.'s work was partially supported by the Q(AI)$^{2}$ project.~H.L., D.W.~and M.W.~acknowledge support from the project J\"ulich Unified Infrastructure for Quantum computing (JUNIQ) that has received funding from the German Federal Ministry of Education and Research (BMBF) and the Ministry of Culture and Science of the State of North Rhine-Westphalia.
\end{acknowledgments}


\appendix

\section{Derivation of a circuit Hamiltonian for flux-tunable transmons with an additional charge drive term}\label{app:DerivationOfACircuitHamiltonianForFluxTunableTransmons}
\begin{figure*}[!tbp]
    \renewcommand{\hold}{1.0}
    \centering
    \includegraphics[scale=\hold]{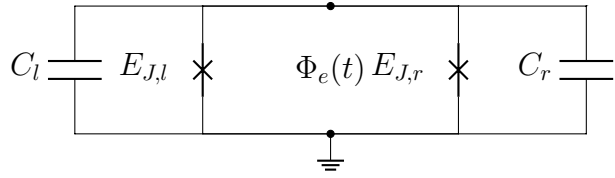}
    \caption{Illustration of a lumped-element circuit with two linear capacitors with capacitances $C_{l}$ (left) and $C_{r}$ (right), two Josephson junctions with Josephson energies $E_{J,l}$ (left) and $E_{J,r}$ (right) and an external flux $\Phi_{e}(t)$ threading through the central loop. As it is common practice, we mark the ground node by a dashed triangle.}\label{fig:FTT_circuit}
\end{figure*}
\newcommand{\PL}{\Phi_{l}(t)}
\newcommand{\PLD}{\dot{\Phi}_{l}(t)}
\newcommand{\ML}{m_{l}}
\newcommand{\PR}{\Phi_{r}(t)}
\newcommand{\MR}{m_{r}}
\newcommand{\PE}{\Phi_{e}(t)}
\newcommand{\PED}{\dot{\Phi}_{e}(t)}
\newcommand{\PV}{\Phi(t)}
\newcommand{\PVD}{\dot{\Phi}(t)}
\newcommand{\PVO}{\op{\Phi}}
\newcommand{\CV}{Q}
\newcommand{\CVO}{\op{Q}}
\newcommand{\MD}{m_{\Delta}}
\newcommand{\CS}{C_{\Sigma}}
\newcommand{\CL}{C_{l}}
\newcommand{\CR}{C_{r}}
\newcommand{\VG}{V_{g}(t)}
\newcommand{\GP}{\beta}

In the main text we use the circuit Hamiltonian \equref{eq:Circuit_Ham} to model flux-tunable transmons with a linear charge drive term.

In this appendix, we derive circuit Hamiltonian \equref{eq:Circuit_Ham} from the assumptions which constitute the lumped-element approximation, see \REF\cite[Section 1.4]{Balanis12}. The external charge variable $n_{g}$(t) provides us with a linear drive term which can be used to implement single-qubit gates. Similarly, the external flux variable $\PE$ provides us with a drive term which can be used to implement two-qubit gates. The following derivation is motivated by the work in \REF\cite{You}, \ie we take into account recent developments in the theory of circuit quantisation. Note that the underlying assumptions we use are slightly different from the ones used in \REF\cite{You}. We use $\hbar=1$ throughout this work.

We begin with the quantisation of the lumped-element circuit illustrated in \figref{fig:FTT_circuit}. Kirchhoff's voltage law yields
\begin{equation}\label{eq:HolonomiConstraint}
  \PL+\PR=\PE,
\end{equation}
for the central loop of the circuit. In our lumped-element model, we treat the external flux $\PE$ as an electromotive force (EMF) $\PED$.

We intend to express the Lagrangian $\mathcal{L}$ of the system in terms of the variable
\begin{equation}
  \PV=\ML\PL+\MR\PR.
\end{equation}
Note that the left
\begin{eqnarray}
  \PL=\frac{\brr{\PV-\MR\PE}}{\MD}
\end{eqnarray}
and the right
\begin{eqnarray}
  \PR=-\frac{\brr{\PV-\ML\PE}}{\MD},
\end{eqnarray}
branch flux variables satisfy \equref{eq:HolonomiConstraint} for all $\ML,\MR \in \mathbb{R}$ without further ado. Here $\MD=\ML-\MR$.

If we make use of the relations $V_{C_{l}}=V_{E_{J_{l}}}$ and $V_{C_{r}}=V_{E_{J_{r}}}$ for the left and right loop in \figref{fig:FTT_circuit}, we can express the Lagrangian as
\begin{equation}
    \begin{split}
      \mathcal{L}&=\frac{C_{l}}{2} \brr{\frac{\brr{\PVD-\MR\PED}}{\MD}}^{2}\\
      &+ \frac{C_{r}}{2} \brr{\frac{\brr{\PVD-\ML\PED}}{\MD}}^{2} - U\brr{\PV},
    \end{split}
\end{equation}
where the potential energy term $U\brr{\PV}$ reads
\begin{equation}
  \begin{split}
    U\brr{\PV}=&-E_{J_{l}} \cos\brr{\frac{2 \pi}{\Phi_{0}} \frac{\brr{\PV-\MR\PE}}{\MD}} \\
    &-E_{J_{r}} \cos\brr{\frac{2 \pi}{\Phi_{0}} \frac{\brr{\PV-\ML\PE}}{\MD}}.
  \end{split}
\end{equation}

In a next step, we evaluate the squares and neglect all factors proportional to $\PED^{2}$, which ultimately only contribute non-measurable global phase factors to the time-evolution of the system. The Lagrangian after this step reads
\begin{equation}
    \mathcal{L}=\frac{\CS}{2\MD^{2}} \PVD^{2} - \frac{\brr{\CL\MR+\CR\ML}}{\MD^{2}}  \PED \PVD - U\brr{\PV}.
\end{equation}
The conjugate variable
\begin{equation}
  Q=\frac{\CS}{\MD^{2}} \PVD - \frac{\brr{\CL\MR+\CR\ML}}{\MD^{2}}\PED,
\end{equation}
can be used to obtain the Hamiltonian function
\begin{equation}
  H=\frac{\MD^{2}}{2\CS} \CV^{2} + \frac{\brr{\CL\MR+\CR\ML}}{\CS}\PED \CV + U\brr{\PV},
\end{equation}
by means of a Legendre transformation, see \REF\cite{Fenchel1949}. Here again we neglect all factors which only contribute a non-measurable global phase factors to the time evolution of the system.

Finally, we can promote the conjugate variables $\Phi$ and $Q$ to the conjugate operators $\PVO$ and $\CVO$ and perform the substitutions
\begin{subequations}\label{eq:substitutions}
  \begin{align}
     \op{\varphi}&=\frac{2 \pi}{\Phi_{0}} \PVO,\\
     \op{n}&=\frac{1}{2 e} \CVO,\\
     \varphi_{e}(t)&=\frac{2 \pi}{\Phi_{0}} \PE,
  \end{align}
\end{subequations}
to obtain the Hamiltonian operator
\renewcommand{\PVO}{\op{\varphi}}
\renewcommand{\CVO}{\op{n}}
\renewcommand{\PE}{\varphi_{e}(t)}
\renewcommand{\PED}{\dot{\varphi}_{e}(t)}
\begin{equation}
  \op{H}=E_{\CS} \CVO^{2} + \frac{\brr{\CL\MR+\CR\ML}}{\CS} \PED \CVO + U\brr{\PE}.
\end{equation}
Here we made use of $\hbar=1$. The parametrisation of the model can be simplified by assuming $\MD=1$, $\CL=\CR$, $\CS=C$ and $\MR=-\GP$. With these assumptions, we can express the Hamiltonian as
\begin{equation}
  \op{H}=E_{C} \CVO^{2} + \brr{\frac{1}{2}-\GP} \PED \CVO + U\brr{\PE}.
\end{equation}

\begin{figure*}[!tbp]
  \centering
   \renewcommand{\hold}{1.0}
    \centering
    \includegraphics[scale=\hold]{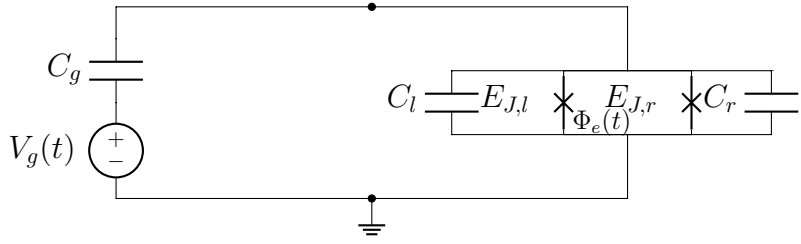}
    \caption{Illustration of a lumped-element circuit. The right branch contains two linear capacitors with capacitances $C_{l}$ (left) and $C_{r}$ (right), two Josephson junctions with Josephson energies $E_{J,l}$ (left) and $E_{J,r}$ (right) and an external flux $\Phi_{e}(t)$ threading through the loop between the two Josephson junctions. The left branch contains a linear capacitor with capacitance $C_{g}$ and a voltage source $V_{g}(t)$. As it is common practice, we mark the ground node by a dashed triangle.}\label{fig:FTT_D_circuit}
\end{figure*}

The next step in the derivation of the circuit Hamiltonian \equref{eq:Circuit_Ham} is to add a linear drive term of the form $n_{g}(t) \op{n}$ to the model. In \figref{fig:FTT_D_circuit} we display a modified circuit. The circuit shown in \figref{fig:FTT_D_circuit}
contains an additional branch with a voltage source modelled with the real-valued function $V_{g}(t)$ and a coupling capacitor with the capacitance $C_{g}$. Note that we model the voltage source also as an EMF.

Kirchhoff's voltage law for the central loop
\begin{equation}
  V_{C_{g}}+V_{g}(t)=V_{C_{l}},
\end{equation}
can be used to obtain the circuit Hamiltonian $\op{H}^{*}$ for the system displayed in \figref{fig:FTT_D_circuit} by making use of the previously obtained results for the circuit displayed in \figref{fig:FTT_circuit}.

The Lagrangian $\mathcal{L}^{*}$ of the modified system reads
\renewcommand{\PE}{\Phi_{e}(t)}
\renewcommand{\PED}{\dot{\Phi}_{e}(t)}
\begin{equation}\label{eq:LagNew}
  \mathcal{L}^{*}=\mathcal{L}+\frac{C_{g}}{2} \brr{\PLD-\VG}^{2}.
\end{equation}
Therefore, $\mathcal{L}^{*}$ can also be expressed in terms of the variable $\PV$. If we evaluate the square in \equref{eq:LagNew}, we find
\begin{equation}
  \mathcal{L}^{*}=\mathcal{L}+\frac{C_{g}}{2\MD^{2}}\PV^{2} - \frac{C_{g}}{\MD^{2}} \PVD \brr{\MR \PED +\MD \VG},
\end{equation}
where we neglect all factors which only contribute a non-measurable global phase factors to the time evolution of the system. As before, we simplify the parametrisation. Assuming $\MD=1$, $\CL=\CR$, $C=\CS$ and $\MR=-\GP$, yields
\begin{equation}
  \begin{split}
    \mathcal{L}^{*}=&\frac{\brr{C + C_{g}}}{2} \PVD^{2} - \brr{\frac{C}{2} - \GP \brr{C + C_{g}}} \PED \PVD \\
    &- C_{g} \VG \PVD - U\brr{\PV},
  \end{split}
\end{equation}
where we also neglected factors which in the end only contribute a non-measurable global phase factors to the time evolution of the system. Next, in an ad hoc manner, we make the assumption that $C+C_{g} \rightarrow C$ in such a way that the system's time evolution can be modelled with the Lagrangian
\begin{equation}
  \begin{split}
    \mathcal{L}^{*}=&\frac{C}{2} \PVD^{2} - C \brr{\frac{1}{2}-\GP} \PED \PVD - C_{g} \VG \PVD\\
                    &- U\brr{\PV}.
  \end{split}
\end{equation}
This allows us to express the conjugate variable as
\begin{equation}
  Q=  C \PVD - C \brr{\frac{1}{2}-\GP} \PED - C_{g} \VG.
\end{equation}
Consequently, the first part of the Hamiltonian function reads
\begin{equation}
  Q\PVD= \frac{Q^{2}}{C} + \brr{\frac{1}{2}-\GP} \PED Q + \frac{C_{g}}{C } \VG Q
\end{equation}
and the second part of the Hamiltonian function can be expressed as
\begin{equation}
  \mathcal{L}^{*}=\frac{Q^{2}}{2C} - U(\PV),
\end{equation}
where all factors which only contribute a non-measurable global phase factors to the time evolution of the system are neglected.
Therefore, adding both parts yields
\begin{equation}
  \begin{split}
      H^{*}=&E_{C} \brr{\frac{\CV}{2 e}}^{2} + \brr{\frac{1}{2}-\GP} (2 e)\PED \brr{\frac{\CV}{2 e}} \\
      &- 2 E_{C} n_{g}(t) \brr{\frac{\CV}{2 e}} + U\brr{\PV},
  \end{split}
\end{equation}
where the real-valued function $n_{g}(t)$ is defined as
\begin{equation}
  n_{g}(t)=- \frac{C_{g} V_{g}(t)}{2 e}.
\end{equation}\renewcommand{\PED}{\dot{\varphi}_{e}(t)}\renewcommand{\PE}{\varphi_{e}(t)}
As before, we use $\hbar=1$ to simplify the Hamiltonian.

If we promote the conjugate variables to conjugate operators, complete the square with regard to the external charge variable, make use of the substitutions in \equref{eq:substitutions}(a-c) and drop all terms which only contribute non-measurable global phase factors to the time evolution of the system, we obtain the result
\begin{equation}\label{eq:cirHamil}
  \begin{split}
      \op{H}^{*}=&E_{C} \brr{\CVO -n_{g}(t) }^{2} + \brr{\frac{1}{2}-\GP} \PED \CVO\\
      &- E_{J,l} \cos\brr{\PVO+\GP \PE}\\
      &- E_{J,r} \cos\brr{\PVO+ (\GP-1) \PE}.
  \end{split}
\end{equation}
In the main text we use the Hamiltonian \equref{eq:cirHamil} to model flux-tunable transmons. Note that in the main text we omit the labels $e$ and $g$ in the real-valued functions $n_{g}(t)$ and $\PE$.

\section{Detailed discussion of the effective Hamiltonian}\label{app:TheEffectiveHamiltonianModel}
In this appendix, we provide a detailed discussion of the effective Hamiltonian \equref{eq:Eff_Ham_def} we use to obtain the results in \secref{sec:InfluenceOfTheAdiabaticApproximationOnGate-errorTrajectories}. Note that we use $\hbar=1$ throughout this work.

The effective Hamiltonian we use to model our NIGQCs, is defined as
\begin{equation}
  \op{H}_{\text{Eff.}}=\op{H}_{\idxRES,\Sigma}+\op{H}_{\idxFTE,\Sigma}+\op{D}_{\idxCD}+\op{\mathcal{D}}_{\idxFD}+\op{W}_{\idxINT}.
\end{equation}
The first term
\begin{equation}
    \op{H}_{\idxRES,\Sigma} =\sum_{k\in K} \omega_{k}^{(R)} \op{a}_{k}^{\dagger}\op{a}_{k},
\end{equation}
describes a collection of non-interacting resonators. Here $K \subseteq \mathbb{N}^{0}$ denotes an index set for the resonators and $\omega_{k}^{(R)}$ refers to the different resonator frequencies. The operators $\op{a}$ and $\op{a}^{\dagger}$ are the bosonic annihilation and creation operators. We use the basis states of the time-independent harmonic oscillator, see \REF\cite[Section 2.5]{Weinberg2015}, as the basis states for our simulations.

The second term
\begin{equation}
  \op{H}_{\idxFTE,\Sigma}=\sum_{j \in J} \omega_{j}^{(q)}(t) \op{b}_{j}^{\dagger}\op{b}_{j} + \frac{\alpha_{j}^{(q)}(t)}{2} \brr{\op{b}_{j}^{\dagger}\op{b}_{j}\brr{\op{b}_{j}^{\dagger}\op{b}_{j}-\op{I}}},
\end{equation}
describes a collection of non-interacting flux-tunable transmons which are modelled as adiabatic, anharmonic oscillators. The operators $\op{b}$ and $\op{b}^{\dagger}$ are the bosonic annihilation and creation operators. The function
\begin{equation}\label{eq:tunable_freq}
  \omega_{j}^{(q)}(t)=\sqrt{2 E_{C_{j}} E_{J_{\text{eff.},j}}(t)} - \frac{E_{C_{j}}}{4} \sum_{n=0}^{24} a_{n} \Xi(t)^{n},
\end{equation}
is used to model the flux-tunable transmon qubit frequency. Here $a_{n}$ are real-valued constants and $E_{C_{j}}$ and $E_{J_{\text{eff.},j}}(t)$ are the capacitive and effective Josephson energies for the transmon qubits, respectively. The latter is defined as
\begin{equation}
  E_{J_{\text{eff.},j}}(t)=E_{\Sigma,j} \sqrt{\cos\brr{ \frac{\varphi_{j}(t)}{2} }^{2}+ d_{j}^{2} \sin\brr{ \frac{\varphi_{j}(t)}{2} }^{2}},
\end{equation}
where $d_{j}=(E_{J_{r,j}}-E_{J_{l,j}})/(E_{J_{r,j}}+E_{J_{l,j}})$. The function $\Xi_j(t)$ in \equref{eq:tunable_freq} is defined as
\begin{equation}
  \Xi_{j}(t)=\sqrt{\frac{E_{C_{j}}}{2 E_{J_{\text{eff.},j}}(t)}}.
\end{equation}
Furthermore, we define the flux-tunable anharmonicity as
\begin{equation}\label{eq:tunable_anharm}
  \alpha_{j}^{(q)}(t)=- \frac{E_{C_{j}}}{4} \sum_{n=0}^{24} b_{n} \Xi(t)^{n},
\end{equation}
where the different $b_{n}$ are real-valued constants. We emphasise that \equaref{eq:tunable_freq}{eq:tunable_anharm} are taken from \REF\cite{Didier}. The functions in \equaref{eq:tunable_freq}{eq:tunable_anharm} are used to approximate the lowest three eigenvalues of the circuit Hamiltonian \equref{eq:Circuit_Ham}. In \REF\cite[Appendix B]{Lagemann21} the authors asses and discuss the accuracy of this approximation.

We use the time-dependent harmonic basis states
\begin{equation}
    \psi^{(m)}(x(t))=\frac{1}{\sqrt{2^{m} m!}} \left(\frac{\xi(t)}{\pi}\right)^{\frac{1}{4}} e^{-\frac{x^{2}(t)}{2}} \mathcal{H}_{m}(x(t)),
\end{equation}
of the time-dependent harmonic oscillator
\begin{equation}
  \op{H}=E_{C} \op{n}^{2} + \frac{E_{J,\text{eff}}(t)}{2} \left(\op{\varphi}-\varphi_{\text{eff}}(t)\right)^{2},
\end{equation}
as the basis states for our simulations of the effective flux-tunable transmons. Additionally, we define the auxiliary functions
\begin{subequations}\label{eq:aux_sub}
  \begin{align}
     \xi(t)&=(E_{J,\text{eff}}(t)/2 E_{C})^{1/2}\\
     x(t)&=\sqrt{\xi(t)}(\varphi-\varphi_{\text{eff}}(t))\\
     \varphi_{\text{eff.}}(t)&=\arctan\brr{d \tan\brr{ \frac{\varphi(t)}{2} }},
  \end{align}
\end{subequations}
and note that $\mathcal{H}_{m}$ denotes the Hermite polynomial of order $m \in \mathbb{N}^{0}$.

The third term
\begin{equation}
  \op{D}_{\idxCD}=\sum_{j \in J} \Omega_{j}(t) \brr{\op{b}_{j}^{\dagger} + \op{b}_{j}},\\
\end{equation}
describes a charge drive. Here $\Omega_{j}(t) \propto -2 E_{C_{j}} n_{j}(t)$ and we approximate the charge operators $\op{n}_{j}$ by effective charge operators $\op{n}_{j,\text{eff.}}$ which can be expressed in terms of the bosonic annihilation and creation operators, see \REF\cite{Koch}.

The fourth term
\begin{equation}\label{eq:app_drive_flux}
  \begin{split}
    \op{\mathcal{D}}_{\idxFD}=&\sum_{j \in J} - i \sqrt{\frac{\xi_{j}(t)}{2}} \dot{\varphi}_{\text{eff.},j}(t) \brr{\op{b}_{j}^{\dagger}- \op{b}_{j}}\\
    +&\sum_{j \in J} \frac{i}{4} \frac{\dot{\xi}_{j}(t)}{\xi_{j}(t)} \brr{\op{b}_{j}^{\dagger} \op{b}_{j}^{\dagger} - \op{b}_{j} \op{b}_{j}},
  \end{split}
\end{equation}
describes a \changetwo{nonadiabatic} flux drive. We find
\begin{equation}
    \dot{\varphi_{\text{eff},j}}(t)=\dot{\varphi}_{j}(t)\frac{d_{j}}{2 \left(\cos\left(\frac{\varphi_{j}(t)}{2}\right)^{2}+d_{j}^{2} \sin\left(\frac{\varphi_{j}(t)}{2}\right)^{2}\right)}
\end{equation}
and
\begin{equation}
    \frac{\dot{\xi}_{j}(t)}{\xi_{j}(t)}=\dot{\varphi}_{j}(t)\frac{(d_{j}^{2}-1) \sin(\varphi_{j}(t))}{8 \left(\cos\left(\frac{\varphi_{j}(t)}{2}\right)^{2}+d_{j}^{2} \sin\left(\frac{\varphi_{j}(t)}{2}\right)^{2}\right)}.
\end{equation}
The term in \equref{eq:app_drive_flux} results from the fact that we model the effective flux-tunable transmon in a time-dependent basis. Therefore, for the \changetwo{time-dependent Schr\"odinger equation} to stay form invariant a time-dependent basis transformation term is needed, see \REF\cite{Lagemann21}.

The fifth term
\begin{equation}
    \op{W}_{\idxINT}=\sum_{(k,j)\in K \times J} g_{k,j}^{(a, b)}(t) \brr{\op{a}_{k}^{\dagger} + \op{a}_{k}} \tens{} \brr{\op{b}_{j}^{\dagger} + \op{b}_{j}},
\end{equation}
describes time-dependent dipole-dipole interactions. As before, the time dependence of the interaction strength
\begin{equation}\label{eq:time_dep_int_strength}
  g_{k,j}^{(a, b)}(t)=G_{k,j} \sqrt[4]{ \frac{ E_{J_{\text{eff.,j}}}(t)}{8 E_{C_{j}}} },
\end{equation}
is a result of the fact that we model the effective flux-tunable transmon in a time-dependent basis, see \REF\cite{Lagemann21}. This time-dependent interaction strength model is motivated by the work in \REF\cite{Koch}.

The relation between the effective and the circuit model is discussed in \REF\cite{Lagemann21}.

\section{Simulations of the parametric coupler device architecture}\label{app:SimulationsOfTheParametricCouplerDeviceArchitecture}
\newcommand{\FctlDf}{\omega^{(D)}}
\newcommand{\FctlFo}{\varphi_{0}}
\newcommand{\FctlTrf}{T_{\text{r/f}} }
\newcommand{\ListParametricCoupler}{\cite{Ganzhorn20,Roth19,Gu21,McKay16,Bengtsson2020}}
\begin{figure}[tbp!]
\renewcommand{\hold}{1.0}
\centering
\includegraphics[scale=\hold]{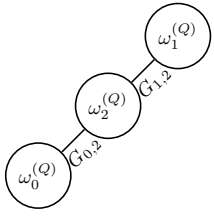}
\caption{Illustrations of fixed-frequency transmon qubits with qubit frequencies $\omega_{0}^{(Q)}$ and $\omega_{1}^{(Q)}$ and a flux-tunable transmon with park frequency $\omega_{2}^{(Q)}$ which constitute a non-ideal gate-based transmon quantum computer (NIGQC) with two qubits. The Hamiltonian we use to model the dynamics of our NIGQC is given by \equref{eq:Eff_Ham_def}. The device parameters we use to specify the Hamiltonian are listed in \tabref{tab:device_parameters_ibm}. The control pulses we use to implement the gate is given by \equref{eq:flux_ctl_sep}.\label{fig:device_sketch_ibm}}
\end{figure}
\begin{table}[tbp!]
\caption{Device parameters for the system illustrated in \figref{fig:device_sketch_ibm}. The units are the same as for the parameters in \tabref{tab:device_parameters}. The device parameters are motivated by the experiments discussed in \REF\cite{Ganzhorn20}.}\label{tab:device_parameters_ibm}
\begin{ruledtabular}
\begin{tabular}{  c c c c c c c c }
$i$&                     $\omega_{i}^{(Q)}/2\pi$&              $\alpha_{i}^{(Q)}/2\pi$&              $E_{C_{i}}/2\pi$&              $E_{J_{l,i}}/2\pi$&              $E_{J_{r,i}}/2\pi$&              $\varphi_{0,i}/2\pi$&                             \\

\hline

0              &                       $5.100$        &              $-0.310$       &              $1.079$        &              $13.446$       &              $0$            &              $0$            &                             \\

1              &                       $6.200$        &              $-0.285$       &              $1.027$        &              $20.371$       &              $0$            &              $0$            &                             \\

2              &                       $8.100$        &              $-0.235$       &              $0.880$        &              $17.905$       &              $21.486$       &              $0.075$        &                             \\

\end{tabular}
\end{ruledtabular}
\end{table}
\begin{table}[tbp!]
\caption{Control pulse parameters for the implementation of a $\CZ_{0,1}$ two-qubit gate on a two-qubit NIGQC as illustrated in \figref{fig:device_sketch_ibm}. We use the parameters to specify the microwave pulse given by \equref{eq:flux_ctl_sep}. The first column shows the gate we implement. The second column shows the rise and fall time $\FctlTrf$ in ns. The third column shows the pulse duration without buffer time $\FctlTp$ in ns. The fourth column shows the pulse duration with buffer time $\FctlTd$ in ns. The fifth column shows the unit-less pulse amplitude $\FctlAp$. The sixth column shows the drive frequency $\FctlDf$ in GHz. We use the \textbf{effective Hamiltonian} \equref{eq:Eff_Ham_def} to obtain the parameters listed in this table.\label{tab:CtlTqgIIHBIBM}}
\begin{ruledtabular}
\begin{tabular}{ c c c c c c}
$\text{Gate}$      &        $\FctlTrf$        &    $\FctlTp$        &    $\FctlTd$        &              $\FctlAp/2\pi$  &              $\FctlDf/2\pi$    \\

\hline

$\CZ_{0,1}$      &     $24.674$       &       $299.406$        &  $304.406$      &              $0.082$        &              $0.808$      \\

\end{tabular}
\end{ruledtabular}
\end{table}

In the main text, in \secref{sec:InfluenceOfTheAdiabaticApproximationOnGate-errorTrajectories}, we investigated the influence of the adiabatic approximation on gate-error trajectories obtained with the effective Hamiltonian \equref{eq:Eff_Ham_def} for the device architecture illustrated in \figsref{fig:device_sketch}(a-c). Here the adiabatic approximation was applied to one effective flux-tunable transmon only. In this appendix, we perform analogous simulations for a different device architecture studied in  \REFS\ListParametricCoupler.
Here we use a flux microwave pulse to implement two-qubit $\CZ$ gates.

\renewcommand{\hold}{0.85}
\begin{figure}[!tbp]
  \centering
  \includegraphics[scale=\hold]{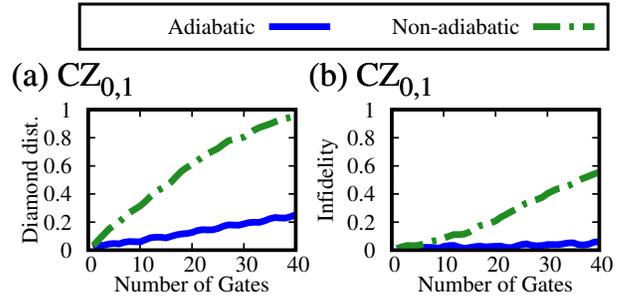}
  \caption{(Color Online) Gate errors as functions of the number of gates for a program which executes forty $\CZ$ gates in a row. In (a) we show the diamond distance and in (b) we show the average infidelity. We run the program on the two-qubit NIGQC illustrated in \figref{fig:device_sketch_ibm}. The results are obtained with the \textbf{effective Hamiltonian} \equref{eq:Circuit_Ham_def} and the device parameters listed in \tabref{tab:device_parameters_ibm}. We run the program two times on each NIGQC. On the first run, we model the flux-tunable transmons as adiabatic qubits. These results are displayed as a blue solid line. On the second run, we model the flux-tunable transmons as \changetwo{nonadiabatic} qubits. These results are displayed as a green dashed line. We can observe that the gate-error trajectories for the two cases can deviate up to $0.8$ for the diamond distance and $0.6$ for the average infidelity. \changethree{Moreover}, we can observe that the gate-error trajectories for both cases exhibit rather different qualitative behaviour.\label{fig:APPROX_SEP}}
\end{figure}

The device architecture is illustrated in \figref{fig:device_sketch_ibm} and the device parameters are listed in \tabref{tab:device_parameters_ibm}. The interaction strength $G$ is set to $85$ MHz for all simulations, see \appref{app:TheEffectiveHamiltonianModel}. Here we couple two fixed-frequency transmons with qubit frequencies $\omega_{0}^{(Q)}$ and $\omega_{1}^{(Q)}$ by means of a flux-tunable transmon with park frequency $\omega_{2}^{(Q)}$. We use the control pulse
\begin{equation}\label{eq:flux_ctl_sep}
\varphi(t)=\FctlFo+\FctlAp e(t) \cos(\FctlDf t),
\end{equation}
for the external flux $\varphi(t)$ to implement a sequence of $\CZ$ gates. The real-valued function $e(t)$ is an envelope function, $\FctlAp$ is the pulse amplitude, $\FctlDf$ is the drive frequency and $\FctlFo$ denotes the flux offset. Since the flux offset defines an operating point for the device, we list the parameter in \tabref{tab:device_parameters_ibm}. The envelope function is defined as
\begin{equation}
e(t) = \begin{cases}
\sin(\lambda t) &\text{if $0 \leq t < \FctlTrf$}\\
1     &\text{if $\FctlTrf \leq t \leq \Delta T $}\\
\sin(\frac{\pi}{2}+\lambda (t-\Delta T)) &\text{if $ \Delta T < t \leq \FctlTp $,}
\end{cases}
\end{equation}
where $\FctlTrf$ denotes the rise and fall time, $\lambda=\pi/(2 \FctlTrf)$ and $\Delta T=\FctlTp-\FctlTrf$. Note that in the computer program we add an additional free time evolution (the buffer time) to the pulse such that the complete pulse duration $\FctlTd$ is $5$ ns longer than $\FctlTp$. The pulse parameters for the pulse we use to obtain the results in this appendix are listed in \tabref{tab:CtlTqgIIHBIBM}. Note that we use single-qubit z-axis rotations $R_{i}^{(z)}(\phi_{i})$ for every qubit to improve the gate performance.

Figures~\ref{fig:APPROX_SEP}(a-b) show the diamond distance $\DNV$(a) and the average infidelity $\IFV$(b) as functions of the number of gates for the two-qubit NIGQC illustrated in \figref{fig:device_sketch_ibm}. Here we executed a program which consists of forty $\CZ$ gates in a row. In each panel we show the results for the two different cases; adiabatic regime (blue solid) and \changetwo{nonadiabatic} regime (green dashed).

As one can see, for the adiabatic case we find smaller gate errors and a smaller increase in gate errors over time. Furthermore, after forty $\CZ$ gate repetitions we find that the diamond distances deviate up to $0.8$ and the average infidelities differ by about $0.6$. Consequently, we find that neglecting the flux drive term in \equref{eq:drive_flux} leads to substantial deviations between the two NIGQC models.

Prototype gate-based quantum computers with similar device parameters are discussed in \REFS\ListParametricCoupler. To the best knowledge of the authors, we find that most models which theoretically describe this device architecture model flux-tunable transmons adiabatically.

\section{Tables with control pulse parameters and gate-error quantifiers}\label{app:ControlPulseParameters}

In this appendix, we provide the control pulse parameters listed in \tabsref{tab:CtlSqgIIMB}{tab:CtlTqgIVHB} which we use to obtain the results in \secref{sec:Results} and the gate-error metrics listed in \tabsref{tab:EMIIMB}{tab:EMIVHB} we use to obtain \figref{fig:metrics}.

The control pulse parameters for the circuit Hamiltonian \equref{eq:Circuit_Ham_def} NIGQC model are listed in \tabsref{tab:CtlSqgIIMB}{tab:CtlTqgIVMB}. Similarly, the control pulse parameters for the effective Hamiltonian \equref{eq:Eff_Ham_def} NIGQC model are listed in \tabsref{tab:CtlSqgIIHB}{tab:CtlTqgIVHB}.

The gate-error metrics for the circuit Hamiltonian \equref{eq:Circuit_Ham_def} NIGQC model are provided in \tabsref{tab:EMIIMB}{tab:EMIVMB}. Similarly, the gate-error metrics for the effective Hamiltonian \equref{eq:Eff_Ham_def} NIGQC model are listed in \tabsref{tab:EMIIHB}{tab:EMIVHB}


\begin{table}[tbp!]
\caption{Control pulse parameters for the implementation of $\ROT(\pi/2)$ rotations on a two-qubit NIGQC as illustrated in \figref{fig:device_sketch}(a). We use the parameters to specify the microwave pulse given by \equref{eq:charge_ctl}. The first column shows the gate we implement. The second column shows the pulse duration $\CctlTd$ in ns. The third column shows the drive frequency $\CctlDf$ in GHz. The fourth column shows the pulse amplitude $\CctlAp$ as a unit-less quantity. The fifth column shows the envelope function parameter $\CctlSg$ in ns. The sixth column shows the DRAG amplitude $\CctlDr$ in ns. We use the \textbf{circuit Hamiltonian} \equref{eq:Circuit_Ham_def} to obtain the parameters listed in this table.\label{tab:CtlSqgIIMB}}
\begin{ruledtabular}
\begin{tabular}{ c c c c c c }
$\text{Gate}$&              $\CctlTd$        &              $\CctlDf/2\pi$&              $\CctlAp$            &              $\CctlSg$       &              $\CctlDr$            \\

\hline

$\ROT_{0}(\pi/2)$      &              $52.250$       &              $4.196$        &              $0.004$        &              $12.082$       &              $0.072$        \\

$\ROT_{1}(\pi/2)$       &              $52.950$       &              $5.195$        &              $0.005$        &              $10.000$       &              $0.070$        \\

\end{tabular}
\end{ruledtabular}
\end{table}
\begin{table}[tbp!]
\caption{Control pulse parameters for the implementation of $\CZ$ gates on a two-qubit NIGQC as illustrated in \figref{fig:device_sketch}(a). We use the parameters to specify the unimodal pulse (UMP) given by \equref{eq:flux_ctl_ump}. The first column shows the gate we implement. The second column shows the pulse type. The third column shows the pulse time parameter $\FctlTp$ in ns. The fourth column shows the pulse duration $\FctlTd$ in ns. The fifth column shows the unit-less pulse amplitude $\FctlAp$. The sixth column shows the rise and fall parameter $\FctlSg$ in ns. We use the \textbf{circuit Hamiltonian} \equref{eq:Circuit_Ham_def} to obtain the parameters listed in this table.\label{tab:CtlTqgIIMB}}
\begin{ruledtabular}
\begin{tabular}{c c c c c c }
$\text{Gate}$ &              $\text{Pulse}$              &              $\FctlTp$        &              $\FctlTd$        &              $\FctlAp/2\pi$  &              $\FctlSg$       \\

\hline

$\CZ_{0,1}$&     $\text{UMP}$         &              $99.835$       &              $125.000$      &              $0.392$        &              $1.313$        \\

\end{tabular}
\end{ruledtabular}
\end{table}

\begin{table}[tbp!]
\caption{Control pulse parameters for the implementation of $\ROT(\pi/2)$ rotations on a three-qubit NIGQC as illustrated in \figref{fig:device_sketch}(b). The units are the same as in \tabref{tab:CtlSqgIIMB}. We use the  \textbf{circuit Hamiltonian} \equref{eq:Circuit_Ham_def} to obtain the parameters listed in this table.\label{tab:CtlSqgIIIMB}}
\begin{ruledtabular}
\begin{tabular}{ c c c c c c }
$\text{Gate}$&              $\CctlTd$        &              $\CctlDf/2\pi$&              $\CctlAp$            &              $\CctlSg$       &              $\CctlDr$            \\

\hline

$\ROT_{0}(\pi/2)$      &              $52.250$       &              $4.196$        &              $0.004$        &              $12.093$       &              $0.168$        \\

$\ROT_{1}(\pi/2)$      &              $52.950$       &              $5.190$        &              $0.004$        &              $9.997$        &              $0.067$        \\

$\ROT_{2}(\pi/2)$      &              $52.950$       &              $5.695$        &              $0.004$        &              $10.011$       &              $0.066$        \\

\end{tabular}
\end{ruledtabular}
\end{table}
\begin{table}[tbp!]
\caption{Control pulse parameters for the implementation of $\CZ$ gates on a three-qubit NIGQC as illustrated in \figref{fig:device_sketch}(b).The units are the same as in \tabref{tab:CtlTqgIIMB}. We use the effective Hamiltonian \equref{eq:Circuit_Ham_def} to obtain the parameters listed in this table.\label{tab:CtlTqgIIIMB}}
\begin{ruledtabular}
\begin{tabular}{c c c c c c }
$\text{Gate}$ &              $\text{Pulse}$              &              $\FctlTp$        &              $\FctlTd$        &              $\FctlAp/2\pi$  &              $\FctlSg$       \\

\hline

$\CZ_{0,1}$&     $\text{UMP}$         &              $96.026$       &              $125.000$      &              $0.391$        &              $1.823$        \\

$\CZ_{1,2}$&     $\text{UMP}$         &              $75.367$       &              $110.000$      &              $0.276$        &              $0.513$        \\

\end{tabular}
\end{ruledtabular}
\end{table}

\begin{table}[tbp!]
\caption{Control pulse parameters for the implementation of $\ROT(\pi/2)$ rotations on a four-qubit NIGQC as illustrated in \figref{fig:device_sketch}(c). The units are the same as in \tabref{tab:CtlSqgIIMB}. We use the  \textbf{circuit Hamiltonian} \equref{eq:Circuit_Ham_def} to obtain the parameters listed in this table.\label{tab:CtlSqgIVMB}}
\begin{ruledtabular}
\begin{tabular}{ c c c c c c }
$\text{Gate}$&              $\CctlTd$        &              $\CctlDf/2\pi$&              $\CctlAp$            &              $\CctlSg$       &              $\CctlDr$            \\

\hline

$\ROT_{0}(\pi/2)$      &              $52.250$       &              $4.193$        &              $0.004$        &              $12.378$       &              $0.047$        \\

$\ROT_{1}(\pi/2)$      &              $52.950$       &              $5.190$        &              $0.004$        &              $10.255$       &              $0.063$        \\

$\ROT_{2}(\pi/2)$      &              $52.950$       &              $5.689$        &              $0.004$        &              $10.312$       &              $0.065$        \\

$\ROT_{3}(\pi/2)$      &              $52.950$       &              $4.951$        &              $0.005$        &              $10.191$       &              $0.012$        \\

\end{tabular}
\end{ruledtabular}
\end{table}
\begin{table}[tbp!]
\caption{Control pulse parameters for the implementation of $\CZ$ gates on a four-qubit NIGQC as illustrated in \figref{fig:device_sketch}(c). The units are the same as in \tabref{tab:CtlTqgIIMB}. We use the  \textbf{circuit Hamiltonian} \equref{eq:Circuit_Ham_def} to obtain the parameters listed in this table.\label{tab:CtlTqgIVMB}}
\begin{ruledtabular}
\begin{tabular}{c c c c c c }
$\text{Gate}$ &              $\text{Pulse}$              &              $\FctlTp$        &              $\FctlTd$        &              $\FctlAp/2\pi$  &              $\FctlSg$       \\

\hline

$\CZ_{0,1}$&     $\text{UMP}$         &              $100.241$      &              $125.000$      &              $0.392$        &              $1.283$        \\

$\CZ_{1,2}$&     $\text{UMP}$         &              $68.046$       &              $90.000$       &              $0.275$        &              $0.182$        \\

$\CZ_{2,3}$&     $\text{UMP}$         &              $80.500$       &              $94.000$       &              $0.320$        &              $0.500$        \\

$\CZ_{0,3}$&     $\text{UMP}$         &              $97.708$       &              $116.000$      &              $0.353$        &              $1.458$        \\

\end{tabular}
\end{ruledtabular}
\end{table}



\begin{table}[tbp!]
\caption{Control pulse parameters for the implementation of $\ROT(\pi/2)$ rotations on a two-qubit NIGQC as illustrated in \figref{fig:device_sketch}(a). We use the parameters to specify the microwave pulse given by \equref{eq:charge_ctl}. The first column shows the gate we implement. The second column shows the pulse duration $\CctlTd$ in ns. The third column shows the drive frequency $\CctlDf$ in GHz. The fourth column shows the pulse amplitude $\CctlAp$ as a unit-less quantity. The fifth column shows the envelope function parameter $\CctlSg$ in ns. The sixth column shows the DRAG amplitude $\CctlDr$ in ns. We use the \textbf{effective Hamiltonian} \equref{eq:Eff_Ham_def} to obtain the parameters listed in this table.\label{tab:CtlSqgIIHB}}
\begin{ruledtabular}
\begin{tabular}{ c c c c c c }
$\text{Gate}$&              $\CctlTd$        &              $\CctlDf/2\pi$&              $\CctlAp$            &              $\CctlSg$       &              $\CctlDr$            \\

\hline

$\ROT_{0}(\pi/2)$     &              $52.250$       &              $4.196$        &              $0.058$        &              $12.082$       &              $0.072$        \\

$\ROT_{1}(\pi/2)$     &              $52.950$       &              $5.195$        &              $0.065$        &              $10.000$       &              $0.070$        \\

\end{tabular}
\end{ruledtabular}
\end{table}
\begin{table}[tbp!]
\caption{Control pulse parameters for the implementation of $\CZ$ gates on a two-qubit NIGQC as illustrated in \figref{fig:device_sketch}(a). We use the parameters to specify the unimodal pulse (UMP) and bimodal pulse (BMP) given by \equaref{eq:flux_ctl_ump}{eq:flux_ctl_bmp}, respectively. The first column shows the gate we implement. The second column shows the pulse type. The third column shows the pulse time parameter $\FctlTp$ in ns. The fourth column shows the pulse duration $\FctlTd$ in ns. The fifth column shows the unit-less pulse amplitude $\FctlAp$. The sixth column shows the rise and fall parameter $\FctlSg$ in ns. We use the \textbf{effective Hamiltonian} \equref{eq:Eff_Ham_def} to obtain the parameters listed in this table.\label{tab:CtlTqgIIHB}}
\begin{ruledtabular}
\begin{tabular}{c c c c c c }
$\text{Gate}$ &              $\text{Pulse}$              &              $\FctlTp$        &              $\FctlTd$        &              $\FctlAp/2\pi$  &              $\FctlSg$       \\

\hline

$\CZ_{0,1}$&     $\text{UMP}$         &$87.258$       &              $95.000$       &              $0.391$        &              $0.459$        \\

$\CZ_{0,1}$&     $\text{BMP}$         &$88.570$       &              $95.000$       &              $0.392$        &              $0.394$        \\

\end{tabular}
\end{ruledtabular}
\end{table}

\begin{table}[tbp!]
\caption{Control pulse parameters for the implementation of $\ROT(\pi/2)$ rotations on a three-qubit NIGQC as illustrated in \figref{fig:device_sketch}(b). The units are the same as in \tabref{tab:CtlSqgIIHB}. We use the \textbf{effective Hamiltonian} \equref{eq:Eff_Ham_def} to obtain the parameters listed in this table.\label{tab:CtlSqgIIIHB}}
\begin{ruledtabular}
\begin{tabular}{ c c c c c c }
$\text{Gate}$&              $\CctlTd$        &              $\CctlDf/2\pi$&              $\CctlAp$            &              $\CctlSg$       &              $\CctlDr$            \\

\hline

$\ROT_{0}(\pi/2)$ &              $52.250$       &              $4.196$        &              $0.058$        &              $12.082$       &              $0.072$        \\

$\ROT_{1}(\pi/2)$     &              $52.950$       &              $5.189$        &              $0.065$        &              $10.000$       &              $0.070$        \\

$\ROT_{2}(\pi/2)$      &              $52.950$       &              $5.694$        &              $0.066$        &              $9.990$        &              $0.032$        \\

\end{tabular}
\end{ruledtabular}
\end{table}
\begin{table}[tbp!]
\caption{Control pulse parameters for the implementation of $\CZ$ gates on a three-qubit NIGQC as illustrated in \figref{fig:device_sketch}(b). The units are the same as in \tabref{tab:CtlTqgIIHB}. We use the effective Hamiltonian \equref{eq:Eff_Ham_def} to obtain the parameters listed in this table.\label{tab:CtlTqgIIIHB}}
\begin{ruledtabular}
\begin{tabular}{c c c c c c }
$\text{Gate}$ &              $\text{Pulse}$              &              $\FctlTp$        &              $\FctlTd$        &              $\FctlAp/2\pi$  &              $\FctlSg$       \\

\hline

$\CZ_{0,1}$&     $\text{UMP}$         &              $87.252$       &              $95.006$       &              $0.391$        &              $0.494$        \\

$\CZ_{0,1}$&     $\text{BMP}$         &             $90.057$       &              $92.188$       &              $0.391$        &              $0.420$        \\

$\CZ_{1,2}$&     $\text{UMP}$         &              $68.831$       &              $80.000$       &              $0.276$        &              $0.554$        \\

\end{tabular}
\end{ruledtabular}
\end{table}

\begin{table}[tbp!]
\caption{Control pulse parameters for the implementation of $\ROT(\pi/2)$ rotations on a four-qubit NIGQC as illustrated in \figref{fig:device_sketch}(c). The units are the same as in \tabref{tab:CtlSqgIIHB}. We use the \textbf{effective Hamiltonian} \equref{eq:Eff_Ham_def} to obtain the parameters listed in this table.\label{tab:CtlSqgIVHB}}
\begin{ruledtabular}
\begin{tabular}{ c c c c c c }
$\text{Gate}$&              $\CctlTd$        &              $\CctlDf/2\pi$&              $\CctlAp$            &              $\CctlSg$       &              $\CctlDr$            \\

\hline

$\ROT_{0}(\pi/2)$       &              $52.250$       &              $4.191$        &              $0.058$        &              $12.082$       &              $0.072$        \\

$\ROT_{1}(\pi/2)$      &              $52.950$       &              $5.189$        &              $0.065$        &              $10.000$       &              $0.070$        \\

$\ROT_{2}(\pi/2)$      &              $52.950$       &              $5.688$        &              $0.066$        &              $9.990$        &              $0.032$        \\

$\ROT_{3}(\pi/2)$      &              $52.950$       &              $4.950$        &              $0.066$        &              $9.990$        &              $0.032$        \\

\end{tabular}
\end{ruledtabular}
\end{table}
\begin{table}[tbp!]
\caption{Control pulse parameters for the implementation of $\CZ$ gates on a four-qubit NIGQC as illustrated in \figref{fig:device_sketch}(c). The units are the same as in \tabref{tab:CtlTqgIIHB}. We use the \textbf{effective Hamiltonian} \equref{eq:Eff_Ham_def} to obtain the parameters listed in this table.\label{tab:CtlTqgIVHB}}
\begin{ruledtabular}
\begin{tabular}{c c c c c c }
$\text{Gate}$ &              $\text{Pulse}$              &              $\FctlTp$        &              $\FctlTd$        &              $\FctlAp/2\pi$  &              $\FctlSg$       \\

\hline

$\CZ_{0,1}$&     $\text{UMP}$         &              $87.254$       &              $95.013$       &              $0.391$        &              $0.453$        \\

$\CZ_{0,1}$&     $\text{BMP}$         &              $89.925$       &              $98.114$       &              $0.392$        &              $0.400$        \\

$\CZ_{1,2}$&     $\text{UMP}$         &              $67.802$       &              $115.238$      &              $0.275$        &              $0.338$        \\

$\CZ_{2,3}$&     $\text{UMP}$         &              $71.620$       &              $98.197$       &              $0.320$        &              $0.543$        \\

$\CZ_{0,3}$&     $\text{UMP}$         &              $92.616$       &              $124.768$      &              $0.353$        &              $1.877$        \\

\end{tabular}
\end{ruledtabular}
\end{table}



\begin{table}[tbp!]
\caption{Gate-error quantifiers for $\ROT(\pi/2)$ rotations and $\CZ$ gates for a two-qubit NIGQC as illustrated in \figref{fig:device_sketch}(a). The gate-error quantifiers are obtained with the \textbf{circuit Hamiltonian} \equref{eq:Circuit_Ham_def}, the device parameters listed in \tabref{tab:device_parameters} and the pulse parameters listed in \tabaref{tab:CtlSqgIIMB}{tab:CtlTqgIIMB}. The first column shows the target gate. The second column shows the pulse type, see \figsref{fig:PulseTimeEvolution}(a-c). The third column shows the diamond distance $\DNV$ given by \equaref{eq:diamond_norm_inf}{eq:diamond_norm_sup}. The fourth column shows the average infidelity $\IFV$ given by \equref{eq:fid_avg}. The fifth column shows the leakage measure $\LNV$ given by \equref{eq:leak}. \label{tab:EMIIMB}}
\begin{ruledtabular}
\begin{tabular}{c c c c c }
$\text{Gate}$ & $\text{Pulse}$ &               $\DNV$&              $\IFV$&              $\LNV$ \\

\hline

$\ROT_{0}(\pi/2)$ & $\text{MP}$ &              $0.0093$       &              $0.0004$       &              $0.0004$       \\

$\ROT_{1}(\pi/2)$ & $\text{MP}$ &              $0.0080$       &              $0.0004$       &              $0.0004$       \\

$\CZ_{0,1}$ & $\text{UMP}$ &              $0.0290$       &              $0.0011$       &              $0.0008$      \\

\end{tabular}
\end{ruledtabular}
\end{table}
\begin{table}[tbp!]
\caption{Gate-error quantifiers for $\ROT(\pi/2)$ rotations and $\CZ$ gates for a three-qubit NIGQC as illustrated in \figref{fig:device_sketch}(b). The gate-error quantifiers are obtained with the \textbf{circuit Hamiltonian} \equref{eq:Circuit_Ham_def}, the device parameters listed in \tabref{tab:device_parameters} and the pulse parameters listed in \tabaref{tab:CtlSqgIIIMB}{tab:CtlTqgIIIMB}. The rows and columns show the same unit-less quantities as \tabref{tab:EMIIMB}.\label{tab:EMIIIMB}}
\begin{ruledtabular}
\begin{tabular}{c c c c c c }
$\text{Gate}$ & $\text{Pulse}$ &               $\DNV$&              $\IFV$&              $\LNV$\\

\hline

$\ROT_{0}(\pi/2)$ & $\text{MP}$ &              $0.044$        &              $0.003$        &              $0.002$        \\

$\ROT_{1}(\pi/2)$ & $\text{MP}$ &              $0.038$        &              $0.002$        &              $0.001$        \\

$\ROT_{2}(\pi/2)$ & $\text{MP}$ &              $0.039$        &              $0.002$        &              $0.001$        \\

$\CZ_{0,1}$ & $\text{UMP}$ &              $0.044$        &              $0.010$        &              $0.010$        \\

$\CZ_{1,2}$ & $\text{UMP}$ &              $0.012$        &              $0.002$        &              $0.002$       \\

\end{tabular}
\end{ruledtabular}
\end{table}
\begin{table}[tbp!]
\caption{Gate-error quantifiers for $\ROT(\pi/2)$ rotations and $\CZ$ gates for a four-qubit NIGQC as illustrated in \figref{fig:device_sketch}(c). The gate-error quantifiers are obtained with the \textbf{circuit Hamiltonian} \equref{eq:Circuit_Ham_def}, the device parameters listed in \tabref{tab:device_parameters} and the pulse parameters listed in \tabaref{tab:CtlSqgIVMB}{tab:CtlTqgIVMB}. The rows and columns show the same unit-less quantities as \tabref{tab:EMIIMB}.\label{tab:EMIVMB}}
\begin{ruledtabular}
\begin{tabular}{c c c c c c }
$\text{Gate}$ & $\text{Pulse}$ &               $\DNV$&              $\IFV$&              $\LNV$\\

\hline

$\ROT_{0}(\pi/2)$ & $\text{MP}$ &              $0.058$        &              $0.004$        &              $0.002$        \\

$\ROT_{1}(\pi/2)$ & $\text{MP}$ &              $0.054$        &              $0.003$        &              $0.002$       \\

$\ROT_{2}(\pi/2)$ & $\text{MP}$ &              $0.053$        &              $0.003$        &              $0.002$        \\

$\ROT_{3}(\pi/2)$ & $\text{MP}$ &              $0.057$        &              $0.004$        &              $0.002$        \\

$\CZ_{0,1}$ & $\text{UMP}$ &              $0.031$        &              $0.005$        &              $0.004$        \\

$\CZ_{1,2}$ & $\text{UMP}$ &              $0.144$        &              $0.029$        &              $0.018$        \\

$\CZ_{2,3}$ & $\text{UMP}$ &              $0.073$        &              $0.008$        &              $0.005$        \\

$\CZ_{0,3}$ & $\text{UMP}$ &              $0.046$        &              $0.005$        &              $0.003$        \\

\end{tabular}
\end{ruledtabular}
\end{table}



\begin{table}[tbp!]
\caption{Gate-error quantifiers for $\ROT(\pi/2)$ rotations and $\CZ$ gates for a two-qubit NIGQC as illustrated in \figref{fig:device_sketch}(a). The gate-error quantifiers are obtained with the \textbf{effective Hamiltonian} \equref{eq:Eff_Ham_def}, the device parameters listed in \tabref{tab:device_parameters} and the pulse parameters listed in \tabaref{tab:CtlSqgIIHB}{tab:CtlTqgIIHB}. The first column shows the target gate. The second column shows the pulse type, see \figsref{fig:PulseTimeEvolution}(a-c). The third column shows the diamond distance $\DNV$ given by \equaref{eq:diamond_norm_inf}{eq:diamond_norm_sup}. The fourth column shows the average infidelity $\IFV$ given by \equref{eq:fid_avg}. The fifth column shows the leakage measure $\LNV$ given by \equref{eq:leak}.\label{tab:EMIIHB}}
\begin{ruledtabular}
\begin{tabular}{c c c c c c }
$\text{Gate}$ & $\text{Pulse}$ &               $\DNV$&              $\IFV$&              $\LNV$\\

\hline

$\ROT_{0}(\pi/2)$ & $\text{MP}$ &              $0.0089$       &              $0.0004$       &              $0.0004$       \\

$\ROT_{1}(\pi/2)$ & $\text{MP}$ &              $0.0090$       &              $0.0004$       &              $0.0004$       \\

$\CZ_{0,1}$ & $\text{UMP}$ &              $0.0424$       &              $0.0012$       &              $0.0005$       \\

$\CZ_{0,1}$ & $\text{BMP}$ &              $0.0167$        &              $0.0006$        &              $0.0005$        \\

\end{tabular}
\end{ruledtabular}
\end{table}
\begin{table}[tbp!]
\caption{Gate-error quantifiers for $\ROT(\pi/2)$ rotations and $\CZ$ gates for a three-qubit NIGQC as illustrated in \figref{fig:device_sketch}(b). The gate-error quantifiers are obtained with the \textbf{effective Hamiltonian} \equref{eq:Eff_Ham_def}, the device parameters listed in \tabref{tab:device_parameters} and the pulse parameters listed in \tabaref{tab:CtlSqgIIIHB}{tab:CtlTqgIIIHB}. The rows and columns show the same unit-less quantities as \tabref{tab:EMIIHB}.\label{tab:EMIIIHB}}
\begin{ruledtabular}
\begin{tabular}{c c c c c c }
$\text{Gate}$ & $\text{Pulse}$ &               $\DNV$&              $\IFV$&              $\LNV$\\

\hline

$\ROT_{0}(\pi/2)$ & $\text{MP}$ &              $0.046$        &              $0.003$        &              $0.002$        \\

$\ROT_{1}(\pi/2)$ & $\text{MP}$ &              $0.040$        &              $0.002$        &              $0.002$        \\

$\ROT_{2}(\pi/2)$ & $\text{MP}$ &              $0.039$        &              $0.002$        &              $0.002$       \\

$\CZ_{0,1}$ & $\text{UMP}$ &              $0.057$        &              $0.003$        &              $0.002$        \\

$\CZ_{0,1}$ & $\text{BMP}$ &              $0.031$        &              $0.004$        &              $0.004$        \\

$\CZ_{1,2}$ & $\text{UMP}$ &              $0.028$        &              $0.006$        &              $0.006$        \\

\end{tabular}
\end{ruledtabular}
\end{table}
\begin{table}[tbp!]
\caption{Gate-error quantifiers for $\ROT(\pi/2)$ rotations and $\CZ$ gates for a four-qubit NIGQC as illustrated in \figref{fig:device_sketch}(c). The gate-error quantifiers are obtained with the \textbf{effective Hamiltonian} \equref{eq:Eff_Ham_def}, the device parameters listed in \tabref{tab:device_parameters} and the pulse parameters listed in \tabaref{tab:CtlSqgIVHB}{tab:CtlTqgIVHB}. The rows and columns show the same unit-less quantities as \tabref{tab:EMIIHB}.\label{tab:EMIVHB}}
\begin{ruledtabular}
\begin{tabular}{c c c c c c }
$\text{Gate}$ & $\text{Pulse}$ &               $\DNV$&              $\IFV$&              $\LNV$\\

\hline

$\ROT_{0}(\pi/2)$ & $\text{MP}$ &              $0.060$        &              $0.004$        &              $0.003$       \\

$\ROT_{1}(\pi/2)$ & $\text{MP}$ &              $0.056$        &              $0.003$        &              $0.002$       \\

$\ROT_{2}(\pi/2)$ & $\text{MP}$ &              $0.054$        &              $0.003$        &              $0.002$       \\

$\ROT_{3}(\pi/2)$ & $\text{MP}$ &              $0.060$        &              $0.004$        &              $0.002$       \\

$\CZ_{0,1}$ & $\text{UMP}$ &              $0.051$        &              $0.004$        &              $0.003$       \\

$\CZ_{0,1}$ & $\text{BMP}$ &              $0.041$        &              $0.004$        &              $0.004$       \\

$\CZ_{1,2}$ & $\text{UMP}$ &              $0.146$        &              $0.015$        &              $0.005$       \\

$\CZ_{2,3}$ & $\text{UMP}$ &              $0.078$        &              $0.010$        &              $0.007$       \\

$\CZ_{0,3}$ & $\text{UMP}$ &              $0.058$        &              $0.005$        &              $0.003$       \\

\end{tabular}
\end{ruledtabular}
\end{table}
\clearpage
\bibliographystyle{apsrev4-2}
\bibliography{ms}

\begin{thebibliography}{65}%
\makeatletter
\providecommand \@ifxundefined [1]{%
 \@ifx{#1\undefined}
}%
\providecommand \@ifnum [1]{%
 \ifnum #1\expandafter \@firstoftwo
 \else \expandafter \@secondoftwo
 \fi
}%
\providecommand \@ifx [1]{%
 \ifx #1\expandafter \@firstoftwo
 \else \expandafter \@secondoftwo
 \fi
}%
\providecommand \natexlab [1]{#1}%
\providecommand \enquote  [1]{``#1''}%
\providecommand \bibnamefont  [1]{#1}%
\providecommand \bibfnamefont [1]{#1}%
\providecommand \citenamefont [1]{#1}%
\providecommand \href@noop [0]{\@secondoftwo}%
\providecommand \href [0]{\begingroup \@sanitize@url \@href}%
\providecommand \@href[1]{\@@startlink{#1}\@@href}%
\providecommand \@@href[1]{\endgroup#1\@@endlink}%
\providecommand \@sanitize@url [0]{\catcode `\\12\catcode `\$12\catcode
  `\&12\catcode `\#12\catcode `\^12\catcode `\_12\catcode `\%12\relax}%
\providecommand \@@startlink[1]{}%
\providecommand \@@endlink[0]{}%
\providecommand \url  [0]{\begingroup\@sanitize@url \@url }%
\providecommand \@url [1]{\endgroup\@href {#1}{\urlprefix }}%
\providecommand \urlprefix  [0]{URL }%
\providecommand \Eprint [0]{\href }%
\providecommand \doibase [0]{https://doi.org/}%
\providecommand \selectlanguage [0]{\@gobble}%
\providecommand \bibinfo  [0]{\@secondoftwo}%
\providecommand \bibfield  [0]{\@secondoftwo}%
\providecommand \translation [1]{[#1]}%
\providecommand \BibitemOpen [0]{}%
\providecommand \bibitemStop [0]{}%
\providecommand \bibitemNoStop [0]{.\EOS\space}%
\providecommand \EOS [0]{\spacefactor3000\relax}%
\providecommand \BibitemShut  [1]{\csname bibitem#1\endcsname}%
\let\auto@bib@innerbib\@empty
\bibitem [{\citenamefont {Nielsen}\ and\ \citenamefont
  {Chuang}(2011)}]{Nielsen:2011:QCQ:1972505}%
  \BibitemOpen
  \bibfield  {author} {\bibinfo {author} {\bibfnamefont {M.~A.}\ \bibnamefont
  {Nielsen}}\ and\ \bibinfo {author} {\bibfnamefont {I.~L.}\ \bibnamefont
  {Chuang}},\ }\href@noop {} {\emph {\bibinfo {title} {Quantum Computation and
  Quantum Information: 10th Anniversary Edition}}}\ (\bibinfo  {publisher}
  {Cambridge University Press},\ \bibinfo {year} {2011})\BibitemShut {NoStop}%
\bibitem [{\citenamefont {Watrous}(2018)}]{Watrous2018}%
  \BibitemOpen
  \bibfield  {author} {\bibinfo {author} {\bibfnamefont {J.}~\bibnamefont
  {Watrous}},\ }\href {https://doi.org/10.1017/9781316848142} {\emph {\bibinfo
  {title} {The Theory of Quantum Information}}}\ (\bibinfo  {publisher}
  {Cambridge University Press},\ \bibinfo {year} {2018})\BibitemShut {NoStop}%
\bibitem [{\citenamefont {Willsch}\ \emph {et~al.}(2017)\citenamefont
  {Willsch}, \citenamefont {Nocon}, \citenamefont {Jin}, \citenamefont
  {DeRaedt},\ and\ \citenamefont {Michielsen}}]{Wi17}%
  \BibitemOpen
  \bibfield  {author} {\bibinfo {author} {\bibfnamefont {D.}~\bibnamefont
  {Willsch}}, \bibinfo {author} {\bibfnamefont {M.}~\bibnamefont {Nocon}},
  \bibinfo {author} {\bibfnamefont {F.}~\bibnamefont {Jin}}, \bibinfo {author}
  {\bibfnamefont {H.}~\bibnamefont {DeRaedt}},\ and\ \bibinfo {author}
  {\bibfnamefont {K.}~\bibnamefont {Michielsen}},\ }\href
  {https://journals.aps.org/pra/abstract/10.1103/PhysRevA.96.062302} {\bibfield
   {journal} {\bibinfo  {journal} {Physical Review A}\ }\textbf {\bibinfo
  {volume} {96}},\ \bibinfo {pages} {062302} (\bibinfo {year}
  {2017})}\BibitemShut {NoStop}%
\bibitem [{\citenamefont {Michielsen}\ \emph {et~al.}(2017)\citenamefont
  {Michielsen}, \citenamefont {Nocon}, \citenamefont {Willsch}, \citenamefont
  {Jin}, \citenamefont {Lippert},\ and\ \citenamefont {{De
  Raedt}}}]{Michielsen17}%
  \BibitemOpen
  \bibfield  {author} {\bibinfo {author} {\bibfnamefont {K.}~\bibnamefont
  {Michielsen}}, \bibinfo {author} {\bibfnamefont {M.}~\bibnamefont {Nocon}},
  \bibinfo {author} {\bibfnamefont {D.}~\bibnamefont {Willsch}}, \bibinfo
  {author} {\bibfnamefont {F.}~\bibnamefont {Jin}}, \bibinfo {author}
  {\bibfnamefont {T.}~\bibnamefont {Lippert}},\ and\ \bibinfo {author}
  {\bibfnamefont {H.}~\bibnamefont {{De Raedt}}},\ }\href
  {https://doi.org/https://doi.org/10.1016/j.cpc.2017.06.011} {\bibfield
  {journal} {\bibinfo  {journal} {Computer Physics Communications}\ }\textbf
  {\bibinfo {volume} {220}},\ \bibinfo {pages} {44} (\bibinfo {year}
  {2017})}\BibitemShut {NoStop}%
\bibitem [{\citenamefont {Nielsen}(2002)}]{Nielsen2002}%
  \BibitemOpen
  \bibfield  {author} {\bibinfo {author} {\bibfnamefont {M.~A.}\ \bibnamefont
  {Nielsen}},\ }\href
  {https://doi.org/https://doi.org/10.1016/S0375-9601(02)01272-0} {\bibfield
  {journal} {\bibinfo  {journal} {Physics Letters A}\ }\textbf {\bibinfo
  {volume} {303}},\ \bibinfo {pages} {249} (\bibinfo {year}
  {2002})}\BibitemShut {NoStop}%
\bibitem [{\citenamefont {Jin}\ \emph {et~al.}(2021)\citenamefont {Jin},
  \citenamefont {Willsch}, \citenamefont {Willsch}, \citenamefont {Lagemann},
  \citenamefont {Michielsen},\ and\ \citenamefont {De~Raedt}}]{Jin21}%
  \BibitemOpen
  \bibfield  {author} {\bibinfo {author} {\bibfnamefont {F.}~\bibnamefont
  {Jin}}, \bibinfo {author} {\bibfnamefont {D.}~\bibnamefont {Willsch}},
  \bibinfo {author} {\bibfnamefont {M.}~\bibnamefont {Willsch}}, \bibinfo
  {author} {\bibfnamefont {H.}~\bibnamefont {Lagemann}}, \bibinfo {author}
  {\bibfnamefont {K.}~\bibnamefont {Michielsen}},\ and\ \bibinfo {author}
  {\bibfnamefont {H.}~\bibnamefont {De~Raedt}},\ }\href
  {https://doi.org/10.7566/JPSJ.90.012001} {\bibfield  {journal} {\bibinfo
  {journal} {Journal of the Physical Society of Japan}\ }\textbf {\bibinfo
  {volume} {90}},\ \bibinfo {pages} {012001} (\bibinfo {year} {2021})},\
  \Eprint {https://arxiv.org/abs/https://doi.org/10.7566/JPSJ.90.012001}
  {https://doi.org/10.7566/JPSJ.90.012001} \BibitemShut {NoStop}%
\bibitem [{\citenamefont {Kitaev}(1997)}]{Kitaev1997}%
  \BibitemOpen
  \bibfield  {author} {\bibinfo {author} {\bibfnamefont {A.~Y.}\ \bibnamefont
  {Kitaev}},\ }\href {https://doi.org/10.1070/rm1997v052n06abeh002155}
  {\bibfield  {journal} {\bibinfo  {journal} {Russian Mathematical Surveys}\
  }\textbf {\bibinfo {volume} {52}},\ \bibinfo {pages} {1191} (\bibinfo {year}
  {1997})}\BibitemShut {NoStop}%
\bibitem [{\citenamefont {Sanders}\ \emph {et~al.}(2015)\citenamefont
  {Sanders}, \citenamefont {Wallman},\ and\ \citenamefont
  {Sanders}}]{Sanders2015}%
  \BibitemOpen
  \bibfield  {author} {\bibinfo {author} {\bibfnamefont {Y.~R.}\ \bibnamefont
  {Sanders}}, \bibinfo {author} {\bibfnamefont {J.~J.}\ \bibnamefont
  {Wallman}},\ and\ \bibinfo {author} {\bibfnamefont {B.~C.}\ \bibnamefont
  {Sanders}},\ }\href {https://doi.org/10.1088/1367-2630/18/1/012002}
  {\bibfield  {journal} {\bibinfo  {journal} {New Journal of Physics}\ }\textbf
  {\bibinfo {volume} {18}},\ \bibinfo {pages} {012002} (\bibinfo {year}
  {2015})}\BibitemShut {NoStop}%
\bibitem [{\citenamefont {Rol}\ \emph {et~al.}(2019)\citenamefont {Rol},
  \citenamefont {Battistel}, \citenamefont {Malinowski}, \citenamefont
  {Bultink}, \citenamefont {Tarasinski}, \citenamefont {Vollmer}, \citenamefont
  {Haider}, \citenamefont {Muthusubramanian}, \citenamefont {Bruno},
  \citenamefont {Terhal},\ and\ \citenamefont {DiCarlo}}]{Rol19}%
  \BibitemOpen
  \bibfield  {author} {\bibinfo {author} {\bibfnamefont {M.~A.}\ \bibnamefont
  {Rol}}, \bibinfo {author} {\bibfnamefont {F.}~\bibnamefont {Battistel}},
  \bibinfo {author} {\bibfnamefont {F.~K.}\ \bibnamefont {Malinowski}},
  \bibinfo {author} {\bibfnamefont {C.~C.}\ \bibnamefont {Bultink}}, \bibinfo
  {author} {\bibfnamefont {B.~M.}\ \bibnamefont {Tarasinski}}, \bibinfo
  {author} {\bibfnamefont {R.}~\bibnamefont {Vollmer}}, \bibinfo {author}
  {\bibfnamefont {N.}~\bibnamefont {Haider}}, \bibinfo {author} {\bibfnamefont
  {N.}~\bibnamefont {Muthusubramanian}}, \bibinfo {author} {\bibfnamefont
  {A.}~\bibnamefont {Bruno}}, \bibinfo {author} {\bibfnamefont {B.~M.}\
  \bibnamefont {Terhal}},\ and\ \bibinfo {author} {\bibfnamefont
  {L.}~\bibnamefont {DiCarlo}},\ }\href
  {https://doi.org/10.1103/PhysRevLett.123.120502} {\bibfield  {journal}
  {\bibinfo  {journal} {Phys. Rev. Lett.}\ }\textbf {\bibinfo {volume} {123}},\
  \bibinfo {pages} {120502} (\bibinfo {year} {2019})}\BibitemShut {NoStop}%
\bibitem [{\citenamefont {Werninghaus}\ \emph {et~al.}(2021)\citenamefont
  {Werninghaus}, \citenamefont {Egger}, \citenamefont {Roy}, \citenamefont
  {Machnes}, \citenamefont {Wilhelm},\ and\ \citenamefont
  {Filipp}}]{Werninghaus2021}%
  \BibitemOpen
  \bibfield  {author} {\bibinfo {author} {\bibfnamefont {M.}~\bibnamefont
  {Werninghaus}}, \bibinfo {author} {\bibfnamefont {D.~J.}\ \bibnamefont
  {Egger}}, \bibinfo {author} {\bibfnamefont {F.}~\bibnamefont {Roy}}, \bibinfo
  {author} {\bibfnamefont {S.}~\bibnamefont {Machnes}}, \bibinfo {author}
  {\bibfnamefont {F.~K.}\ \bibnamefont {Wilhelm}},\ and\ \bibinfo {author}
  {\bibfnamefont {S.}~\bibnamefont {Filipp}},\ }\href
  {https://doi.org/10.1038/s41534-020-00346-2} {\bibfield  {journal} {\bibinfo
  {journal} {npj Quantum Information}\ }\textbf {\bibinfo {volume} {7}},\
  \bibinfo {pages} {14} (\bibinfo {year} {2021})}\BibitemShut {NoStop}%
\bibitem [{\citenamefont {Krinner}\ \emph {et~al.}(2019)\citenamefont
  {Krinner}, \citenamefont {Storz}, \citenamefont {Kurpiers}, \citenamefont
  {Magnard}, \citenamefont {Heinsoo}, \citenamefont {Keller}, \citenamefont
  {L{\"u}tolf}, \citenamefont {Eichler},\ and\ \citenamefont
  {Wallraff}}]{Krinner2019}%
  \BibitemOpen
  \bibfield  {author} {\bibinfo {author} {\bibfnamefont {S.}~\bibnamefont
  {Krinner}}, \bibinfo {author} {\bibfnamefont {S.}~\bibnamefont {Storz}},
  \bibinfo {author} {\bibfnamefont {P.}~\bibnamefont {Kurpiers}}, \bibinfo
  {author} {\bibfnamefont {P.}~\bibnamefont {Magnard}}, \bibinfo {author}
  {\bibfnamefont {J.}~\bibnamefont {Heinsoo}}, \bibinfo {author} {\bibfnamefont
  {R.}~\bibnamefont {Keller}}, \bibinfo {author} {\bibfnamefont
  {J.}~\bibnamefont {L{\"u}tolf}}, \bibinfo {author} {\bibfnamefont
  {C.}~\bibnamefont {Eichler}},\ and\ \bibinfo {author} {\bibfnamefont
  {A.}~\bibnamefont {Wallraff}},\ }\href
  {https://doi.org/10.1140/epjqt/s40507-019-0072-0} {\bibfield  {journal}
  {\bibinfo  {journal} {EPJ Quantum Technology}\ }\textbf {\bibinfo {volume}
  {6}},\ \bibinfo {pages} {2} (\bibinfo {year} {2019})}\BibitemShut {NoStop}%
\bibitem [{\citenamefont {Burnett}\ \emph {et~al.}(2019)\citenamefont
  {Burnett}, \citenamefont {Bengtsson}, \citenamefont {Scigliuzzo},
  \citenamefont {Niepce}, \citenamefont {Kudra}, \citenamefont {Delsing},\ and\
  \citenamefont {Bylander}}]{Burnett2019}%
  \BibitemOpen
  \bibfield  {author} {\bibinfo {author} {\bibfnamefont {J.~J.}\ \bibnamefont
  {Burnett}}, \bibinfo {author} {\bibfnamefont {A.}~\bibnamefont {Bengtsson}},
  \bibinfo {author} {\bibfnamefont {M.}~\bibnamefont {Scigliuzzo}}, \bibinfo
  {author} {\bibfnamefont {D.}~\bibnamefont {Niepce}}, \bibinfo {author}
  {\bibfnamefont {M.}~\bibnamefont {Kudra}}, \bibinfo {author} {\bibfnamefont
  {P.}~\bibnamefont {Delsing}},\ and\ \bibinfo {author} {\bibfnamefont
  {J.}~\bibnamefont {Bylander}},\ }\href
  {https://doi.org/10.1038/s41534-019-0168-5} {\bibfield  {journal} {\bibinfo
  {journal} {npj Quantum Information}\ }\textbf {\bibinfo {volume} {5}},\
  \bibinfo {pages} {54} (\bibinfo {year} {2019})}\BibitemShut {NoStop}%
\bibitem [{\citenamefont {McEwen}\ \emph {et~al.}(2022)\citenamefont {McEwen},
  \citenamefont {Faoro}, \citenamefont {Arya}, \citenamefont {Dunsworth},
  \citenamefont {Huang}, \citenamefont {Kim}, \citenamefont {Burkett},
  \citenamefont {Fowler}, \citenamefont {Arute}, \citenamefont {Bardin},
  \citenamefont {Bengtsson}, \citenamefont {Bilmes}, \citenamefont {Buckley},
  \citenamefont {Bushnell}, \citenamefont {Chen}, \citenamefont {Collins},
  \citenamefont {Demura}, \citenamefont {Derk}, \citenamefont {Erickson},
  \citenamefont {Giustina}, \citenamefont {Harrington}, \citenamefont {Hong},
  \citenamefont {Jeffrey}, \citenamefont {Kelly}, \citenamefont {Klimov},
  \citenamefont {Kostritsa}, \citenamefont {Laptev}, \citenamefont {Locharla},
  \citenamefont {Mi}, \citenamefont {Miao}, \citenamefont {Montazeri},
  \citenamefont {Mutus}, \citenamefont {Naaman}, \citenamefont {Neeley},
  \citenamefont {Neill}, \citenamefont {Opremcak}, \citenamefont {Quintana},
  \citenamefont {Redd}, \citenamefont {Roushan}, \citenamefont {Sank},
  \citenamefont {Satzinger}, \citenamefont {Shvarts}, \citenamefont {White},
  \citenamefont {Yao}, \citenamefont {Yeh}, \citenamefont {Yoo}, \citenamefont
  {Chen}, \citenamefont {Smelyanskiy}, \citenamefont {Martinis}, \citenamefont
  {Neven}, \citenamefont {Megrant}, \citenamefont {Ioffe},\ and\ \citenamefont
  {Barends}}]{McEwen22}%
  \BibitemOpen
  \bibfield  {author} {\bibinfo {author} {\bibfnamefont {M.}~\bibnamefont
  {McEwen}}, \bibinfo {author} {\bibfnamefont {L.}~\bibnamefont {Faoro}},
  \bibinfo {author} {\bibfnamefont {K.}~\bibnamefont {Arya}}, \bibinfo {author}
  {\bibfnamefont {A.}~\bibnamefont {Dunsworth}}, \bibinfo {author}
  {\bibfnamefont {T.}~\bibnamefont {Huang}}, \bibinfo {author} {\bibfnamefont
  {S.}~\bibnamefont {Kim}}, \bibinfo {author} {\bibfnamefont {B.}~\bibnamefont
  {Burkett}}, \bibinfo {author} {\bibfnamefont {A.}~\bibnamefont {Fowler}},
  \bibinfo {author} {\bibfnamefont {F.}~\bibnamefont {Arute}}, \bibinfo
  {author} {\bibfnamefont {J.~C.}\ \bibnamefont {Bardin}}, \bibinfo {author}
  {\bibfnamefont {A.}~\bibnamefont {Bengtsson}}, \bibinfo {author}
  {\bibfnamefont {A.}~\bibnamefont {Bilmes}}, \bibinfo {author} {\bibfnamefont
  {B.~B.}\ \bibnamefont {Buckley}}, \bibinfo {author} {\bibfnamefont
  {N.}~\bibnamefont {Bushnell}}, \bibinfo {author} {\bibfnamefont
  {Z.}~\bibnamefont {Chen}}, \bibinfo {author} {\bibfnamefont {R.}~\bibnamefont
  {Collins}}, \bibinfo {author} {\bibfnamefont {S.}~\bibnamefont {Demura}},
  \bibinfo {author} {\bibfnamefont {A.~R.}\ \bibnamefont {Derk}}, \bibinfo
  {author} {\bibfnamefont {C.}~\bibnamefont {Erickson}}, \bibinfo {author}
  {\bibfnamefont {M.}~\bibnamefont {Giustina}}, \bibinfo {author}
  {\bibfnamefont {S.~D.}\ \bibnamefont {Harrington}}, \bibinfo {author}
  {\bibfnamefont {S.}~\bibnamefont {Hong}}, \bibinfo {author} {\bibfnamefont
  {E.}~\bibnamefont {Jeffrey}}, \bibinfo {author} {\bibfnamefont
  {J.}~\bibnamefont {Kelly}}, \bibinfo {author} {\bibfnamefont {P.~V.}\
  \bibnamefont {Klimov}}, \bibinfo {author} {\bibfnamefont {F.}~\bibnamefont
  {Kostritsa}}, \bibinfo {author} {\bibfnamefont {P.}~\bibnamefont {Laptev}},
  \bibinfo {author} {\bibfnamefont {A.}~\bibnamefont {Locharla}}, \bibinfo
  {author} {\bibfnamefont {X.}~\bibnamefont {Mi}}, \bibinfo {author}
  {\bibfnamefont {K.~C.}\ \bibnamefont {Miao}}, \bibinfo {author}
  {\bibfnamefont {S.}~\bibnamefont {Montazeri}}, \bibinfo {author}
  {\bibfnamefont {J.}~\bibnamefont {Mutus}}, \bibinfo {author} {\bibfnamefont
  {O.}~\bibnamefont {Naaman}}, \bibinfo {author} {\bibfnamefont
  {M.}~\bibnamefont {Neeley}}, \bibinfo {author} {\bibfnamefont
  {C.}~\bibnamefont {Neill}}, \bibinfo {author} {\bibfnamefont
  {A.}~\bibnamefont {Opremcak}}, \bibinfo {author} {\bibfnamefont
  {C.}~\bibnamefont {Quintana}}, \bibinfo {author} {\bibfnamefont
  {N.}~\bibnamefont {Redd}}, \bibinfo {author} {\bibfnamefont {P.}~\bibnamefont
  {Roushan}}, \bibinfo {author} {\bibfnamefont {D.}~\bibnamefont {Sank}},
  \bibinfo {author} {\bibfnamefont {K.~J.}\ \bibnamefont {Satzinger}}, \bibinfo
  {author} {\bibfnamefont {V.}~\bibnamefont {Shvarts}}, \bibinfo {author}
  {\bibfnamefont {T.}~\bibnamefont {White}}, \bibinfo {author} {\bibfnamefont
  {Z.~J.}\ \bibnamefont {Yao}}, \bibinfo {author} {\bibfnamefont
  {P.}~\bibnamefont {Yeh}}, \bibinfo {author} {\bibfnamefont {J.}~\bibnamefont
  {Yoo}}, \bibinfo {author} {\bibfnamefont {Y.}~\bibnamefont {Chen}}, \bibinfo
  {author} {\bibfnamefont {V.}~\bibnamefont {Smelyanskiy}}, \bibinfo {author}
  {\bibfnamefont {J.~M.}\ \bibnamefont {Martinis}}, \bibinfo {author}
  {\bibfnamefont {H.}~\bibnamefont {Neven}}, \bibinfo {author} {\bibfnamefont
  {A.}~\bibnamefont {Megrant}}, \bibinfo {author} {\bibfnamefont
  {L.}~\bibnamefont {Ioffe}},\ and\ \bibinfo {author} {\bibfnamefont
  {R.}~\bibnamefont {Barends}},\ }\href
  {https://doi.org/10.1038/s41567-021-01432-8} {\bibfield  {journal} {\bibinfo
  {journal} {Nature Physics}\ }\textbf {\bibinfo {volume} {18}},\ \bibinfo
  {pages} {107} (\bibinfo {year} {2022})}\BibitemShut {NoStop}%
\bibitem [{\citenamefont {Lacroix}\ \emph {et~al.}(2020)\citenamefont
  {Lacroix}, \citenamefont {Hellings}, \citenamefont {Andersen}, \citenamefont
  {Di~Paolo}, \citenamefont {Remm}, \citenamefont {Lazar}, \citenamefont
  {Krinner}, \citenamefont {Norris}, \citenamefont {Gabureac}, \citenamefont
  {Heinsoo}, \citenamefont {Blais}, \citenamefont {Eichler},\ and\
  \citenamefont {Wallraff}}]{Lacroix2020}%
  \BibitemOpen
  \bibfield  {author} {\bibinfo {author} {\bibfnamefont {N.}~\bibnamefont
  {Lacroix}}, \bibinfo {author} {\bibfnamefont {C.}~\bibnamefont {Hellings}},
  \bibinfo {author} {\bibfnamefont {C.~K.}\ \bibnamefont {Andersen}}, \bibinfo
  {author} {\bibfnamefont {A.}~\bibnamefont {Di~Paolo}}, \bibinfo {author}
  {\bibfnamefont {A.}~\bibnamefont {Remm}}, \bibinfo {author} {\bibfnamefont
  {S.}~\bibnamefont {Lazar}}, \bibinfo {author} {\bibfnamefont
  {S.}~\bibnamefont {Krinner}}, \bibinfo {author} {\bibfnamefont {G.~J.}\
  \bibnamefont {Norris}}, \bibinfo {author} {\bibfnamefont {M.}~\bibnamefont
  {Gabureac}}, \bibinfo {author} {\bibfnamefont {J.}~\bibnamefont {Heinsoo}},
  \bibinfo {author} {\bibfnamefont {A.}~\bibnamefont {Blais}}, \bibinfo
  {author} {\bibfnamefont {C.}~\bibnamefont {Eichler}},\ and\ \bibinfo {author}
  {\bibfnamefont {A.}~\bibnamefont {Wallraff}},\ }\href
  {https://doi.org/10.1103/PRXQuantum.1.020304} {\bibfield  {journal} {\bibinfo
   {journal} {PRX Quantum}\ }\textbf {\bibinfo {volume} {1}},\ \bibinfo {pages}
  {110304} (\bibinfo {year} {2020})}\BibitemShut {NoStop}%
\bibitem [{\citenamefont {Krinner}\ \emph {et~al.}(2020)\citenamefont
  {Krinner}, \citenamefont {Lazar}, \citenamefont {Remm}, \citenamefont
  {Andersen}, \citenamefont {Lacroix}, \citenamefont {Norris}, \citenamefont
  {Hellings}, \citenamefont {Gabureac}, \citenamefont {Eichler},\ and\
  \citenamefont {Wallraff}}]{Krinner2020}%
  \BibitemOpen
  \bibfield  {author} {\bibinfo {author} {\bibfnamefont {S.}~\bibnamefont
  {Krinner}}, \bibinfo {author} {\bibfnamefont {S.}~\bibnamefont {Lazar}},
  \bibinfo {author} {\bibfnamefont {A.}~\bibnamefont {Remm}}, \bibinfo {author}
  {\bibfnamefont {C.}~\bibnamefont {Andersen}}, \bibinfo {author}
  {\bibfnamefont {N.}~\bibnamefont {Lacroix}}, \bibinfo {author} {\bibfnamefont
  {G.}~\bibnamefont {Norris}}, \bibinfo {author} {\bibfnamefont
  {C.}~\bibnamefont {Hellings}}, \bibinfo {author} {\bibfnamefont
  {M.}~\bibnamefont {Gabureac}}, \bibinfo {author} {\bibfnamefont
  {C.}~\bibnamefont {Eichler}},\ and\ \bibinfo {author} {\bibfnamefont
  {A.}~\bibnamefont {Wallraff}},\ }\href
  {https://doi.org/10.1103/PhysRevApplied.14.024042} {\bibfield  {journal}
  {\bibinfo  {journal} {Phys. Rev. Applied}\ }\textbf {\bibinfo {volume}
  {14}},\ \bibinfo {pages} {024042} (\bibinfo {year} {2020})}\BibitemShut
  {NoStop}%
\bibitem [{\citenamefont {Andersen}\ \emph {et~al.}(2020)\citenamefont
  {Andersen}, \citenamefont {Remm}, \citenamefont {Lazar}, \citenamefont
  {Krinner}, \citenamefont {Lacroix}, \citenamefont {Norris}, \citenamefont
  {Gabureac}, \citenamefont {Eichler},\ and\ \citenamefont
  {Wallraff}}]{Andersen20}%
  \BibitemOpen
  \bibfield  {author} {\bibinfo {author} {\bibfnamefont {C.~K.}\ \bibnamefont
  {Andersen}}, \bibinfo {author} {\bibfnamefont {A.}~\bibnamefont {Remm}},
  \bibinfo {author} {\bibfnamefont {S.}~\bibnamefont {Lazar}}, \bibinfo
  {author} {\bibfnamefont {S.}~\bibnamefont {Krinner}}, \bibinfo {author}
  {\bibfnamefont {N.}~\bibnamefont {Lacroix}}, \bibinfo {author} {\bibfnamefont
  {G.~J.}\ \bibnamefont {Norris}}, \bibinfo {author} {\bibfnamefont
  {M.}~\bibnamefont {Gabureac}}, \bibinfo {author} {\bibfnamefont
  {C.}~\bibnamefont {Eichler}},\ and\ \bibinfo {author} {\bibfnamefont
  {A.}~\bibnamefont {Wallraff}},\ }\href
  {https://doi.org/10.1038/s41567-020-0920-y} {\bibfield  {journal} {\bibinfo
  {journal} {Nature Physics}\ }\textbf {\bibinfo {volume} {16}},\ \bibinfo
  {pages} {875} (\bibinfo {year} {2020})}\BibitemShut {NoStop}%
\bibitem [{\citenamefont {Balanis}(2012)}]{Balanis12}%
  \BibitemOpen
  \bibfield  {author} {\bibinfo {author} {\bibfnamefont {C.~A.}\ \bibnamefont
  {Balanis}},\ }\href@noop {} {\emph {\bibinfo {title} {Advanced Engineering
  Electromagnetics}}}\ (\bibinfo  {publisher} {Wiley},\ \bibinfo {year}
  {2012})\BibitemShut {NoStop}%
\bibitem [{\citenamefont {Lagemann}\ \emph {et~al.}(2022)\citenamefont
  {Lagemann}, \citenamefont {Willsch}, \citenamefont {Willsch}, \citenamefont
  {Jin}, \citenamefont {De~Raedt},\ and\ \citenamefont
  {Michielsen}}]{Lagemann21}%
  \BibitemOpen
  \bibfield  {author} {\bibinfo {author} {\bibfnamefont {H.}~\bibnamefont
  {Lagemann}}, \bibinfo {author} {\bibfnamefont {D.}~\bibnamefont {Willsch}},
  \bibinfo {author} {\bibfnamefont {M.}~\bibnamefont {Willsch}}, \bibinfo
  {author} {\bibfnamefont {F.}~\bibnamefont {Jin}}, \bibinfo {author}
  {\bibfnamefont {H.}~\bibnamefont {De~Raedt}},\ and\ \bibinfo {author}
  {\bibfnamefont {K.}~\bibnamefont {Michielsen}},\ }\href
  {https://doi.org/10.1103/PhysRevA.106.022615} {\bibfield  {journal} {\bibinfo
   {journal} {Phys. Rev. A}\ }\textbf {\bibinfo {volume} {106}},\ \bibinfo
  {pages} {022615} (\bibinfo {year} {2022})}\BibitemShut {NoStop}%
\bibitem [{\citenamefont {DeRaedt}(1987)}]{DeRaedt87}%
  \BibitemOpen
  \bibfield  {author} {\bibinfo {author} {\bibfnamefont {H.}~\bibnamefont
  {DeRaedt}},\ }\href
  {https://www.sciencedirect.com/science/article/abs/pii/0167797787900025}
  {\bibfield  {journal} {\bibinfo  {journal} {Computer Physics Reports}\
  }\textbf {\bibinfo {volume} {7}},\ \bibinfo {pages} {1} (\bibinfo {year}
  {1987})}\BibitemShut {NoStop}%
\bibitem [{\citenamefont {Huyghebaert}\ and\ \citenamefont {{De
  Raedt}}(1990)}]{Huyghebaert90}%
  \BibitemOpen
  \bibfield  {author} {\bibinfo {author} {\bibfnamefont {J.}~\bibnamefont
  {Huyghebaert}}\ and\ \bibinfo {author} {\bibfnamefont {H.}~\bibnamefont {{De
  Raedt}}},\ }\href {https://doi.org/10.1088/0305-4470/23/24/019} {\bibfield
  {journal} {\bibinfo  {journal} {J. Phys. A: Math. Gen.}\ }\textbf {\bibinfo
  {volume} {23}},\ \bibinfo {pages} {5777} (\bibinfo {year}
  {1990})}\BibitemShut {NoStop}%
\bibitem [{\citenamefont {Vion}\ \emph {et~al.}(2002)\citenamefont {Vion},
  \citenamefont {Aassime}, \citenamefont {Cottet}, \citenamefont {Joyez},
  \citenamefont {Pothier}, \citenamefont {Urbina}, \citenamefont {Esteve},\
  and\ \citenamefont {Devoret}}]{Vion2002CPBqubitsQuantronium}%
  \BibitemOpen
  \bibfield  {author} {\bibinfo {author} {\bibfnamefont {D.}~\bibnamefont
  {Vion}}, \bibinfo {author} {\bibfnamefont {A.}~\bibnamefont {Aassime}},
  \bibinfo {author} {\bibfnamefont {A.}~\bibnamefont {Cottet}}, \bibinfo
  {author} {\bibfnamefont {P.}~\bibnamefont {Joyez}}, \bibinfo {author}
  {\bibfnamefont {H.}~\bibnamefont {Pothier}}, \bibinfo {author} {\bibfnamefont
  {C.}~\bibnamefont {Urbina}}, \bibinfo {author} {\bibfnamefont
  {D.}~\bibnamefont {Esteve}},\ and\ \bibinfo {author} {\bibfnamefont {M.~H.}\
  \bibnamefont {Devoret}},\ }\href {https://doi.org/10.1126/science.1069372}
  {\bibfield  {journal} {\bibinfo  {journal} {Science}\ }\textbf {\bibinfo
  {volume} {296}},\ \bibinfo {pages} {886} (\bibinfo {year} {2002})},\ \Eprint
  {https://arxiv.org/abs/http://science.sciencemag.org/content/296/5569/886.full.pdf}
  {http://science.sciencemag.org/content/296/5569/886.full.pdf} \BibitemShut
  {NoStop}%
\bibitem [{\citenamefont {Koch}\ \emph {et~al.}(2007)\citenamefont {Koch},
  \citenamefont {Yu}, \citenamefont {Gambetta}, \citenamefont {Houck},
  \citenamefont {Schuster}, \citenamefont {Majer}, \citenamefont {Blais},
  \citenamefont {Devoret}, \citenamefont {Girvin},\ and\ \citenamefont
  {Schoelkopf}}]{Koch}%
  \BibitemOpen
  \bibfield  {author} {\bibinfo {author} {\bibfnamefont {J.}~\bibnamefont
  {Koch}}, \bibinfo {author} {\bibfnamefont {T.~M.}\ \bibnamefont {Yu}},
  \bibinfo {author} {\bibfnamefont {J.}~\bibnamefont {Gambetta}}, \bibinfo
  {author} {\bibfnamefont {A.~A.}\ \bibnamefont {Houck}}, \bibinfo {author}
  {\bibfnamefont {D.~I.}\ \bibnamefont {Schuster}}, \bibinfo {author}
  {\bibfnamefont {J.}~\bibnamefont {Majer}}, \bibinfo {author} {\bibfnamefont
  {A.}~\bibnamefont {Blais}}, \bibinfo {author} {\bibfnamefont {M.~H.}\
  \bibnamefont {Devoret}}, \bibinfo {author} {\bibfnamefont {S.~M.}\
  \bibnamefont {Girvin}},\ and\ \bibinfo {author} {\bibfnamefont {R.~J.}\
  \bibnamefont {Schoelkopf}},\ }\href
  {https://doi.org/10.1103/PhysRevA.76.042319} {\bibfield  {journal} {\bibinfo
  {journal} {Phys. Rev. A}\ }\textbf {\bibinfo {volume} {76}},\ \bibinfo
  {pages} {042319} (\bibinfo {year} {2007})}\BibitemShut {NoStop}%
\bibitem [{\citenamefont {You}\ \emph {et~al.}(2019)\citenamefont {You},
  \citenamefont {Sauls},\ and\ \citenamefont {Koch}}]{You}%
  \BibitemOpen
  \bibfield  {author} {\bibinfo {author} {\bibfnamefont {X.}~\bibnamefont
  {You}}, \bibinfo {author} {\bibfnamefont {J.~A.}\ \bibnamefont {Sauls}},\
  and\ \bibinfo {author} {\bibfnamefont {J.}~\bibnamefont {Koch}},\ }\href
  {https://doi.org/10.1103/PhysRevB.99.174512} {\bibfield  {journal} {\bibinfo
  {journal} {Phys. Rev. B}\ }\textbf {\bibinfo {volume} {99}},\ \bibinfo
  {pages} {174512} (\bibinfo {year} {2019})}\BibitemShut {NoStop}%
\bibitem [{\citenamefont {Riwar}\ and\ \citenamefont
  {DiVincenzo}(2022)}]{Riwar21}%
  \BibitemOpen
  \bibfield  {author} {\bibinfo {author} {\bibfnamefont {R.-P.}\ \bibnamefont
  {Riwar}}\ and\ \bibinfo {author} {\bibfnamefont {D.~P.}\ \bibnamefont
  {DiVincenzo}},\ }\href {https://doi.org/10.1038/s41534-022-00539-x}
  {\bibfield  {journal} {\bibinfo  {journal} {npj Quantum Information}\
  }\textbf {\bibinfo {volume} {8}},\ \bibinfo {pages} {36} (\bibinfo {year}
  {2022})}\BibitemShut {NoStop}%
\bibitem [{\citenamefont {Didier}\ \emph {et~al.}(2018)\citenamefont {Didier},
  \citenamefont {Sete}, \citenamefont {da~Silva},\ and\ \citenamefont
  {Rigetti}}]{Didier}%
  \BibitemOpen
  \bibfield  {author} {\bibinfo {author} {\bibfnamefont {N.}~\bibnamefont
  {Didier}}, \bibinfo {author} {\bibfnamefont {E.~A.}\ \bibnamefont {Sete}},
  \bibinfo {author} {\bibfnamefont {M.~P.}\ \bibnamefont {da~Silva}},\ and\
  \bibinfo {author} {\bibfnamefont {C.}~\bibnamefont {Rigetti}},\ }\href
  {https://doi.org/10.1103/PhysRevA.97.022330} {\bibfield  {journal} {\bibinfo
  {journal} {Phys. Rev. A}\ }\textbf {\bibinfo {volume} {97}},\ \bibinfo
  {pages} {022330} (\bibinfo {year} {2018})}\BibitemShut {NoStop}%
\bibitem [{\citenamefont {Motzoi}\ \emph {et~al.}(2009)\citenamefont {Motzoi},
  \citenamefont {Gambetta}, \citenamefont {Rebentrost},\ and\ \citenamefont
  {Wilhelm}}]{Motzoi09}%
  \BibitemOpen
  \bibfield  {author} {\bibinfo {author} {\bibfnamefont {F.}~\bibnamefont
  {Motzoi}}, \bibinfo {author} {\bibfnamefont {J.~M.}\ \bibnamefont
  {Gambetta}}, \bibinfo {author} {\bibfnamefont {P.}~\bibnamefont
  {Rebentrost}},\ and\ \bibinfo {author} {\bibfnamefont {F.~K.}\ \bibnamefont
  {Wilhelm}},\ }\href {https://doi.org/10.1103/PhysRevLett.103.110501}
  {\bibfield  {journal} {\bibinfo  {journal} {Phys. Rev. Lett.}\ }\textbf
  {\bibinfo {volume} {103}},\ \bibinfo {pages} {110501} (\bibinfo {year}
  {2009})}\BibitemShut {NoStop}%
\bibitem [{\citenamefont {McKay}\ \emph {et~al.}(2017)\citenamefont {McKay},
  \citenamefont {Wood}, \citenamefont {Sheldon}, \citenamefont {Chow},\ and\
  \citenamefont {Gambetta}}]{McKay17}%
  \BibitemOpen
  \bibfield  {author} {\bibinfo {author} {\bibfnamefont {D.~C.}\ \bibnamefont
  {McKay}}, \bibinfo {author} {\bibfnamefont {C.~J.}\ \bibnamefont {Wood}},
  \bibinfo {author} {\bibfnamefont {S.}~\bibnamefont {Sheldon}}, \bibinfo
  {author} {\bibfnamefont {J.~M.}\ \bibnamefont {Chow}},\ and\ \bibinfo
  {author} {\bibfnamefont {J.~M.}\ \bibnamefont {Gambetta}},\ }\href
  {https://doi.org/10.1103/PhysRevA.96.022330} {\bibfield  {journal} {\bibinfo
  {journal} {Phys. Rev. A}\ }\textbf {\bibinfo {volume} {96}},\ \bibinfo
  {pages} {022330} (\bibinfo {year} {2017})}\BibitemShut {NoStop}%
\bibitem [{\citenamefont {Blais}\ \emph {et~al.}(2021)\citenamefont {Blais},
  \citenamefont {Grimsmo}, \citenamefont {Girvin},\ and\ \citenamefont
  {Wallraff}}]{Blais2020circuit}%
  \BibitemOpen
  \bibfield  {author} {\bibinfo {author} {\bibfnamefont {A.}~\bibnamefont
  {Blais}}, \bibinfo {author} {\bibfnamefont {A.~L.}\ \bibnamefont {Grimsmo}},
  \bibinfo {author} {\bibfnamefont {S.~M.}\ \bibnamefont {Girvin}},\ and\
  \bibinfo {author} {\bibfnamefont {A.}~\bibnamefont {Wallraff}},\ }\href
  {https://doi.org/10.1103/RevModPhys.93.025005} {\bibfield  {journal}
  {\bibinfo  {journal} {Rev. Mod. Phys.}\ }\textbf {\bibinfo {volume} {93}},\
  \bibinfo {pages} {025005} (\bibinfo {year} {2021})}\BibitemShut {NoStop}%
\bibitem [{\citenamefont {Willsch}(2020)}]{Willsch2020}%
  \BibitemOpen
  \bibfield  {author} {\bibinfo {author} {\bibfnamefont {D.}~\bibnamefont
  {Willsch}},\ }\emph {\bibinfo {title} {{S}upercomputer simulations of
  transmon quantum computers}},\ \href
  {https://juser.fz-juelich.de/record/885927} {\bibinfo {type}
  {Dissertation}},\ \bibinfo  {school} {RWTH Aachen University} (\bibinfo
  {year} {2020})\BibitemShut {NoStop}%
\bibitem [{\citenamefont {{J\"{u}lich Supercomputing Centre}}(2019)}]{JUWELS}%
  \BibitemOpen
  \bibfield  {author} {\bibinfo {author} {\bibnamefont {{J\"{u}lich
  Supercomputing Centre}}},\ }\href {https://doi.org/10.17815/jlsrf-5-171}
  {\bibfield  {journal} {\bibinfo  {journal} {Journal of large-scale research
  facilities}\ }\textbf {\bibinfo {volume} {5}},\ \bibinfo {pages} {A135}
  (\bibinfo {year} {2019})}\BibitemShut {NoStop}%
\bibitem [{\citenamefont {Johnston}\ \emph {et~al.}(2009)\citenamefont
  {Johnston}, \citenamefont {Kribs},\ and\ \citenamefont
  {Paulsen}}]{Johnston09}%
  \BibitemOpen
  \bibfield  {author} {\bibinfo {author} {\bibfnamefont {N.}~\bibnamefont
  {Johnston}}, \bibinfo {author} {\bibfnamefont {D.}~\bibnamefont {Kribs}},\
  and\ \bibinfo {author} {\bibfnamefont {V.}~\bibnamefont {Paulsen}},\ }\href
  {https://www.rintonpress.com/journals/qiconline.html#v9n12} {\bibfield
  {journal} {\bibinfo  {journal} {Quantum Information and Computation}\
  }\textbf {\bibinfo {volume} {9}} (\bibinfo {year} {2009})}\BibitemShut
  {NoStop}%
\bibitem [{\citenamefont {Corporation}()}]{MKL09}%
  \BibitemOpen
  \bibfield  {author} {\bibinfo {author} {\bibfnamefont {I.}~\bibnamefont
  {Corporation}},\ }\href
  {https://software.intel.com/content/www/us/en/develop/tools/oneapi/components/onemkl.html#gs.axn2um}
  {\bibinfo {title} {Mkl:math kernel library}}\BibitemShut {NoStop}%
\bibitem [{\citenamefont {DiCarlo}\ \emph {et~al.}(2019)\citenamefont
  {DiCarlo}, \citenamefont {Chow}, \citenamefont {Gambetta}, \citenamefont
  {Bishop}, \citenamefont {Johnson}, \citenamefont {Schuster}, \citenamefont
  {Majer}, \citenamefont {Blais}, \citenamefont {Frunzio}, \citenamefont
  {Girvin},\ and\ \citenamefont {Schoelkopf}}]{DiCarlo2009}%
  \BibitemOpen
  \bibfield  {author} {\bibinfo {author} {\bibfnamefont {L.}~\bibnamefont
  {DiCarlo}}, \bibinfo {author} {\bibfnamefont {J.~M.}\ \bibnamefont {Chow}},
  \bibinfo {author} {\bibfnamefont {J.~M.}\ \bibnamefont {Gambetta}}, \bibinfo
  {author} {\bibfnamefont {L.~S.}\ \bibnamefont {Bishop}}, \bibinfo {author}
  {\bibfnamefont {B.~R.}\ \bibnamefont {Johnson}}, \bibinfo {author}
  {\bibfnamefont {D.~I.}\ \bibnamefont {Schuster}}, \bibinfo {author}
  {\bibfnamefont {J.}~\bibnamefont {Majer}}, \bibinfo {author} {\bibfnamefont
  {A.}~\bibnamefont {Blais}}, \bibinfo {author} {\bibfnamefont
  {L.}~\bibnamefont {Frunzio}}, \bibinfo {author} {\bibfnamefont {S.~M.}\
  \bibnamefont {Girvin}},\ and\ \bibinfo {author} {\bibfnamefont {R.~J.}\
  \bibnamefont {Schoelkopf}},\ }\href {https://doi.org/10.1038/nature08121}
  {\bibfield  {journal} {\bibinfo  {journal} {Nature}\ }\textbf {\bibinfo
  {volume} {460}},\ \bibinfo {pages} {120502} (\bibinfo {year}
  {2019})}\BibitemShut {NoStop}%
\bibitem [{\citenamefont {DiCarlo}\ \emph {et~al.}(2010)\citenamefont
  {DiCarlo}, \citenamefont {Reed}, \citenamefont {Sun}, \citenamefont
  {Johnson}, \citenamefont {Chow}, \citenamefont {Gambetta}, \citenamefont
  {Frunzio}, \citenamefont {Girvin}, \citenamefont {Devoret},\ and\
  \citenamefont {Schoelkopf}}]{DiCarlo2010}%
  \BibitemOpen
  \bibfield  {author} {\bibinfo {author} {\bibfnamefont {L.}~\bibnamefont
  {DiCarlo}}, \bibinfo {author} {\bibfnamefont {M.~D.}\ \bibnamefont {Reed}},
  \bibinfo {author} {\bibfnamefont {L.}~\bibnamefont {Sun}}, \bibinfo {author}
  {\bibfnamefont {B.~R.}\ \bibnamefont {Johnson}}, \bibinfo {author}
  {\bibfnamefont {J.~M.}\ \bibnamefont {Chow}}, \bibinfo {author}
  {\bibfnamefont {J.~M.}\ \bibnamefont {Gambetta}}, \bibinfo {author}
  {\bibfnamefont {L.}~\bibnamefont {Frunzio}}, \bibinfo {author} {\bibfnamefont
  {S.~M.}\ \bibnamefont {Girvin}}, \bibinfo {author} {\bibfnamefont {M.~H.}\
  \bibnamefont {Devoret}},\ and\ \bibinfo {author} {\bibfnamefont {R.~J.}\
  \bibnamefont {Schoelkopf}},\ }\href {https://doi.org/10.1038/nature09416}
  {\bibfield  {journal} {\bibinfo  {journal} {Nature}\ }\textbf {\bibinfo
  {volume} {467}},\ \bibinfo {pages} {574} (\bibinfo {year}
  {2010})}\BibitemShut {NoStop}%
\bibitem [{\citenamefont {Weinberg}(2015)}]{Weinberg2015}%
  \BibitemOpen
  \bibfield  {author} {\bibinfo {author} {\bibfnamefont {S.}~\bibnamefont
  {Weinberg}},\ }\href {https://doi.org/10.1017/CBO9781316276105} {\emph
  {\bibinfo {title} {Lectures on Quantum Mechanics}}},\ \bibinfo {edition}
  {2nd}\ ed.\ (\bibinfo  {publisher} {Cambridge University Press},\ \bibinfo
  {year} {2015})\BibitemShut {NoStop}%
\bibitem [{\citenamefont {Hund}(1927)}]{Hund1927}%
  \BibitemOpen
  \bibfield  {author} {\bibinfo {author} {\bibfnamefont {F.}~\bibnamefont
  {Hund}},\ }\href {https://doi.org/10.1007/BF01400234} {\bibfield  {journal}
  {\bibinfo  {journal} {Zeitschrift f{\"u}r Physik}\ }\textbf {\bibinfo
  {volume} {40}},\ \bibinfo {pages} {742} (\bibinfo {year} {1927})}\BibitemShut
  {NoStop}%
\bibitem [{\citenamefont {von Neumann}\ and\ \citenamefont
  {Wigner}(1993)}]{vonNeumann1993}%
  \BibitemOpen
  \bibfield  {author} {\bibinfo {author} {\bibfnamefont {J.}~\bibnamefont {von
  Neumann}}\ and\ \bibinfo {author} {\bibfnamefont {E.~P.}\ \bibnamefont
  {Wigner}},\ }\bibinfo {title} {{\"U}ber merkw{\"u}rdige diskrete
  eigenwerte},\ in\ \href {https://doi.org/10.1007/978-3-662-02781-3_19} {\emph
  {\bibinfo {booktitle} {The Collected Works of Eugene Paul Wigner: Part A: The
  Scientific Papers}}},\ \bibinfo {editor} {edited by\ \bibinfo {editor}
  {\bibfnamefont {A.~S.}\ \bibnamefont {Wightman}}}\ (\bibinfo  {publisher}
  {Springer Berlin Heidelberg},\ \bibinfo {address} {Berlin, Heidelberg},\
  \bibinfo {year} {1993})\ pp.\ \bibinfo {pages} {291--293}\BibitemShut
  {NoStop}%
\bibitem [{\citenamefont {Uhlig}(2020)}]{Uhlig2020}%
  \BibitemOpen
  \bibfield  {author} {\bibinfo {author} {\bibfnamefont {F.}~\bibnamefont
  {Uhlig}},\ }\href {https://doi.org/10.48550/ARXIV.2002.01274} {\bibinfo
  {title} {Coalescing eigenvalues and crossing eigencurves of 1-parameter
  matrix flows}} (\bibinfo {year} {2020})\BibitemShut {NoStop}%
\bibitem [{\citenamefont {Srinivasan}\ and\ \citenamefont
  {Kidambi}(2020)}]{Srinivasan2020}%
  \BibitemOpen
  \bibfield  {author} {\bibinfo {author} {\bibfnamefont {U.}~\bibnamefont
  {Srinivasan}}\ and\ \bibinfo {author} {\bibfnamefont {R.}~\bibnamefont
  {Kidambi}},\ }\href {https://doi.org/10.48550/ARXIV.2006.14254} {\bibinfo
  {title} {A sorting algorithm for complex eigenvalues}} (\bibinfo {year}
  {2020})\BibitemShut {NoStop}%
\bibitem [{\citenamefont {Krinner}\ \emph {et~al.}(2022)\citenamefont
  {Krinner}, \citenamefont {Lacroix}, \citenamefont {Remm}, \citenamefont
  {Di~Paolo}, \citenamefont {Genois}, \citenamefont {Leroux}, \citenamefont
  {Hellings}, \citenamefont {Lazar}, \citenamefont {Swiadek}, \citenamefont
  {Herrmann}, \citenamefont {Norris}, \citenamefont {Andersen}, \citenamefont
  {M{\"u}ller}, \citenamefont {Blais}, \citenamefont {Eichler},\ and\
  \citenamefont {Wallraff}}]{Krinner21}%
  \BibitemOpen
  \bibfield  {author} {\bibinfo {author} {\bibfnamefont {S.}~\bibnamefont
  {Krinner}}, \bibinfo {author} {\bibfnamefont {N.}~\bibnamefont {Lacroix}},
  \bibinfo {author} {\bibfnamefont {A.}~\bibnamefont {Remm}}, \bibinfo {author}
  {\bibfnamefont {A.}~\bibnamefont {Di~Paolo}}, \bibinfo {author}
  {\bibfnamefont {E.}~\bibnamefont {Genois}}, \bibinfo {author} {\bibfnamefont
  {C.}~\bibnamefont {Leroux}}, \bibinfo {author} {\bibfnamefont
  {C.}~\bibnamefont {Hellings}}, \bibinfo {author} {\bibfnamefont
  {S.}~\bibnamefont {Lazar}}, \bibinfo {author} {\bibfnamefont
  {F.}~\bibnamefont {Swiadek}}, \bibinfo {author} {\bibfnamefont
  {J.}~\bibnamefont {Herrmann}}, \bibinfo {author} {\bibfnamefont {G.~J.}\
  \bibnamefont {Norris}}, \bibinfo {author} {\bibfnamefont {C.~K.}\
  \bibnamefont {Andersen}}, \bibinfo {author} {\bibfnamefont {M.}~\bibnamefont
  {M{\"u}ller}}, \bibinfo {author} {\bibfnamefont {A.}~\bibnamefont {Blais}},
  \bibinfo {author} {\bibfnamefont {C.}~\bibnamefont {Eichler}},\ and\ \bibinfo
  {author} {\bibfnamefont {A.}~\bibnamefont {Wallraff}},\ }\href
  {https://doi.org/10.1038/s41586-022-04566-8} {\bibfield  {journal} {\bibinfo
  {journal} {Nature}\ }\textbf {\bibinfo {volume} {605}},\ \bibinfo {pages}
  {669} (\bibinfo {year} {2022})}\BibitemShut {NoStop}%
\bibitem [{\citenamefont {Arute}\ \emph {et~al.}(2019)\citenamefont {Arute},
  \citenamefont {Arya}, \citenamefont {Babbush}, \citenamefont {Bacon},
  \citenamefont {Bardin}, \citenamefont {Barends}, \citenamefont {Biswas},
  \citenamefont {Boixo}, \citenamefont {Brandao}, \citenamefont {Buell},
  \citenamefont {Burkett}, \citenamefont {Chen}, \citenamefont {Chen},
  \citenamefont {Chiaro}, \citenamefont {Collins}, \citenamefont {Courtney},
  \citenamefont {Dunsworth}, \citenamefont {Farhi}, \citenamefont {Foxen},
  \citenamefont {Fowler}, \citenamefont {Gidney}, \citenamefont {Giustina},
  \citenamefont {Graff}, \citenamefont {Guerin}, \citenamefont {Habegger},
  \citenamefont {Harrigan}, \citenamefont {Hartmann}, \citenamefont {Ho},
  \citenamefont {Hoffmann}, \citenamefont {Huang}, \citenamefont {Humble},
  \citenamefont {Isakov}, \citenamefont {Jeffrey}, \citenamefont {Jiang},
  \citenamefont {Kafri}, \citenamefont {Kechedzhi}, \citenamefont {Kelly},
  \citenamefont {Klimov}, \citenamefont {Knysh}, \citenamefont {Korotkov},
  \citenamefont {Kostritsa}, \citenamefont {Landhuis}, \citenamefont
  {Lindmark}, \citenamefont {Lucero}, \citenamefont {Lyakh}, \citenamefont
  {Mandrà}, \citenamefont {McClean}, \citenamefont {McEwen}, \citenamefont
  {Megrant}, \citenamefont {Mi}, \citenamefont {Michielsen}, \citenamefont
  {Mohseni}, \citenamefont {Mutus}, \citenamefont {Naaman}, \citenamefont
  {Neeley}, \citenamefont {Neill}, \citenamefont {Niu}, \citenamefont {Ostby},
  \citenamefont {Petukhov}, \citenamefont {Platt}, \citenamefont {Quintana},
  \citenamefont {Rieffel}, \citenamefont {Roushan}, \citenamefont {Rubin},
  \citenamefont {Sank}, \citenamefont {Satzinger}, \citenamefont {Smelyanskiy},
  \citenamefont {Sung}, \citenamefont {Trevithick}, \citenamefont
  {Vainsencher}, \citenamefont {Villalonga}, \citenamefont {White},
  \citenamefont {Yao}, \citenamefont {Yeh}, \citenamefont {Zalcman},
  \citenamefont {Neven},\ and\ \citenamefont {Martinis}}]{Arute19}%
  \BibitemOpen
  \bibfield  {author} {\bibinfo {author} {\bibfnamefont {F.}~\bibnamefont
  {Arute}}, \bibinfo {author} {\bibfnamefont {K.}~\bibnamefont {Arya}},
  \bibinfo {author} {\bibfnamefont {R.}~\bibnamefont {Babbush}}, \bibinfo
  {author} {\bibfnamefont {D.}~\bibnamefont {Bacon}}, \bibinfo {author}
  {\bibfnamefont {J.}~\bibnamefont {Bardin}}, \bibinfo {author} {\bibfnamefont
  {R.}~\bibnamefont {Barends}}, \bibinfo {author} {\bibfnamefont
  {R.}~\bibnamefont {Biswas}}, \bibinfo {author} {\bibfnamefont
  {S.}~\bibnamefont {Boixo}}, \bibinfo {author} {\bibfnamefont
  {F.}~\bibnamefont {Brandao}}, \bibinfo {author} {\bibfnamefont
  {D.}~\bibnamefont {Buell}}, \bibinfo {author} {\bibfnamefont
  {B.}~\bibnamefont {Burkett}}, \bibinfo {author} {\bibfnamefont
  {Y.}~\bibnamefont {Chen}}, \bibinfo {author} {\bibfnamefont {J.}~\bibnamefont
  {Chen}}, \bibinfo {author} {\bibfnamefont {B.}~\bibnamefont {Chiaro}},
  \bibinfo {author} {\bibfnamefont {R.}~\bibnamefont {Collins}}, \bibinfo
  {author} {\bibfnamefont {W.}~\bibnamefont {Courtney}}, \bibinfo {author}
  {\bibfnamefont {A.}~\bibnamefont {Dunsworth}}, \bibinfo {author}
  {\bibfnamefont {E.}~\bibnamefont {Farhi}}, \bibinfo {author} {\bibfnamefont
  {B.}~\bibnamefont {Foxen}}, \bibinfo {author} {\bibfnamefont
  {A.}~\bibnamefont {Fowler}}, \bibinfo {author} {\bibfnamefont {C.~M.}\
  \bibnamefont {Gidney}}, \bibinfo {author} {\bibfnamefont {M.}~\bibnamefont
  {Giustina}}, \bibinfo {author} {\bibfnamefont {R.}~\bibnamefont {Graff}},
  \bibinfo {author} {\bibfnamefont {K.}~\bibnamefont {Guerin}}, \bibinfo
  {author} {\bibfnamefont {S.}~\bibnamefont {Habegger}}, \bibinfo {author}
  {\bibfnamefont {M.}~\bibnamefont {Harrigan}}, \bibinfo {author}
  {\bibfnamefont {M.}~\bibnamefont {Hartmann}}, \bibinfo {author}
  {\bibfnamefont {A.}~\bibnamefont {Ho}}, \bibinfo {author} {\bibfnamefont
  {M.~R.}\ \bibnamefont {Hoffmann}}, \bibinfo {author} {\bibfnamefont
  {T.}~\bibnamefont {Huang}}, \bibinfo {author} {\bibfnamefont
  {T.}~\bibnamefont {Humble}}, \bibinfo {author} {\bibfnamefont
  {S.}~\bibnamefont {Isakov}}, \bibinfo {author} {\bibfnamefont
  {E.}~\bibnamefont {Jeffrey}}, \bibinfo {author} {\bibfnamefont
  {Z.}~\bibnamefont {Jiang}}, \bibinfo {author} {\bibfnamefont
  {D.}~\bibnamefont {Kafri}}, \bibinfo {author} {\bibfnamefont
  {K.}~\bibnamefont {Kechedzhi}}, \bibinfo {author} {\bibfnamefont
  {J.}~\bibnamefont {Kelly}}, \bibinfo {author} {\bibfnamefont
  {P.}~\bibnamefont {Klimov}}, \bibinfo {author} {\bibfnamefont
  {S.}~\bibnamefont {Knysh}}, \bibinfo {author} {\bibfnamefont
  {A.}~\bibnamefont {Korotkov}}, \bibinfo {author} {\bibfnamefont
  {F.}~\bibnamefont {Kostritsa}}, \bibinfo {author} {\bibfnamefont
  {D.}~\bibnamefont {Landhuis}}, \bibinfo {author} {\bibfnamefont
  {M.}~\bibnamefont {Lindmark}}, \bibinfo {author} {\bibfnamefont
  {E.}~\bibnamefont {Lucero}}, \bibinfo {author} {\bibfnamefont
  {D.}~\bibnamefont {Lyakh}}, \bibinfo {author} {\bibfnamefont
  {S.}~\bibnamefont {Mandrà}}, \bibinfo {author} {\bibfnamefont {J.~R.}\
  \bibnamefont {McClean}}, \bibinfo {author} {\bibfnamefont {M.}~\bibnamefont
  {McEwen}}, \bibinfo {author} {\bibfnamefont {A.}~\bibnamefont {Megrant}},
  \bibinfo {author} {\bibfnamefont {X.}~\bibnamefont {Mi}}, \bibinfo {author}
  {\bibfnamefont {K.}~\bibnamefont {Michielsen}}, \bibinfo {author}
  {\bibfnamefont {M.}~\bibnamefont {Mohseni}}, \bibinfo {author} {\bibfnamefont
  {J.}~\bibnamefont {Mutus}}, \bibinfo {author} {\bibfnamefont
  {O.}~\bibnamefont {Naaman}}, \bibinfo {author} {\bibfnamefont
  {M.}~\bibnamefont {Neeley}}, \bibinfo {author} {\bibfnamefont
  {C.}~\bibnamefont {Neill}}, \bibinfo {author} {\bibfnamefont {M.~Y.}\
  \bibnamefont {Niu}}, \bibinfo {author} {\bibfnamefont {E.}~\bibnamefont
  {Ostby}}, \bibinfo {author} {\bibfnamefont {A.}~\bibnamefont {Petukhov}},
  \bibinfo {author} {\bibfnamefont {J.}~\bibnamefont {Platt}}, \bibinfo
  {author} {\bibfnamefont {C.}~\bibnamefont {Quintana}}, \bibinfo {author}
  {\bibfnamefont {E.~G.}\ \bibnamefont {Rieffel}}, \bibinfo {author}
  {\bibfnamefont {P.}~\bibnamefont {Roushan}}, \bibinfo {author} {\bibfnamefont
  {N.}~\bibnamefont {Rubin}}, \bibinfo {author} {\bibfnamefont
  {D.}~\bibnamefont {Sank}}, \bibinfo {author} {\bibfnamefont {K.~J.}\
  \bibnamefont {Satzinger}}, \bibinfo {author} {\bibfnamefont {V.}~\bibnamefont
  {Smelyanskiy}}, \bibinfo {author} {\bibfnamefont {K.~J.}\ \bibnamefont
  {Sung}}, \bibinfo {author} {\bibfnamefont {M.}~\bibnamefont {Trevithick}},
  \bibinfo {author} {\bibfnamefont {A.}~\bibnamefont {Vainsencher}}, \bibinfo
  {author} {\bibfnamefont {B.}~\bibnamefont {Villalonga}}, \bibinfo {author}
  {\bibfnamefont {T.}~\bibnamefont {White}}, \bibinfo {author} {\bibfnamefont
  {Z.~J.}\ \bibnamefont {Yao}}, \bibinfo {author} {\bibfnamefont
  {P.}~\bibnamefont {Yeh}}, \bibinfo {author} {\bibfnamefont {A.}~\bibnamefont
  {Zalcman}}, \bibinfo {author} {\bibfnamefont {H.}~\bibnamefont {Neven}},\
  and\ \bibinfo {author} {\bibfnamefont {J.}~\bibnamefont {Martinis}},\ }\href
  {https://www.nature.com/articles/s41586-019-1666-5} {\bibfield  {journal}
  {\bibinfo  {journal} {Nature}\ }\textbf {\bibinfo {volume} {574}},\ \bibinfo
  {pages} {505–510} (\bibinfo {year} {2019})}\BibitemShut {NoStop}%
\bibitem [{\citenamefont {Baker}\ \emph {et~al.}(2022)\citenamefont {Baker},
  \citenamefont {Huber}, \citenamefont {Glaser}, \citenamefont {Roy},
  \citenamefont {Tsitsilin}, \citenamefont {Filipp},\ and\ \citenamefont
  {Hartmann}}]{Baker22}%
  \BibitemOpen
  \bibfield  {author} {\bibinfo {author} {\bibfnamefont {A.~J.}\ \bibnamefont
  {Baker}}, \bibinfo {author} {\bibfnamefont {G.~B.~P.}\ \bibnamefont {Huber}},
  \bibinfo {author} {\bibfnamefont {N.~J.}\ \bibnamefont {Glaser}}, \bibinfo
  {author} {\bibfnamefont {F.}~\bibnamefont {Roy}}, \bibinfo {author}
  {\bibfnamefont {I.}~\bibnamefont {Tsitsilin}}, \bibinfo {author}
  {\bibfnamefont {S.}~\bibnamefont {Filipp}},\ and\ \bibinfo {author}
  {\bibfnamefont {M.~J.}\ \bibnamefont {Hartmann}},\ }\href
  {https://doi.org/10.1063/5.0077443} {\bibfield  {journal} {\bibinfo
  {journal} {Applied Physics Letters}\ }\textbf {\bibinfo {volume} {120}},\
  \bibinfo {pages} {054002} (\bibinfo {year} {2022})},\ \Eprint
  {https://arxiv.org/abs/https://doi.org/10.1063/5.0077443}
  {https://doi.org/10.1063/5.0077443} \BibitemShut {NoStop}%
\bibitem [{\citenamefont {Cohen}\ \emph {et~al.}(2022)\citenamefont {Cohen},
  \citenamefont {Petrescu}, \citenamefont {Shillito},\ and\ \citenamefont
  {Blais}}]{Cohen2022}%
  \BibitemOpen
  \bibfield  {author} {\bibinfo {author} {\bibfnamefont {J.}~\bibnamefont
  {Cohen}}, \bibinfo {author} {\bibfnamefont {A.}~\bibnamefont {Petrescu}},
  \bibinfo {author} {\bibfnamefont {R.}~\bibnamefont {Shillito}},\ and\
  \bibinfo {author} {\bibfnamefont {A.}~\bibnamefont {Blais}},\ }\href@noop {}
  {\bibinfo {title} {Reminiscence of classical chaos in driven transmons}}
  (\bibinfo {year} {2022}),\ \Eprint {https://arxiv.org/abs/2207.09361}
  {arXiv:2207.09361 [quant-ph]} \BibitemShut {NoStop}%
\bibitem [{\citenamefont {Hertzberg}\ \emph {et~al.}(2021)\citenamefont
  {Hertzberg}, \citenamefont {Zhang}, \citenamefont {Rosenblatt}, \citenamefont
  {Magesan}, \citenamefont {Smolin}, \citenamefont {Yau}, \citenamefont
  {Adiga}, \citenamefont {Sandberg}, \citenamefont {Brink}, \citenamefont
  {Chow},\ and\ \citenamefont
  {Orcutt}}]{Hertzberg2021FrequencyCrowdingProblem}%
  \BibitemOpen
  \bibfield  {author} {\bibinfo {author} {\bibfnamefont {J.~B.}\ \bibnamefont
  {Hertzberg}}, \bibinfo {author} {\bibfnamefont {E.~J.}\ \bibnamefont
  {Zhang}}, \bibinfo {author} {\bibfnamefont {S.}~\bibnamefont {Rosenblatt}},
  \bibinfo {author} {\bibfnamefont {E.}~\bibnamefont {Magesan}}, \bibinfo
  {author} {\bibfnamefont {J.~A.}\ \bibnamefont {Smolin}}, \bibinfo {author}
  {\bibfnamefont {J.-B.}\ \bibnamefont {Yau}}, \bibinfo {author} {\bibfnamefont
  {V.~P.}\ \bibnamefont {Adiga}}, \bibinfo {author} {\bibfnamefont
  {M.}~\bibnamefont {Sandberg}}, \bibinfo {author} {\bibfnamefont
  {M.}~\bibnamefont {Brink}}, \bibinfo {author} {\bibfnamefont {J.~M.}\
  \bibnamefont {Chow}},\ and\ \bibinfo {author} {\bibfnamefont {J.~S.}\
  \bibnamefont {Orcutt}},\ }\href {https://doi.org/10.1038/s41534-021-00464-5}
  {\bibfield  {journal} {\bibinfo  {journal} {npj Quantum Inf.}\ }\textbf
  {\bibinfo {volume} {7}},\ \bibinfo {pages} {129} (\bibinfo {year}
  {2021})}\BibitemShut {NoStop}%
\bibitem [{\citenamefont {Johnson}()}]{NLopt}%
  \BibitemOpen
  \bibfield  {author} {\bibinfo {author} {\bibfnamefont {S.~G.}\ \bibnamefont
  {Johnson}},\ }\href {http://github.com/stevengj/nlopt} {\bibinfo {title} {The
  nlopt nonlinear-optimization package}}\BibitemShut {NoStop}%
\bibitem [{\citenamefont {Gu}\ \emph {et~al.}(2021)\citenamefont {Gu},
  \citenamefont {Fern\'andez-Pend\'as}, \citenamefont {Vikst\aa{}l},
  \citenamefont {Abad}, \citenamefont {Warren}, \citenamefont {Bengtsson},
  \citenamefont {Tancredi}, \citenamefont {Shumeiko}, \citenamefont {Bylander},
  \citenamefont {Johansson},\ and\ \citenamefont {Kockum}}]{Gu21}%
  \BibitemOpen
  \bibfield  {author} {\bibinfo {author} {\bibfnamefont {X.}~\bibnamefont
  {Gu}}, \bibinfo {author} {\bibfnamefont {J.}~\bibnamefont
  {Fern\'andez-Pend\'as}}, \bibinfo {author} {\bibfnamefont {P.}~\bibnamefont
  {Vikst\aa{}l}}, \bibinfo {author} {\bibfnamefont {T.}~\bibnamefont {Abad}},
  \bibinfo {author} {\bibfnamefont {C.}~\bibnamefont {Warren}}, \bibinfo
  {author} {\bibfnamefont {A.}~\bibnamefont {Bengtsson}}, \bibinfo {author}
  {\bibfnamefont {G.}~\bibnamefont {Tancredi}}, \bibinfo {author}
  {\bibfnamefont {V.}~\bibnamefont {Shumeiko}}, \bibinfo {author}
  {\bibfnamefont {J.}~\bibnamefont {Bylander}}, \bibinfo {author}
  {\bibfnamefont {G.}~\bibnamefont {Johansson}},\ and\ \bibinfo {author}
  {\bibfnamefont {A.~F.}\ \bibnamefont {Kockum}},\ }\href
  {https://doi.org/10.1103/PRXQuantum.2.040348} {\bibfield  {journal} {\bibinfo
   {journal} {PRX Quantum}\ }\textbf {\bibinfo {volume} {2}},\ \bibinfo {pages}
  {040348} (\bibinfo {year} {2021})}\BibitemShut {NoStop}%
\bibitem [{\citenamefont {Wittler}\ \emph {et~al.}(2021)\citenamefont
  {Wittler}, \citenamefont {Roy}, \citenamefont {Pack}, \citenamefont
  {Werninghaus}, \citenamefont {Roy}, \citenamefont {Egger}, \citenamefont
  {Filipp}, \citenamefont {Wilhelm},\ and\ \citenamefont
  {Machnes}}]{Wittler21}%
  \BibitemOpen
  \bibfield  {author} {\bibinfo {author} {\bibfnamefont {N.}~\bibnamefont
  {Wittler}}, \bibinfo {author} {\bibfnamefont {F.}~\bibnamefont {Roy}},
  \bibinfo {author} {\bibfnamefont {K.}~\bibnamefont {Pack}}, \bibinfo {author}
  {\bibfnamefont {M.}~\bibnamefont {Werninghaus}}, \bibinfo {author}
  {\bibfnamefont {A.~S.}\ \bibnamefont {Roy}}, \bibinfo {author} {\bibfnamefont
  {D.~J.}\ \bibnamefont {Egger}}, \bibinfo {author} {\bibfnamefont
  {S.}~\bibnamefont {Filipp}}, \bibinfo {author} {\bibfnamefont {F.~K.}\
  \bibnamefont {Wilhelm}},\ and\ \bibinfo {author} {\bibfnamefont
  {S.}~\bibnamefont {Machnes}},\ }\href
  {https://doi.org/10.1103/PhysRevApplied.15.034080} {\bibfield  {journal}
  {\bibinfo  {journal} {Phys. Rev. Applied}\ }\textbf {\bibinfo {volume}
  {15}},\ \bibinfo {pages} {034080} (\bibinfo {year} {2021})}\BibitemShut
  {NoStop}%
\bibitem [{\citenamefont {McKay}\ \emph {et~al.}(2016)\citenamefont {McKay},
  \citenamefont {Filipp}, \citenamefont {Mezzacapo}, \citenamefont {Magesan},
  \citenamefont {Chow},\ and\ \citenamefont {Gambetta}}]{McKay16}%
  \BibitemOpen
  \bibfield  {author} {\bibinfo {author} {\bibfnamefont {D.~C.}\ \bibnamefont
  {McKay}}, \bibinfo {author} {\bibfnamefont {S.}~\bibnamefont {Filipp}},
  \bibinfo {author} {\bibfnamefont {A.}~\bibnamefont {Mezzacapo}}, \bibinfo
  {author} {\bibfnamefont {E.}~\bibnamefont {Magesan}}, \bibinfo {author}
  {\bibfnamefont {J.~M.}\ \bibnamefont {Chow}},\ and\ \bibinfo {author}
  {\bibfnamefont {J.~M.}\ \bibnamefont {Gambetta}},\ }\href
  {https://doi.org/10.1103/PhysRevApplied.6.064007} {\bibfield  {journal}
  {\bibinfo  {journal} {Phys. Rev. Applied}\ }\textbf {\bibinfo {volume} {6}},\
  \bibinfo {pages} {064007} (\bibinfo {year} {2016})}\BibitemShut {NoStop}%
\bibitem [{\citenamefont {Roth}\ \emph {et~al.}(2017)\citenamefont {Roth},
  \citenamefont {Ganzhorn}, \citenamefont {Moll}, \citenamefont {Filipp},
  \citenamefont {Salis},\ and\ \citenamefont {Schmidt}}]{Roth19}%
  \BibitemOpen
  \bibfield  {author} {\bibinfo {author} {\bibfnamefont {M.}~\bibnamefont
  {Roth}}, \bibinfo {author} {\bibfnamefont {M.}~\bibnamefont {Ganzhorn}},
  \bibinfo {author} {\bibfnamefont {N.}~\bibnamefont {Moll}}, \bibinfo {author}
  {\bibfnamefont {S.}~\bibnamefont {Filipp}}, \bibinfo {author} {\bibfnamefont
  {G.}~\bibnamefont {Salis}},\ and\ \bibinfo {author} {\bibfnamefont
  {S.}~\bibnamefont {Schmidt}},\ }\href
  {https://doi.org/10.1103/PhysRevA.96.062323} {\bibfield  {journal} {\bibinfo
  {journal} {Phys. Rev. A}\ }\textbf {\bibinfo {volume} {96}},\ \bibinfo
  {pages} {062323} (\bibinfo {year} {2017})}\BibitemShut {NoStop}%
\bibitem [{\citenamefont {Yan}\ \emph {et~al.}(2018)\citenamefont {Yan},
  \citenamefont {Krantz}, \citenamefont {Sung}, \citenamefont {Kjaergaard},
  \citenamefont {Campbell}, \citenamefont {Orlando}, \citenamefont
  {Gustavsson},\ and\ \citenamefont {Oliver}}]{Yan18}%
  \BibitemOpen
  \bibfield  {author} {\bibinfo {author} {\bibfnamefont {F.}~\bibnamefont
  {Yan}}, \bibinfo {author} {\bibfnamefont {P.}~\bibnamefont {Krantz}},
  \bibinfo {author} {\bibfnamefont {Y.}~\bibnamefont {Sung}}, \bibinfo {author}
  {\bibfnamefont {M.}~\bibnamefont {Kjaergaard}}, \bibinfo {author}
  {\bibfnamefont {D.~L.}\ \bibnamefont {Campbell}}, \bibinfo {author}
  {\bibfnamefont {T.~P.}\ \bibnamefont {Orlando}}, \bibinfo {author}
  {\bibfnamefont {S.}~\bibnamefont {Gustavsson}},\ and\ \bibinfo {author}
  {\bibfnamefont {W.~D.}\ \bibnamefont {Oliver}},\ }\href
  {https://doi.org/10.1103/PhysRevApplied.10.054062} {\bibfield  {journal}
  {\bibinfo  {journal} {Phys. Rev. Applied}\ }\textbf {\bibinfo {volume}
  {10}},\ \bibinfo {pages} {054062} (\bibinfo {year} {2018})}\BibitemShut
  {NoStop}%
\bibitem [{\citenamefont {Willsch}\ \emph {et~al.}(2023)\citenamefont
  {Willsch}, \citenamefont {Rieger}, \citenamefont {Winkel}, \citenamefont
  {Willsch}, \citenamefont {Dickel}, \citenamefont {Krause}, \citenamefont
  {Ando}, \citenamefont {Lescanne}, \citenamefont {Leghtas}, \citenamefont
  {Bronn}, \citenamefont {Deb}, \citenamefont {Lanes}, \citenamefont {Minev},
  \citenamefont {Dennig}, \citenamefont {Geisert}, \citenamefont {Günzler},
  \citenamefont {Ihssen}, \citenamefont {Paluch}, \citenamefont {Reisinger},
  \citenamefont {Hanna}, \citenamefont {Bae}, \citenamefont {Schüffelgen},
  \citenamefont {Grützmacher}, \citenamefont {Buimaga-Iarinca}, \citenamefont
  {Morari}, \citenamefont {Wernsdorfer}, \citenamefont {DiVincenzo},
  \citenamefont {Michielsen}, \citenamefont {Catelani},\ and\ \citenamefont
  {Pop}}]{willsch2023josephsonharmonics}%
  \BibitemOpen
  \bibfield  {author} {\bibinfo {author} {\bibfnamefont {D.}~\bibnamefont
  {Willsch}}, \bibinfo {author} {\bibfnamefont {D.}~\bibnamefont {Rieger}},
  \bibinfo {author} {\bibfnamefont {P.}~\bibnamefont {Winkel}}, \bibinfo
  {author} {\bibfnamefont {M.}~\bibnamefont {Willsch}}, \bibinfo {author}
  {\bibfnamefont {C.}~\bibnamefont {Dickel}}, \bibinfo {author} {\bibfnamefont
  {J.}~\bibnamefont {Krause}}, \bibinfo {author} {\bibfnamefont
  {Y.}~\bibnamefont {Ando}}, \bibinfo {author} {\bibfnamefont {R.}~\bibnamefont
  {Lescanne}}, \bibinfo {author} {\bibfnamefont {Z.}~\bibnamefont {Leghtas}},
  \bibinfo {author} {\bibfnamefont {N.~T.}\ \bibnamefont {Bronn}}, \bibinfo
  {author} {\bibfnamefont {P.}~\bibnamefont {Deb}}, \bibinfo {author}
  {\bibfnamefont {O.}~\bibnamefont {Lanes}}, \bibinfo {author} {\bibfnamefont
  {Z.~K.}\ \bibnamefont {Minev}}, \bibinfo {author} {\bibfnamefont
  {B.}~\bibnamefont {Dennig}}, \bibinfo {author} {\bibfnamefont
  {S.}~\bibnamefont {Geisert}}, \bibinfo {author} {\bibfnamefont
  {S.}~\bibnamefont {Günzler}}, \bibinfo {author} {\bibfnamefont
  {S.}~\bibnamefont {Ihssen}}, \bibinfo {author} {\bibfnamefont
  {P.}~\bibnamefont {Paluch}}, \bibinfo {author} {\bibfnamefont
  {T.}~\bibnamefont {Reisinger}}, \bibinfo {author} {\bibfnamefont
  {R.}~\bibnamefont {Hanna}}, \bibinfo {author} {\bibfnamefont {J.~H.}\
  \bibnamefont {Bae}}, \bibinfo {author} {\bibfnamefont {P.}~\bibnamefont
  {Schüffelgen}}, \bibinfo {author} {\bibfnamefont {D.}~\bibnamefont
  {Grützmacher}}, \bibinfo {author} {\bibfnamefont {L.}~\bibnamefont
  {Buimaga-Iarinca}}, \bibinfo {author} {\bibfnamefont {C.}~\bibnamefont
  {Morari}}, \bibinfo {author} {\bibfnamefont {W.}~\bibnamefont {Wernsdorfer}},
  \bibinfo {author} {\bibfnamefont {D.~P.}\ \bibnamefont {DiVincenzo}},
  \bibinfo {author} {\bibfnamefont {K.}~\bibnamefont {Michielsen}}, \bibinfo
  {author} {\bibfnamefont {G.}~\bibnamefont {Catelani}},\ and\ \bibinfo
  {author} {\bibfnamefont {I.~M.}\ \bibnamefont {Pop}},\ }\href@noop {}
  {\bibinfo {title} {Observation of josephson harmonics in tunnel junctions}}
  (\bibinfo {year} {2023}),\ \Eprint {https://arxiv.org/abs/2302.09192}
  {arXiv:2302.09192 [quant-ph]} \BibitemShut {NoStop}%
\bibitem [{\citenamefont {Roth}\ \emph {et~al.}(2022)\citenamefont {Roth},
  \citenamefont {Ma},\ and\ \citenamefont {Chew}}]{Roth22}%
  \BibitemOpen
  \bibfield  {author} {\bibinfo {author} {\bibfnamefont {T.}~\bibnamefont
  {Roth}}, \bibinfo {author} {\bibfnamefont {R.}~\bibnamefont {Ma}},\ and\
  \bibinfo {author} {\bibfnamefont {W.~C.}\ \bibnamefont {Chew}},\ }\href
  {https://doi.org/10.1109/MAP.2022.3176593} {\bibfield  {journal} {\bibinfo
  {journal} {IEEE Antennas and Propagation Magazine}\ ,\ \bibinfo {pages} {2}}
  (\bibinfo {year} {2022})}\BibitemShut {NoStop}%
\bibitem [{\citenamefont {Foxen}\ \emph {et~al.}(2020)\citenamefont {Foxen},
  \citenamefont {Neill}, \citenamefont {Dunsworth}, \citenamefont {Roushan},
  \citenamefont {Chiaro}, \citenamefont {Megrant}, \citenamefont {Kelly},
  \citenamefont {Chen}, \citenamefont {Satzinger}, \citenamefont {Barends},
  \citenamefont {Arute}, \citenamefont {Arya}, \citenamefont {Babbush},
  \citenamefont {Bacon}, \citenamefont {Bardin}, \citenamefont {Boixo},
  \citenamefont {Buell}, \citenamefont {Burkett}, \citenamefont {Chen},
  \citenamefont {Collins}, \citenamefont {Farhi}, \citenamefont {Fowler},
  \citenamefont {Gidney}, \citenamefont {Giustina}, \citenamefont {Graff},
  \citenamefont {Harrigan}, \citenamefont {Huang}, \citenamefont {Isakov},
  \citenamefont {Jeffrey}, \citenamefont {Jiang}, \citenamefont {Kafri},
  \citenamefont {Kechedzhi}, \citenamefont {Klimov}, \citenamefont {Korotkov},
  \citenamefont {Kostritsa}, \citenamefont {Landhuis}, \citenamefont {Lucero},
  \citenamefont {McClean}, \citenamefont {McEwen}, \citenamefont {Mi},
  \citenamefont {Mohseni}, \citenamefont {Mutus}, \citenamefont {Naaman},
  \citenamefont {Neeley}, \citenamefont {Niu}, \citenamefont {Petukhov},
  \citenamefont {Quintana}, \citenamefont {Rubin}, \citenamefont {Sank},
  \citenamefont {Smelyanskiy}, \citenamefont {Vainsencher}, \citenamefont
  {White}, \citenamefont {Yao}, \citenamefont {Yeh}, \citenamefont {Zalcman},
  \citenamefont {Neven},\ and\ \citenamefont {Martinis}}]{Foxen20}%
  \BibitemOpen
  \bibfield  {author} {\bibinfo {author} {\bibfnamefont {B.}~\bibnamefont
  {Foxen}}, \bibinfo {author} {\bibfnamefont {C.}~\bibnamefont {Neill}},
  \bibinfo {author} {\bibfnamefont {A.}~\bibnamefont {Dunsworth}}, \bibinfo
  {author} {\bibfnamefont {P.}~\bibnamefont {Roushan}}, \bibinfo {author}
  {\bibfnamefont {B.}~\bibnamefont {Chiaro}}, \bibinfo {author} {\bibfnamefont
  {A.}~\bibnamefont {Megrant}}, \bibinfo {author} {\bibfnamefont
  {J.}~\bibnamefont {Kelly}}, \bibinfo {author} {\bibfnamefont
  {Z.}~\bibnamefont {Chen}}, \bibinfo {author} {\bibfnamefont {K.}~\bibnamefont
  {Satzinger}}, \bibinfo {author} {\bibfnamefont {R.}~\bibnamefont {Barends}},
  \bibinfo {author} {\bibfnamefont {F.}~\bibnamefont {Arute}}, \bibinfo
  {author} {\bibfnamefont {K.}~\bibnamefont {Arya}}, \bibinfo {author}
  {\bibfnamefont {R.}~\bibnamefont {Babbush}}, \bibinfo {author} {\bibfnamefont
  {D.}~\bibnamefont {Bacon}}, \bibinfo {author} {\bibfnamefont {J.~C.}\
  \bibnamefont {Bardin}}, \bibinfo {author} {\bibfnamefont {S.}~\bibnamefont
  {Boixo}}, \bibinfo {author} {\bibfnamefont {D.}~\bibnamefont {Buell}},
  \bibinfo {author} {\bibfnamefont {B.}~\bibnamefont {Burkett}}, \bibinfo
  {author} {\bibfnamefont {Y.}~\bibnamefont {Chen}}, \bibinfo {author}
  {\bibfnamefont {R.}~\bibnamefont {Collins}}, \bibinfo {author} {\bibfnamefont
  {E.}~\bibnamefont {Farhi}}, \bibinfo {author} {\bibfnamefont
  {A.}~\bibnamefont {Fowler}}, \bibinfo {author} {\bibfnamefont
  {C.}~\bibnamefont {Gidney}}, \bibinfo {author} {\bibfnamefont
  {M.}~\bibnamefont {Giustina}}, \bibinfo {author} {\bibfnamefont
  {R.}~\bibnamefont {Graff}}, \bibinfo {author} {\bibfnamefont
  {M.}~\bibnamefont {Harrigan}}, \bibinfo {author} {\bibfnamefont
  {T.}~\bibnamefont {Huang}}, \bibinfo {author} {\bibfnamefont {S.~V.}\
  \bibnamefont {Isakov}}, \bibinfo {author} {\bibfnamefont {E.}~\bibnamefont
  {Jeffrey}}, \bibinfo {author} {\bibfnamefont {Z.}~\bibnamefont {Jiang}},
  \bibinfo {author} {\bibfnamefont {D.}~\bibnamefont {Kafri}}, \bibinfo
  {author} {\bibfnamefont {K.}~\bibnamefont {Kechedzhi}}, \bibinfo {author}
  {\bibfnamefont {P.}~\bibnamefont {Klimov}}, \bibinfo {author} {\bibfnamefont
  {A.}~\bibnamefont {Korotkov}}, \bibinfo {author} {\bibfnamefont
  {F.}~\bibnamefont {Kostritsa}}, \bibinfo {author} {\bibfnamefont
  {D.}~\bibnamefont {Landhuis}}, \bibinfo {author} {\bibfnamefont
  {E.}~\bibnamefont {Lucero}}, \bibinfo {author} {\bibfnamefont
  {J.}~\bibnamefont {McClean}}, \bibinfo {author} {\bibfnamefont
  {M.}~\bibnamefont {McEwen}}, \bibinfo {author} {\bibfnamefont
  {X.}~\bibnamefont {Mi}}, \bibinfo {author} {\bibfnamefont {M.}~\bibnamefont
  {Mohseni}}, \bibinfo {author} {\bibfnamefont {J.~Y.}\ \bibnamefont {Mutus}},
  \bibinfo {author} {\bibfnamefont {O.}~\bibnamefont {Naaman}}, \bibinfo
  {author} {\bibfnamefont {M.}~\bibnamefont {Neeley}}, \bibinfo {author}
  {\bibfnamefont {M.}~\bibnamefont {Niu}}, \bibinfo {author} {\bibfnamefont
  {A.}~\bibnamefont {Petukhov}}, \bibinfo {author} {\bibfnamefont
  {C.}~\bibnamefont {Quintana}}, \bibinfo {author} {\bibfnamefont
  {N.}~\bibnamefont {Rubin}}, \bibinfo {author} {\bibfnamefont
  {D.}~\bibnamefont {Sank}}, \bibinfo {author} {\bibfnamefont {V.}~\bibnamefont
  {Smelyanskiy}}, \bibinfo {author} {\bibfnamefont {A.}~\bibnamefont
  {Vainsencher}}, \bibinfo {author} {\bibfnamefont {T.~C.}\ \bibnamefont
  {White}}, \bibinfo {author} {\bibfnamefont {Z.}~\bibnamefont {Yao}}, \bibinfo
  {author} {\bibfnamefont {P.}~\bibnamefont {Yeh}}, \bibinfo {author}
  {\bibfnamefont {A.}~\bibnamefont {Zalcman}}, \bibinfo {author} {\bibfnamefont
  {H.}~\bibnamefont {Neven}},\ and\ \bibinfo {author} {\bibfnamefont {J.~M.}\
  \bibnamefont {Martinis}} (\bibinfo {collaboration} {Google AI Quantum}),\
  }\href {https://doi.org/10.1103/PhysRevLett.125.120504} {\bibfield  {journal}
  {\bibinfo  {journal} {Phys. Rev. Lett.}\ }\textbf {\bibinfo {volume} {125}},\
  \bibinfo {pages} {120504} (\bibinfo {year} {2020})}\BibitemShut {NoStop}%
\bibitem [{\citenamefont {Ganzhorn}\ \emph {et~al.}(2020)\citenamefont
  {Ganzhorn}, \citenamefont {Salis}, \citenamefont {Egger}, \citenamefont
  {Fuhrer}, \citenamefont {Mergenthaler}, \citenamefont {M\"uller},
  \citenamefont {M\"uller}, \citenamefont {Paredes}, \citenamefont {Pechal},
  \citenamefont {Werninghaus},\ and\ \citenamefont {Filipp}}]{Ganzhorn20}%
  \BibitemOpen
  \bibfield  {author} {\bibinfo {author} {\bibfnamefont {M.}~\bibnamefont
  {Ganzhorn}}, \bibinfo {author} {\bibfnamefont {G.}~\bibnamefont {Salis}},
  \bibinfo {author} {\bibfnamefont {D.~J.}\ \bibnamefont {Egger}}, \bibinfo
  {author} {\bibfnamefont {A.}~\bibnamefont {Fuhrer}}, \bibinfo {author}
  {\bibfnamefont {M.}~\bibnamefont {Mergenthaler}}, \bibinfo {author}
  {\bibfnamefont {C.}~\bibnamefont {M\"uller}}, \bibinfo {author}
  {\bibfnamefont {P.}~\bibnamefont {M\"uller}}, \bibinfo {author}
  {\bibfnamefont {S.}~\bibnamefont {Paredes}}, \bibinfo {author} {\bibfnamefont
  {M.}~\bibnamefont {Pechal}}, \bibinfo {author} {\bibfnamefont
  {M.}~\bibnamefont {Werninghaus}},\ and\ \bibinfo {author} {\bibfnamefont
  {S.}~\bibnamefont {Filipp}},\ }\href
  {https://doi.org/10.1103/PhysRevResearch.2.033447} {\bibfield  {journal}
  {\bibinfo  {journal} {Phys. Rev. Research}\ }\textbf {\bibinfo {volume}
  {2}},\ \bibinfo {pages} {033447} (\bibinfo {year} {2020})}\BibitemShut
  {NoStop}%
\bibitem [{\citenamefont {Alicki}\ \emph {et~al.}(2002)\citenamefont {Alicki},
  \citenamefont {Horodecki}, \citenamefont {Horodecki},\ and\ \citenamefont
  {Horodecki}}]{Alicki2002}%
  \BibitemOpen
  \bibfield  {author} {\bibinfo {author} {\bibfnamefont {R.}~\bibnamefont
  {Alicki}}, \bibinfo {author} {\bibfnamefont {M.}~\bibnamefont {Horodecki}},
  \bibinfo {author} {\bibfnamefont {P.}~\bibnamefont {Horodecki}},\ and\
  \bibinfo {author} {\bibfnamefont {R.}~\bibnamefont {Horodecki}},\ }\href
  {https://doi.org/10.1103/PhysRevA.65.062101} {\bibfield  {journal} {\bibinfo
  {journal} {Phys. Rev. A}\ }\textbf {\bibinfo {volume} {65}},\ \bibinfo
  {pages} {062101} (\bibinfo {year} {2002})}\BibitemShut {NoStop}%
\bibitem [{\citenamefont {Alicki}\ \emph {et~al.}(2006)\citenamefont {Alicki},
  \citenamefont {Lidar},\ and\ \citenamefont {Zanardi}}]{Alicki2006}%
  \BibitemOpen
  \bibfield  {author} {\bibinfo {author} {\bibfnamefont {R.}~\bibnamefont
  {Alicki}}, \bibinfo {author} {\bibfnamefont {D.~A.}\ \bibnamefont {Lidar}},\
  and\ \bibinfo {author} {\bibfnamefont {P.}~\bibnamefont {Zanardi}},\ }\href
  {https://doi.org/10.1103/PhysRevA.73.052311} {\bibfield  {journal} {\bibinfo
  {journal} {Phys. Rev. A}\ }\textbf {\bibinfo {volume} {73}},\ \bibinfo
  {pages} {052311} (\bibinfo {year} {2006})}\BibitemShut {NoStop}%
\bibitem [{\citenamefont {Terhal}\ and\ \citenamefont
  {Burkard}(2005)}]{Terhal2005}%
  \BibitemOpen
  \bibfield  {author} {\bibinfo {author} {\bibfnamefont {B.~M.}\ \bibnamefont
  {Terhal}}\ and\ \bibinfo {author} {\bibfnamefont {G.}~\bibnamefont
  {Burkard}},\ }\href {https://doi.org/10.1103/PhysRevA.71.012336} {\bibfield
  {journal} {\bibinfo  {journal} {Phys. Rev. A}\ }\textbf {\bibinfo {volume}
  {71}},\ \bibinfo {pages} {012336} (\bibinfo {year} {2005})}\BibitemShut
  {NoStop}%
\bibitem [{\citenamefont {Lagemann}(2023)}]{Lagemann2022}%
  \BibitemOpen
  \bibfield  {author} {\bibinfo {author} {\bibfnamefont {H.~A.}\ \bibnamefont
  {Lagemann}},\ }\emph {\bibinfo {title} {{R}eal-time simulations of transmon
  systems with time-dependent {H}amiltonian models}},\ \href
  {https://doi.org/10.18154/RWTH-2023-02693} {\bibinfo {type} {Dissertation}},\
  \bibinfo  {school} {RWTH Aachen University}, \bibinfo {address} {Aachen}
  (\bibinfo {year} {2023}),\ \bibinfo {note} {veröffentlicht auf dem
  Publikationsserver der RWTH Aachen University; Dissertation, RWTH Aachen
  University, 2023}\BibitemShut {NoStop}%
\bibitem [{\citenamefont {Lagemann}(2019)}]{LagemannMSCThesis}%
  \BibitemOpen
  \bibfield  {author} {\bibinfo {author} {\bibfnamefont {H.}~\bibnamefont
  {Lagemann}},\ }\emph {\bibinfo {title} {{D}evelopment and implementation of a
  gate-based quantum computer simulator for high-dimensional {H}ilbert
  spaces}},\ \href {https://juser.fz-juelich.de/record/868377} {\bibinfo {type}
  {Masterarbeit}},\ \bibinfo  {school} {RWTH Aachen} (\bibinfo {year} {2019}),\
  \bibinfo {note} {masterarbeit, RWTH Aachen, 2019}\BibitemShut {NoStop}%
\bibitem [{\citenamefont {Murray}(2021)}]{Murray2021}%
  \BibitemOpen
  \bibfield  {author} {\bibinfo {author} {\bibfnamefont {C.~E.}\ \bibnamefont
  {Murray}},\ }\href
  {https://doi.org/https://doi.org/10.1016/j.mser.2021.100646} {\bibfield
  {journal} {\bibinfo  {journal} {Materials Science and Engineering: R:
  Reports}\ }\textbf {\bibinfo {volume} {146}},\ \bibinfo {pages} {100646}
  (\bibinfo {year} {2021})}\BibitemShut {NoStop}%
\bibitem [{\citenamefont {Martinis}\ and\ \citenamefont
  {Megrant}(2014)}]{Martinis14}%
  \BibitemOpen
  \bibfield  {author} {\bibinfo {author} {\bibfnamefont {J.~M.}\ \bibnamefont
  {Martinis}}\ and\ \bibinfo {author} {\bibfnamefont {A.}~\bibnamefont
  {Megrant}},\ }\href@noop {} {\bibinfo {title} {Ucsb final report for the csq
  program: Review of decoherence and materials physics for superconducting
  qubits}} (\bibinfo {year} {2014}),\ \Eprint {https://arxiv.org/abs/1410.5793}
  {arXiv:1410.5793 [quant-ph]} \BibitemShut {NoStop}%
\bibitem [{\citenamefont {Farhi}\ \emph {et~al.}(2014)\citenamefont {Farhi},
  \citenamefont {Goldstone},\ and\ \citenamefont {Gutmann}}]{farhi2014quantum}%
  \BibitemOpen
  \bibfield  {author} {\bibinfo {author} {\bibfnamefont {E.}~\bibnamefont
  {Farhi}}, \bibinfo {author} {\bibfnamefont {J.}~\bibnamefont {Goldstone}},\
  and\ \bibinfo {author} {\bibfnamefont {S.}~\bibnamefont {Gutmann}},\
  }\href@noop {} {\bibinfo {title} {A quantum approximate optimization
  algorithm}} (\bibinfo {year} {2014}),\ \Eprint
  {https://arxiv.org/abs/1411.4028} {arXiv:1411.4028 [quant-ph]} \BibitemShut
  {NoStop}%
\bibitem [{\citenamefont {Willsch}\ \emph {et~al.}(2022)\citenamefont
  {Willsch}, \citenamefont {Willsch}, \citenamefont {Jin}, \citenamefont
  {Michielsen},\ and\ \citenamefont {{De Raedt}}}]{WILLSCHM2022}%
  \BibitemOpen
  \bibfield  {author} {\bibinfo {author} {\bibfnamefont {D.}~\bibnamefont
  {Willsch}}, \bibinfo {author} {\bibfnamefont {M.}~\bibnamefont {Willsch}},
  \bibinfo {author} {\bibfnamefont {F.}~\bibnamefont {Jin}}, \bibinfo {author}
  {\bibfnamefont {K.}~\bibnamefont {Michielsen}},\ and\ \bibinfo {author}
  {\bibfnamefont {H.}~\bibnamefont {{De Raedt}}},\ }\href
  {https://doi.org/https://doi.org/10.1016/j.cpc.2022.108411} {\bibfield
  {journal} {\bibinfo  {journal} {Computer Physics Communications}\ }\textbf
  {\bibinfo {volume} {278}},\ \bibinfo {pages} {108411} (\bibinfo {year}
  {2022})}\BibitemShut {NoStop}%
\bibitem [{\citenamefont {Fenchel}(1949)}]{Fenchel1949}%
  \BibitemOpen
  \bibfield  {author} {\bibinfo {author} {\bibfnamefont {W.}~\bibnamefont
  {Fenchel}},\ }\href {https://doi.org/10.4153/CJM-1949-007-x} {\bibfield
  {journal} {\bibinfo  {journal} {Canadian Journal of Mathematics}\ }\textbf
  {\bibinfo {volume} {1}},\ \bibinfo {pages} {73–77} (\bibinfo {year}
  {1949})}\BibitemShut {NoStop}%
\bibitem [{\citenamefont {Bengtsson}\ \emph {et~al.}(2020)\citenamefont
  {Bengtsson}, \citenamefont {Vikst\aa{}l}, \citenamefont {Warren},
  \citenamefont {Svensson}, \citenamefont {Gu}, \citenamefont {Kockum},
  \citenamefont {Krantz}, \citenamefont {Kri\ifmmode~\check{z}\else
  \v{z}\fi{}an}, \citenamefont {Shiri}, \citenamefont {Svensson}, \citenamefont
  {Tancredi}, \citenamefont {Johansson}, \citenamefont {Delsing}, \citenamefont
  {Ferrini},\ and\ \citenamefont {Bylander}}]{Bengtsson2020}%
  \BibitemOpen
  \bibfield  {author} {\bibinfo {author} {\bibfnamefont {A.}~\bibnamefont
  {Bengtsson}}, \bibinfo {author} {\bibfnamefont {P.}~\bibnamefont
  {Vikst\aa{}l}}, \bibinfo {author} {\bibfnamefont {C.}~\bibnamefont {Warren}},
  \bibinfo {author} {\bibfnamefont {M.}~\bibnamefont {Svensson}}, \bibinfo
  {author} {\bibfnamefont {X.}~\bibnamefont {Gu}}, \bibinfo {author}
  {\bibfnamefont {A.~F.}\ \bibnamefont {Kockum}}, \bibinfo {author}
  {\bibfnamefont {P.}~\bibnamefont {Krantz}}, \bibinfo {author} {\bibfnamefont
  {C.}~\bibnamefont {Kri\ifmmode~\check{z}\else \v{z}\fi{}an}}, \bibinfo
  {author} {\bibfnamefont {D.}~\bibnamefont {Shiri}}, \bibinfo {author}
  {\bibfnamefont {I.-M.}\ \bibnamefont {Svensson}}, \bibinfo {author}
  {\bibfnamefont {G.}~\bibnamefont {Tancredi}}, \bibinfo {author}
  {\bibfnamefont {G.}~\bibnamefont {Johansson}}, \bibinfo {author}
  {\bibfnamefont {P.}~\bibnamefont {Delsing}}, \bibinfo {author} {\bibfnamefont
  {G.}~\bibnamefont {Ferrini}},\ and\ \bibinfo {author} {\bibfnamefont
  {J.}~\bibnamefont {Bylander}},\ }\href
  {https://doi.org/10.1103/PhysRevApplied.14.034010} {\bibfield  {journal}
  {\bibinfo  {journal} {Phys. Rev. Applied}\ }\textbf {\bibinfo {volume}
  {14}},\ \bibinfo {pages} {034010} (\bibinfo {year} {2020})}\BibitemShut
  {NoStop}%
\end{thebibliography}%
\end{document}